\documentclass[aps,prb,showpacs,twocolumn]{revtex4}

\usepackage[T1]{fontenc}
\usepackage{amssymb}
\usepackage{amsbsy}
\usepackage{amsmath}             
\usepackage{epsfig}

\begin{document}

\title{Berezinskii-Kosterlitz-Thouless transition and BCS-Bose
  crossover in the two-dimensional attractive Hubbard model }  
\author{N. Dupuis}
\affiliation{ Laboratoire de Physique des Solides, CNRS UMR 8502, \\
  Universit\'e Paris-Sud, 91405 Orsay, France }
\date{June 7, 2004}


\begin{abstract} 
We study the two-dimensional attractive Hubbard model using the
mapping onto the half-filled repulsive Hubbard model in a uniform
magnetic field coupled to the fermion spins. The low-energy effective
action for charge and pairing fluctuations is obtained in
the hydrodynamic regime. We recover the action of a Bose superfluid
where half the fermion density is identified as the conjugate
variable of the phase of the superconducting order parameter.
By integrating out charge fluctuations, we obtain
a phase-only action. In the zero-temperature superconducting state, 
this action describes
a collective phase mode smoothly evolving from the Anderson-Bogoliubov
mode at weak coupling to the Bogoliubov mode of a Bose superfluid
at strong coupling. At finite temperature, the phase-only action can
be used to extract an effective XY model and thus obtain the
Berezinskii-Kosterlitz-Thouless (BKT) phase transition temperature. We also
identify a renormalized classical regime of superconducting
fluctuations above the BKT phase transition, and a regime of incoherent pairs
at higher temperature. Special care is devoted to the
nearly half-filled case where the symmetry of the order parameter is
enlarged to SO(3) due to strong ${\bf q}=(\pi,\pi)$ charge
fluctuations. The low-energy effective action is then an SO(3)
non-linear sigma model with a (symmetry breaking) magnetic field
proportional to the doping. In the strong-coupling limit,
the attractive Hubbard model can be mapped onto the Heisenberg
model in a magnetic field, which reduces to the
quantum XY model (except for a weak magnetic field, i.e. in the
low-density limit of the attractive model). In the low-density
limit, the Heisenberg model allows to recover the action of a Bose
superfluid, including the $(\boldsymbol{\nabla}\rho)^2$ term
(with $\rho$ the density), and in turn the Gross-Pitaevskii equation.  
\end{abstract}


\maketitle

\section{Introduction} 
\label{sec:introduction}

Many superconducting systems such as high-$T_c$ superconductors, organic
conductors, heavy fermions systems, as well as ultracold
atomic Fermi gases cannot be understood within the BCS
theory. For instance, in high-$T_c$ superconductors, the low
dimensionality reinforces the role of phase
fluctuations. \cite{Loktev01,Emery95} 
The short coherence length in these systems also 
suggests that they might be in an intermediate regime between the
weak-coupling BCS limit of superfluid fermions and the strong-coupling
limit of  condensed composite bosons. \cite{Randeria95,BCS-Bose} In
ultracold atomic Fermi gases that are now  experimentally available,
it appears possible to monitor the evolution from the BCS  to the Bose
limit. \cite{trap} 

A superconducting system at low temperature is conveniently described
by a low-energy effective action written in terms of a few relevant
(bosonic) variables. The minimum description requires to consider the
phase of the superconducting order parameter, but other variables such
as the amplitude of the order parameter or the charge density may also
be included. The effective action (or the corresponding equations of
motion) is sufficient to describe macroscopic
quantum phenomena such as the Meissner effect, the flux quantization,
the Josephson effect, or the vortex dynamics. \cite{Feynman72,Ao95}
In a Bose superfluid, the low-energy effective action leads to the
Gross-Pitaevskii equation, \cite{Gross61,Pitaevskii61} i.e. a non-linear 
Schr\"odinger equation for the complex (superfluid) order parameter
$\Psi$ where $|\Psi|^2=\rho$ is the condensate density. In a 
Fermi system,  there is in general 
no  simple relation between the amplitude of the superconducting order
parameter and the density. In the strong-coupling limit, where
fermions form tightly bound pairs which behave as bosons, we expect
the Gross-Pitaevskii equation to hold. Moreover, since
Fermi and Bose superfluids should behave similarly in
many respects, a Fermi
superfluid should be described by a non-linear Schr\"odinger equation
similar to the Gross-Pitaevskii equation even in the weak-coupling
limit. \cite{Feynman72} This is the conclusion reached by previous
works, \cite{Stone95,Aitchison95,DePalo99} although the
``wavefunction'' in this case is not the superfluid order parameter
(except in the strong-coupling limit). 

The low-energy effective
action provides a convenient framework to discuss the BCS-Bose
crossover between the weak-coupling BCS limit and the Bose limit of
preformed pairs. \cite{DePalo99,Drechsler92,deMelo93,Pistolesi96,Stintzing97,Babaev99,Benfatto03} 
In two dimensions (2D), there is an additional motivation to introduce 
an effective action written in terms of the phase of the order
parameter. The Berezinskii-Kosterlitz-Thouless (BKT) phase
transition\cite{Berezinskii70,Kosterlitz73}  which
takes place in a 2D Fermi superfluid is clearly out of reach of
fermionic approximations based on diagram resummations like the
$T$-matrix approximation. \cite{Randeria95}

While the 2D attractive Hubbard model is not an appropriate
microscopic model for most superconducting systems of interest, it can
be used to understand a number of general issues relevant to many
cases. The main characteristics of this model are well-known. 
\cite{Scalettar89,Moreo91} Away from half-filling, there is a BKT
phase transition to a  
low-temperature superconducting phase. Long-range order sets in at
zero temperature and breaks SO(2) symmetry. At half-filling, ${\bf
  q}=0$ pairing and ${\bf q}=(\pi,\pi)$ charge fluctuations combine to form
an order parameter with SO(3) symmetry. The superconducting transition
then occurs at zero temperature and breaks SO(3) symmetry. In the 
weak-coupling limit, superconductivity is due to strongly overlapping
Cooper pairs, and 
Bogoliubov quasi-particle excitations dominate the low-energy
physics. In the strong-coupling limit, fermions form tightly bound pairs that
Bose condense at low temperature thus giving rise to
superfluidity. Pair-breaking excitations are not possible at low
energy and the thermodynamics is controlled by (collective)
phase fluctuations.

In this paper, we derive the low-energy effective action for charge
and pairing fluctuations in the 2D attractive Hubbard model on a
square lattice. We discuss the BCS-Bose 
crossover and the phase diagram. We use the mapping of the attractive
model onto the repulsive {\it half-filled} model in a
magnetic field which couples to the fermion spins. In this mapping,
the charge and pairing fields 
transform into the three components of the spin-density field. At
half-filling, the magnetic field vanishes in the repulsive model. This
case has been studied in detail in Refs.~\onlinecite{KB1,KB2}. At zero
temperature, the system evolves from a Slater to a Mott-Heisenberg
antiferromagnet as the interaction strength (i.e. the on-site Coulomb
repulsion) increases. Because of
the SO(3) symmetry of the order parameter, the N\'eel temperature vanishes 
in agreement with the Mermin-Wagner\cite{Mermin66} theorem. At finite 
temperature, the system is in a renormalized classical (RC)
regime with an exponentially large antiferromagnetic correlation
length. Away from half-filling, the magnetic field reduces the
symmetry to SO(2) so that a BKT phase transition occurs at finite
temperature. In the attractive model, this corresponds to the
suppression of ${\bf q}=(\pi,\pi)$ charge fluctuations at low energy.   
The main advantage of studying the repulsive model is that
the SO(3) symmetry of the order parameter at or near
half-filling can be easily handled. 

In Sec.~\ref{sec:O2}, we map the attractive model onto the repulsive
one by a canonical particle-hole transformation\cite{Micnas90} and then 
obtain the effective action for the spin fluctuations in the presence
of a finite magnetic field (i.e. away from half-filling in the
attractive model). Collective bosonic fields are introduced by means of a
Hubbard-Stratonovich decoupling of the Hubbard interaction. This
crucial step in our approach differs from the usual decoupling in
two respects. First, we write the interaction in an explicit SO(3)
spin-rotation form by introducing a unit vector ${\bf\Omega_r}$ at each
site and time. \cite{Weng91,KB1,KB2,Schulz95} This allows to recover
the Hartree-Fock (HF) theory at the saddle-point level, while
maintaining SO(3) spin-rotation symmetry (in the absence 
of the magnetic field). Second, we introduce two auxiliary real
fields. One is simply the Hubbard-Stratonovich field $m^{\rm HS}_{\bf
  r}$ for the amplitude of the spin density at site ${\bf r}$. The
other one, $m_{\bf r}$, is directly connected to the actual amplitude
of the spin density. \cite{note0} While these two  
fields are proportional at the saddle-point (i.e. Hartree-Fock) level,
they differ when fluctuations are taken into account. In the weak- and
strong-coupling limits, $\frac{m_{\bf r}}{2}{\bf\Omega_r}$ 
can be identified with the spin density, which allows a direct
interpretation of the low-energy action $S[m,{\bf\Omega}]$ in terms of
physical quantities. Because of the magnetic field, the spin component
along the field, $S^z$, takes a finite value and its fluctuations are
small at low energy. Since spin amplitude fluctuations are also small, the
important fluctuations correspond to rotations around the magnetic
field axis. We derive the effective action $S[m,{\bf\Omega}]$ in this
case. $S[m,{\bf\Omega}]$ takes a simple form in the 
weak-coupling (Slater) and  strong-coupling (Mott-Heisenberg)
limits. These two limits differ in the role of the Berry phase term.

In Sec.~\ref{subsec:leea1}, we deduce the effective action
$S[\rho,\Delta]$ of the charge ($\rho$) and pairing
($\Delta=|\Delta|e^{i\Theta}$) fluctuations in the attractive
model. By integrating out amplitude fluctuations ($|\Delta|$), we
recover the action $S[\rho,\Theta]$ of a Bose superfluid where
half the fermion density is identified as the conjugate variable of
the phase $\Theta$ of the superconducting order parameter. This action
is parametrized by the mass $m_b$ of the ``bosons'' and the amplitude
$g$ of the repulsive interaction between ``bosons''. $m_b$ and $g$ are
computed as a 
function of particle density and interaction strength. We then
analyze in more detail the weak (BCS) and strong (Bose) coupling 
limits. In the BCS limit, we find that the action is a function of charge
and phase fluctuations only, since amplitude fluctuations of the
superconducting order parameter decouple. In the
strong-coupling limit, the amplitude of the superconducting order
parameter and the fermion density $\rho$ satisfy the relation
$|\Delta_{\bf r}|=\frac{1}{2}\sqrt{\rho_{\bf r}(2-\rho_{\bf
    r})}$. The Bose superfluid action $S[\rho,\Theta]\equiv
S[|\Delta|,\Theta]$ then entirely describes the dynamics of the
superconducting order parameter $\Delta$. In
Sec.~\ref{subsec:phase-only}, we derive the 
phase-only action $S[\Theta]$ by integrating out charge
fluctuations. $S[\Theta]$ corresponds to an O(2) sigma model with an
additional term proportional to the first-order time derivative
of $\Theta$. The phase stiffness and the velocity of the O(2) sigma
model are obtained as a function of particle density and interaction 
strength. At zero temperature, superconducting long-range order gives
rise to a gapless (Goldstone) phase mode smoothly
evolving from the Anderson-Bogoliubov mode\cite{Anderson58} at weak
coupling to the Bogoliubov mode\cite{Bogoliubov58} of a Bose
superfluid at strong coupling. In 
Sec.~\ref{subsec:kt}, we show how we can extract from the phase-only
action an effective (classical) XY model whose phase stiffness is a
function of density, interaction strength and temperature. This allows 
us to determine the value of the BKT phase transition temperature as a
function of density and interaction strength. The XY model also
yields an estimate of the crossover temperature $T_X$ below which the
system enters a RC regime of phase fluctuations. At higher
temperature, for $T_X \leq T \leq T_{\rm
  pair}$, there is a regime of incoherent pairs [Cooper (local) pairs
at weak (strong) coupling] with no superconducting short-range
order. $T_{\rm pair}$ is estimated from the HF transition
temperature. 

In the vicinity of
half-filling, the analysis of Sec.~\ref{sec:O2} is not sufficient
since ${\bf q}=(\pi,\pi)$ charge fluctuations (in the attractive
model) are not considered. For a weak magnetic field (in the
repulsive model), there are strong fluctuations of $S^z$, and the
analysis of Sec.~\ref{sec:O2} breaks down. This case is dealt with in
Sec.~\ref{sec:O3}. We show that the dynamics of spin fluctuations in
the repulsive model is governed by an SO(3) non-linear-sigma model
with a (symmetry-breaking) magnetic field proportional to the
doping. The magnetic field
defines a characteristic temperature $T_{\rm SO(3\to 2)}$ above which
the SO(3) spin-rotation symmetry is restored. Below $T_{\rm SO(3\to
  2)}$, the system enters a RC regime of spin
fluctuations with SO(2) symmetry. The global phase diagram, as a
function of density, interaction strength and temperature, is
discussed in Sec.~\ref{subsec:diag}. [For clarity, we discuss the
phase diagram at the end of Sec.~\ref{sec:O2} and postpone the
technical analysis of the SO(3)$\to$SO(2) crossover near half-filling
to Sec.~\ref{sec:O3}.]

In Sec.~\ref{sec:scr}, we consider the strong-coupling limit in more
detail. First we show that the repulsive Hubbard model reduces to the
Heisenberg model to leading order in $1/U$. This allows to obtain the
collective modes beyond the long-wavelength approximation. We then
show that except for a strong magnetic field (i.e. in the low density 
limit of the attractive model), the Heisenberg model in 
a magnetic field reduces to the quantum XY model. In the
low-density limit, we recover from the Heisenberg model the usual
action of a Bose superfluid (for bosons of mass $1/J=U/4t^2$
and density $\rho/2$),
including the terms proportional to $(\boldsymbol{\nabla}\rho_{\bf
  r})^2$ that were omitted in Sec.~\ref{sec:O2}. The classical
equation of motion derived from this action is nothing but 
the Gross-Pitaevskii equation. We therefore obtain a
correspondence between the Gross-Pitaevskii equation in the attractive
model and the semiclassical spin dynamics in the repulsive model. 

To our knowledge, there is no systematic study of the 2D repulsive
Hubbard model in a magnetic field. When translated in
the language of the attractive model, our results reproduce, in a
unique framework, a number of previously known results. We also obtain
new results, in particular regarding the BCS-Bose crossover
and the phase diagram.

\section{Away from half-filling}
\label{sec:O2} 

The attractive Hubbard model on a square lattice is defined by the
Hamiltonian   
\begin{equation}
H = - \sum_{\bf r} 
c^\dagger_{\bf r} (\hat t+\mu) c_{\bf r} 
-U \sum_{\bf r} n_{{\bf r}\uparrow} n_{{\bf r}\downarrow} 
\label{ham}
\end{equation}
where $\hat t$ is the nearest-neighbor hopping operator:
\begin{equation}
\hat t c_{\bf r} = t(c_{{\bf r}+\hat {\bf x}}+c_{{\bf r}-\hat {\bf x}}
+c_{{\bf r}+\hat {\bf y}}+c_{{\bf r}-\hat {\bf y}}).
\end{equation}
$\hat {\bf x}$ and $\hat {\bf y}$ denote unit vectors along the $x$
and $y$ axis. With the notation of Eq.~(\ref{ham}), $-U$ is the
on-site interaction with $U\geq 0$ in the 
attractive case.  The operator $c^\dagger_{{\bf r}\sigma}$ ($c_{{\bf
    r}\sigma}$) creates (annihilates) a fermion of spin $\sigma$
at the lattice site ${\bf r}$, $c_{\bf r}=(c_{{\bf r}\uparrow},c_{{\bf
r}\downarrow})^T$, and  $n_{{\bf r}\sigma}=
c^\dagger_{{\bf r}\sigma} c_{{\bf r}\sigma}$. At half-filling,
particle-hole symmetry implies that the chemical potential
$\mu$ equals $-U/2$. In the following, we will consider
only hole doping so that $\mu\leq -U/2$. We take the lattice spacing
equal to unity and $\hbar=k_B=1$. 

Under the canonical particle-hole transformation\cite{Micnas90}
\begin{eqnarray}
c_{{\bf r}\downarrow} \to (-1)^{\bf r} c^\dagger_{{\bf r}\downarrow},
\,\,\,\,\,\,\, 
c^\dagger_{{\bf r}\downarrow} \to (-1)^{\bf r} c_{{\bf
    r}\downarrow},
\label{phtr}
\end{eqnarray}
the Hamiltonian becomes (omitting a constant term)
\begin{equation}
H = - \sum_{\bf r} 
c^\dagger_{\bf r} \left[\hat t+\left(\mu+\frac{U}{2}\right)  \sigma^z
  + \frac{U}{2} \right] c_{\bf r}  
+U \sum_{\bf r}n_{{\bf r}\uparrow} n_{{\bf r}\downarrow} ,
\label{hamtr}
\end{equation}
where $(\sigma^x,\sigma^y,\sigma^z)$ denotes the Pauli matrices. The
transformed Hamiltonian (\ref{hamtr}) corresponds to the repulsive
half-filled Hubbard model in a magnetic field 
\begin{equation}
h_0 \equiv \mu + \frac{U}{2} 
\end{equation}
along the $z$ axis coupled to the fermion spins. In the attractive
model, the chemical potential 
$\mu$ is fixed by the condition $\langle c^\dagger_{\bf r}c_{\bf
  r}\rangle=\rho_0=1-x$ where $\rho_0$ is the mean density and $x$ the
doping. Under the particle-hole transformation 
(\ref{phtr}), $c^\dagger_{\bf r}c_{\bf r}\to c^\dagger_{\bf r} \sigma^z c_{\bf
  r}+1$. In the repulsive model, the magnetic
field $h_0$ is then determined by 
\begin{equation} 
\langle c^\dagger_{\bf r} \sigma^z c_{\bf r} \rangle =-x.
\label{h0def}
\end{equation}

The charge-density and pairing operators transform as
\begin{eqnarray}
\rho_{\bf r} = c^\dagger_{\bf r} c_{\bf r} &\to&
  (2S^z_{\bf r}+1) , \nonumber  \\ 
 \Delta_{\bf r} = c_{{\bf r}\downarrow} c_{{\bf r}\uparrow} &\to&
  (-1)^{\bf r} S^-_{\bf r}, \nonumber \\
 \Delta^\dagger_{\bf r} = c^\dagger_{{\bf r}\uparrow} c^\dagger_{{\bf
  r}\downarrow} &\to& (-1)^{\bf r} S^+_{\bf r}, 
\label{transformation}
\end{eqnarray}
where $S^\nu_{\bf r}=c^\dagger_{\bf r}\frac{\sigma^\nu}{2} c_{\bf r}$
($\nu=x,y,z$) and $S^\pm_{\bf r}=S^x_{\bf r}\pm iS^y_{\bf r}$. 
In the repulsive model, spin fluctuations are clearly the collective (bosonic)
fluctuations of interest. Accordingly, in the attractive model, both
pairing and charge fluctuations should be considered on equal
footing. One of the motivations to study the repulsive model is that
one has to consider only the particle-hole channel.  On
the contrary, a direct study of the attractive model 
would require to consider both the particle-hole and particle-particle
channels in order to take into account the charge and pairing
fluctuations. The simultaneous introduction of auxiliary 
Hubbard-Stratonovich fields in these two channels is not without
problem if one requires the saddle-point approximation to recover the
HF (or mean-field) theory. \cite{DePalo99} 
Furthermore, in the attractive model at half-filling, ${\bf q}=0$
pairing and ${\bf q}=(\pi,\pi)$ charge fluctuations combine to form an order
parameter with SO(3) symmetry. The SO(3) symmetry of the order 
parameter is much more easily handled in the repulsive model, where
the distance from half-filling determines the (symmetry-breaking)
magnetic field $h_0$ (see Sec.~\ref{sec:O3}). 

We can write the partition function of the repulsive model as a path
integral over Grassmann fields $c^*_{{\bf r}\sigma},c_{{\bf
    r}\sigma}$, with the action 
\begin{equation}
S = \int_0^\beta d\tau \Bigl\lbrace \sum_{\bf r} 
c^\dagger_{\bf r}\partial_\tau c_{\bf r}+H[c^\dagger,c] \Bigr\rbrace ,
\end{equation}
where $\tau$ is an imaginary time and $\beta=1/T$ the inverse
temperature. $H[c^\dagger,c]$ is obtained from the Hamiltonian (\ref{hamtr}) by
replacing the operators by the corresponding Grassmann fields. We now
introduce auxiliary bosonic fields for the collective spin fluctuations in a
way that fulfills the three following requirements: i) the
HF (or mean-field) approximation is recovered from a
saddle-point approximation; ii) the SO(2) (or SO(3) if $h_0=0$)
spin-rotation symmetry is maintained; iii) the auxiliary fields
correspond to the spin-density field ${\bf S_r}=c^\dagger_{\bf
  r}\frac{\boldsymbol{\sigma}}{2} c_{\bf r}$ not only at the
saddle-point level, but also when fluctuations are taken into
account [$\boldsymbol{\sigma}=(\sigma^x,\sigma^y,\sigma^z)$]. 

We start from the identity\cite{Weng91,KB1,KB2}
\begin{equation}
n_{{\bf r}\uparrow} n_{{\bf r}\downarrow} =
\frac{1}{4} \left[\left(c^\dagger_{\bf r}c_{\bf r}\right)^2-
\left(c^\dagger_{\bf r}\boldsymbol{\sigma}\cdot {\bf\Omega_r} c_{\bf
  r}\right)^2\right], 
\label{decomp}
\end{equation}
where ${\bf\Omega_r}$ is an arbitrary site- and time-dependent unit
vector. ${\bf\Omega_r}$ is defined by its polar and azimuthal angles
$\theta_{\bf r},\varphi_{\bf r}$. Spin-rotation invariance is made
explicit by performing an 
angular integration over ${\bf\Omega_r}$ at each site and time (with a
measure normalized to unity). The charge term $(c^\dagger_{\bf
  r}c_{\bf r})^2$ is decoupled by means of an auxiliary (real) field
$\Delta_{c{\bf r}}$. In order to decouple the spin term $(c^\dagger_{\bf
  r} \boldsymbol{\sigma}\cdot {\bf\Omega}_{\bf r} c_{\bf r})^2$, we
introduce in the path integral the unit factor
\begin{eqnarray}
1 &=& \int {\cal D}[m] \prod_{{\bf r},\tau} \delta (m_{\bf r}- c^\dagger_{\bf
  r} \boldsymbol{\sigma}\cdot {\bf\Omega}_{\bf r} c_{\bf r}) \nonumber
  \\ &=& \int {\cal D}[m,m^{\rm HS}] e^{-\int_0^\beta d\tau \sum_{\bf
  r} im^{\rm HS}_{\bf r} (m_{\bf r}- c^\dagger_{\bf
  r} \boldsymbol{\sigma}\cdot {\bf\Omega}_{\bf r} c_{\bf r}) } ,
\label{decomp1}
\end{eqnarray}
where $m^{\rm HS}_{\bf r}$ is a Lagrange multiplier field which
imposes the constraint $m_{\bf r}= 2 {\bf S_r} \cdot {\bf\Omega_r}$.
Both $m$ and $m^{\rm HS}$ are real fields. 
Note that integrating out the $m_{\bf r}$ field
[see Eqs.~(\ref{action1a}-\ref{action1b}) below], one obtains the action
$S[c^\dagger,c,\Delta_c,m^{\rm HS}]$ where $\Delta_{c{\bf
    r}}$ and $m^{\rm HS}$ are the Hubbard-Stratonovich fields which
decouple the interaction term (\ref{decomp}). In general, 
$\frac{m_{\bf r}}{2}{\bf\Omega_r}$ cannot be directly identified with the spin
density ${\bf S_r}$, but the identification turns out to be
correct in the hydrodynamic regime both at weak ($U\ll 4t$) and strong
($U\gg 4t$) coupling.  

\begin{figure}
\epsfxsize 5.5cm
\epsffile[40 450 230 670]{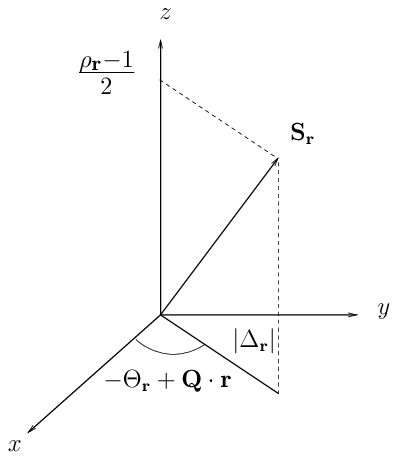}
\caption{Correspondence between the spin density ${\bf S_r}$ in the
  repulsive model, and the charge density $\rho_{\bf r}$ and
  superconducting field $\Delta_{\bf r}=|\Delta_{\bf
  r}|e^{i\Theta_{\bf r}}$ in the attractive model [see
  Eqs.~(\ref{transformation})].  } 
\label{fig:correspondence}
\end{figure}

Using (\ref{decomp}) and (\ref{decomp1}), we write the action as
$S=S_0+S_{\rm int}$ with 
\begin{eqnarray}
S_0 &=& \int_0^\beta d\tau  \sum_{\bf r} c^\dagger_{\bf r} 
\left(\partial_\tau-\hat t -h_0\sigma^z-\frac{U}{2} \right)c_{\bf r} , 
\label{action1a}
\\ 
S_{\rm int} &=& \int_0^\beta d\tau \sum_{\bf r} \biggl[
\frac{\Delta^2_{c{\bf r}}}{U} -\frac{U}{4} m^2_{\bf r} +i m_{\bf r}m^{\rm
    HS}_{\bf r} \nonumber \\ &&  
-c^\dagger_{\bf r}(i\Delta_{c{\bf r}}+im^{\rm HS}_{\bf r}
\boldsymbol{\sigma} \cdot 
{\bf\Omega_r}) c_{\bf r} \biggr] .
\label{action1b}
\end{eqnarray}
In the following, we shall consider the charge field $\Delta_{c{\bf
    r}}$ only at the saddle-point level, i.e. $\Delta_{c{\bf
r}}=i(U/2)\langle c^\dagger_{\bf r}c_{\bf r}\rangle =iU/2$. The term
$-ic^\dagger_{\bf r} \Delta_{c{\bf r}} c_{\bf r}$ in
(\ref{action1b}) cancels the chemical potential term
$-(U/2)c^\dagger_{\bf r}c_{\bf r}$ in (\ref{action1a}). 

Eqs.~(\ref{action1a}-\ref{action1b}) are the starting point of our analysis. In
Sec.~\ref{subsec:HF}, we show that the HF theory is recovered from a
saddle-point approximation over the auxiliary fields $m_{\bf
    r}$, $m^{\rm HS}_{\bf r}$ and ${\bf\Omega_r}$. In the following
sections, we go beyond the HF theory and
derive a low-energy effective action $S[m,{\bf\Omega}]$ for spin
fluctuations, and deduce the effective action
$S[\rho,\Delta]$ of charge and pairing fluctuations in the
attractive model.

\subsection{Hartree-Fock theory}
\label{subsec:HF}

In the presence of the uniform magnetic field along the $z$ axis, we
expect the ground state to exhibit AF order in the $(x,y)$ plane and a
ferromagnetic order along the $z$ axis. We therefore consider a static
saddle-point approximation with $m_{\bf r}=m_0$, $m^{\rm HS}_{\bf
  r}=m_0^{\rm HS}$, and a classical configuration of 
the unit vector field ${\bf\Omega_r}$ given by
\begin{equation}
{\bf\Omega}_{\bf r}^{\rm cl} = (-1)^{\bf r}\sin\theta_0 \hat {\bf x} 
+\cos\theta_0 \hat {\bf z}.
\end{equation}
The HF action then reads
\begin{eqnarray}
S_{\rm HF} &=& \beta N \Bigl(-\frac{U}{4}m_0^2+im_0m^{\rm HS}_0 \Bigr) 
\nonumber \\ && 
+ \int_0^\beta d\tau \sum_{\bf r} c^\dagger_{\bf r} 
[\partial_\tau -\hat t -h \sigma^z-\Delta_0^{\rm HS}(-1)^{\bf r} \sigma^x]
c_{\bf r}, \nonumber \\ & & 
\label{HFaction}
\end{eqnarray}
where $N$ is the total number of lattice sites, and 
\begin{eqnarray}
\Delta_0^{\rm HS} &=& im_0^{\rm HS} \sin\theta_0 , \nonumber \\
h &=& h_0 + im_0^{\rm HS}\cos\theta_0 . \label{delta_h_def}
\end{eqnarray}
The HF action is quadratic and can be easily diagonalized. The
single-particle Green's functions are given by
\begin{eqnarray}
G_\sigma({\bf k},i\omega) &=& -\langle c_\sigma({\bf k},i\omega)
c^*_\sigma({\bf k},i\omega) \rangle \nonumber \\ 
&=& \frac{-i\omega -\epsilon_{{\bf k}\sigma}}{\omega^2+E^2_{{\bf
      k}\sigma}} , 
\nonumber \\
F_\sigma({\bf k},i\omega) &=& -\langle c_\sigma({\bf k},i\omega)
c^*_{\bar\sigma}({\bf k}+{\bf Q},i\omega) \rangle \nonumber \\ 
&=& \frac{\Delta_0^{\rm HS}}{\omega^2+E^2_{{\bf k}\sigma}} ,
\label{GF}
\end{eqnarray}
where 
\begin{eqnarray}
\epsilon_{{\bf k}\sigma} &=& \epsilon_{\bf k}-\sigma h , \nonumber \\ 
E_{{\bf k}\sigma} &=& \sqrt{\epsilon^2_{{\bf k}\sigma}+{\Delta^{\rm HS}_0}^2} .
\end{eqnarray}
$\epsilon_{\bf k}=-2t(\cos k_x+\cos k_y)$ is the energy of the free
fermions on the square lattice. 
$\bar\sigma=-\sigma$, ${\bf Q}=(\pi,\pi)$, and $\omega=\pi T(2n+1)$
($n$ integer) is a fermionic Matsubara frequency. $c_\sigma({\bf
  k},i\omega)$ is the Fourier transformed field of $c_{{\bf
    r}\sigma}$. 

The saddle-point
equations are obtained from $\partial Z_{\rm HF}/\partial
m_0=\partial Z_{\rm HF}/\partial \Delta_0^{\rm HS}=\partial Z_{\rm
  HF} /\partial h=0$, where $Z_{\rm HF}$ is the partition function in
the HF approximation:
\begin{eqnarray}
m_0 &=& \frac{2}{U} im^{\rm HS}_0 , \\ 
\Delta_0 &\equiv& \frac{\Delta_0^{\rm HS}}{U} \nonumber \\ &=&  
\frac{m_0}{2} \sin\theta_0 \nonumber \\ &=& 
\frac{(-1)^{\bf r}}{2} \langle c^\dagger_{\bf r} \sigma^x c_{\bf r} \rangle , 
\label{spteq1}  \\
m_0 \cos\theta_0 &=& \frac{2}{U} (h-h_0) \nonumber \\ &=& \langle
c^\dagger_{\bf r} \sigma^z c_{\bf r} \rangle . 
\label{spteq2}
\end{eqnarray}
Eq.~(\ref{h0def}) then implies 
\begin{eqnarray}
m_0 \cos\theta_0 &=& \frac{2}{U} (h-h_0) \nonumber \\ &=& -x. 
\label{spteq2bis}
\end{eqnarray}
$\Delta_0=(-1)^{\bf r}\langle S^x_{\bf r}\rangle$ is the AF order
parameter in the repulsive model, and the superconducting order
parameter in the attractive model. $h$ is an effective magnetic field
which takes into account the 
mean ferromagnetic magnetization $\langle c^\dagger_{\bf r}\sigma^z
c_{\bf r}\rangle $ along the $z$ axis. In the attractive model,
$h=h_0-xU/2=\mu+\rho_0 U/2$ corresponds to the chemical potential
renormalized by the Hartree self-energy.  

\begin{figure}
\epsfxsize 6.cm
\epsffile[50 460 245 590]{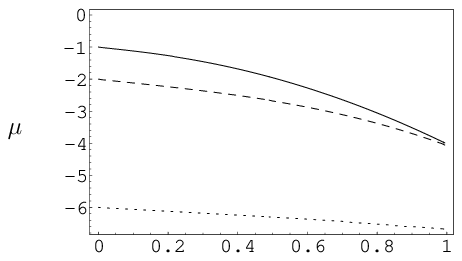}
\epsfxsize 6.cm
\epsffile[50 460 245 590]{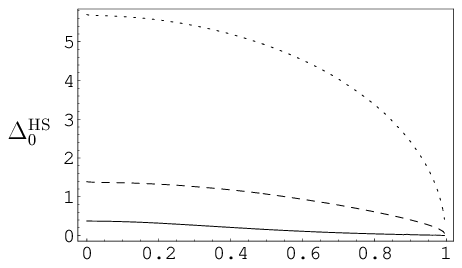}
\epsfxsize 6.cm
\epsffile[50 460 245 590]{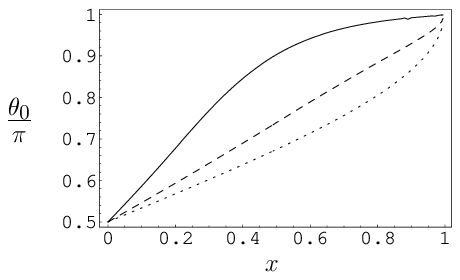}
\caption{Chemical potential $\mu=h_0-U/2$,  order parameter
  $\Delta_0^{\rm HS}$ and $\theta_0$  in the $T=0$ HF state 
{\it vs} doping $x=1-\rho_0$ for 
  $U=2t$, $4t$  and $12t$ (solid, dashed, and dotted
  lines, respectively). The
  energies are measured in units of $t$.  }  
\label{fig:mu_delta_theta_U_fixe}
\end{figure}

\begin{figure}
\epsfxsize 6.cm
\epsffile[50 460 245 590]{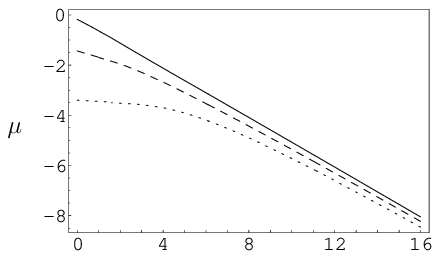}
\epsfxsize 6.cm
\epsffile[50 460 245 590]{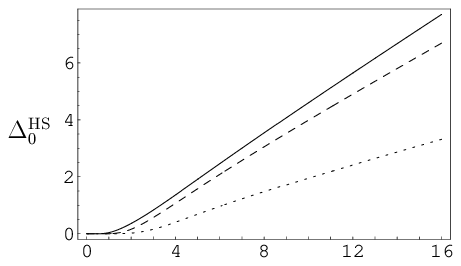}
\epsfxsize 6.cm
\epsffile[50 460 245 590]{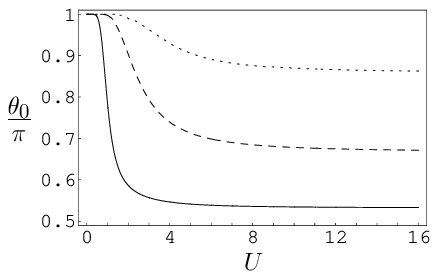}
\caption{Chemical potential $\mu=h_0-U/2$,  order parameter $\Delta_0^{\rm HS}$
  and $\theta_0$ in the $T=0$ HF state 
{\it vs} $U$ for $x=0.1$, $0.5$ and $0.9$ (solid,
  dashed, and dotted lines, respectively). Here and in the following
  figures, we use the analytical results in the weak-coupling regime
  ($U\lesssim t$) where the numerics becomes difficult because of the
  exponentially small value of $\Delta_0^{\rm HS}$. } 
\label{fig:mu_delta_theta_dopage_fixe}
\end{figure}

Using Eqs.~(\ref{GF}), we rewrite the saddle-point
equations (\ref{spteq1}-\ref{spteq2bis}) as 
\begin{eqnarray}
&& \frac{2}{U} =  \int_{\bf k}
\frac{\tanh(\beta E_{{\bf k}\uparrow}/2)}{E_{{\bf k}\uparrow}} , 
\label{gapeq1} \\
&& x =  \int_{\bf k} \epsilon_{{\bf k}\uparrow}
\frac{ \tanh(\beta E_{{\bf
      k}\uparrow}/2)}{E_{{\bf k}\uparrow}} ,
\label{gapeq2} 
\end{eqnarray}
where $\int_{\bf k}=\int_{-\pi}^{\pi}\frac{dk_x}{2\pi}
\int_{-\pi}^{\pi}\frac{dk_y}{2\pi}$.  
At $T=0$, Eqs.~(\ref{gapeq1}-\ref{gapeq2}) determine the
superconducting order parameter 
$\Delta_0^{\rm HS}$ (or, equivalently, $\Delta_0$) and the
renormalized chemical potential $h$. Using 
(\ref{delta_h_def},\ref{spteq2bis}), we then obtain $h_0$,
$im_0^{\rm HS}$ and 
$\theta_0$ as a function of $U$ and $x$
[Figs.~\ref{fig:mu_delta_theta_U_fixe}-\ref{fig:mu_delta_theta_dopage_fixe}].
Eqs.~(\ref{gapeq1}-\ref{gapeq2}) also  
determine the HF transition temperature $T_c^{\rm HF}$ where
AF long-range  order (i.e. superconducting order in the attractive
model) sets in.  At the transition, we have $\Delta_0=\Delta_0^{\rm HS}=0^+$,
$\theta_0=\pi$ and $im_0^{\rm HS}=xU/2$. 

Partial analytical results can be obtained in the limits of weak and strong
couplings. 

\subsubsection{Weak coupling ($U\ll 4t$)}

At half-filling ($x=0$), $h=h_0=0$ and $\theta_0=\pi/2$. The
zero-temperature order parameter and the transition temperature are given by
\begin{eqnarray}
\Delta_0^{\rm HS} &\simeq& 32t e^{-2\pi \sqrt{t/U}} , \nonumber \\
T_c^{\rm HF} &\sim& t e^{-2\pi\sqrt{t/U}} .
\end{eqnarray}
Note that at half-filling, the SO(3) symmetry is restored and the AF order
parameter can have a component along the $z$ axis. The latter
corresponds to a ${\bf q}={\bf Q}$ charge-density-wave order in the
attractive model. The choice $\theta_0=\pi/2$ corresponds to a state
with only superconducting order. 

Away from half-filling ($h_0<0$), $\theta_0\simeq \pi$ and $im_0^{\rm
  HS} \simeq 
xU/2$ since the superconducting order parameter is exponentially
small. At $T=0$, the superconducting order parameter is determined by 
\begin{eqnarray}
\frac{2}{U} &=& \int_{-4t}^{4t} \frac{d\epsilon}{E} {\cal N}_0(\epsilon) 
\nonumber \\ &\simeq & 
{\cal N}_0(h) \int_{-4t}^{4t} \frac{d\epsilon}{E}  ,
\label{gap1}
\end{eqnarray}
where $E=[(\epsilon-h)^2+{\Delta_0^{\rm HS}}^2]^{1/2}$. ${\cal
  N}_0(\epsilon)=(2\pi^2t)^{-1}K[(1-\epsilon^2/16t^2)^{1/2}]$, with
$\epsilon\in [-4t,4t]$, is the density of states of free fermions on 
a square lattice [$K$ is the complete elliptic
integral of the first kind]. Since the integral in (\ref{gap1}) is peaked
around $\epsilon=h$ for $\Delta_0^{\rm HS}\to 0$, we have replaced the density
of states ${\cal N}_0(\epsilon)$ by its value at the renormalized
  chemical potential 
$h$. In the weak-coupling limit, we can also neglect the effect of the
order parameter on the chemical potential (i.e. set $\Delta_0^{\rm HS}=0$ in
(\ref{gapeq2})). We then have $h\simeq 
\epsilon_F$ where $\epsilon_F$ is the Fermi energy of the
non-interacting system. With these approximations, we obtain (for
  $U\to 0$ and $x$ fixed) 
\begin{eqnarray}
\Delta_0^{\rm HS} &\simeq& 2[(4t)^2-\epsilon_F^2]^{1/2} e^{-\frac{1}{U{\cal
      N}_0(\epsilon_F)}} , \nonumber \\
x &\simeq& 
\left \lbrace 
\begin{array}{ll}
\frac{1}{\pi^2 t} |\epsilon_F| \ln \frac{16et}{|\epsilon_F|} & {\rm if} \,\,\,
 x\ll 1,  \nonumber \\ 
 1-\frac{4t-|\epsilon_F|}{2\pi t} & {\rm if} \,\,\, 1-x\ll 1 ,
\end{array} \right. 
\nonumber \\ 
\mu &\simeq& \epsilon_F - \rho_0 \frac{U}{2}. 
\label{deltabcs}
\end{eqnarray}
Since $h\simeq \epsilon_F$ belongs to the non-interacting band
$[-4t,4t]$, the excitation gap (i.e. the minimum energy required to
break a pair) equals $2\Delta_0^{\rm HS}$. 

Similar arguments show that the transition temperature is given by
\begin{equation}
T_c^{\rm HF} \simeq \frac{2\gamma}{\pi} 
[(4t)^2-\epsilon_F^2]^{1/2} e^{-\frac{1}{U{\cal N}_0(\epsilon_F)}},
\label{Tbcs}
\end{equation}
where $\gamma\simeq 1.78$ is the exponential of the Euler constant. We
recover the results of the BCS theory with
$[(4t)^2-\epsilon_F^2]^{1/2}$ playing the role of the cutoff energy
[Eqs.~(\ref{deltabcs},\ref{Tbcs})].

\subsubsection{Strong coupling ($U\gg 4t$)}
\label{subsubsec:sc1}

To leading order in $1/U$ (and $x$ fixed), at $T=0$ we find 
$im_0^{\rm HS}=U/2$ (or $m_0=1$) and 
$\cos\theta_0=-x=h_0/2J$. This gives 
\begin{eqnarray}
\Delta_0^{\rm HS} &=& \frac{U}{2} (1-x^2)^{1/2} , \nonumber \\
\mu &=& -\frac{U}{2} - 2Jx . 
\end{eqnarray}
In the strong-coupling limit, the excitation gap equals
$2(h^2+{\Delta_0^{\rm HS}}^2)^{1/2}=U$ to leading order in $1/U$. 
For $x\ll 1$, $T_c^{\rm HF} \simeq U/4$.

\subsection{Effective action $S[m,{\bf\Omega}]$}
\label{subsec:leea}

In 2D, the HF theory breaks down at finite temperature since it
predicts AF long-range order below $T_c^{\rm HF}$.
Nevertheless, the HF transition temperature bears a physical
meaning as a crossover temperature below which the
amplitude $\Delta_0$ of the AF order parameter takes a finite
value. This can be interpreted as the appearance of local moments
perpendicular to the magnetic field and with an amplitude
$\Delta_0=\Delta_0^{\rm HS}/U$. Note that at weak-coupling these
``local'' moments can be defined only at length scales of order
$\xi_0 \sim t/\Delta_0^{\rm HS}\gg 1$, which corresponds to the size
of bound particle-hole pairs in the HF state. Thus, {\it stricto
  sensu}, local moments form only in the strong-coupling limit when
$\xi_0 \sim 1$.  

Below $T_c^{\rm HF}$, the fluctuations of the fields $m$, $m^{\rm HS}$ and
$\theta$ around their HF values are therefore expected to be
small. Below a crossover temperature $T_X\leq T_c^{\rm 
  HF}$, AF short-range order sets in. The AF correlation length
becomes much larger than the lattice spacing, and the effective action
for AF fluctuations (which correspond to rotations of ${\bf\Omega_r}$
about the $z$ axis) can be
derived within a gradient expansion. In this section, we derive the
low-energy effective action $S[m,{\bf\Omega}]$ for temperatures below
the crossover temperature $T_X$. For
$T\ll T_X\leq T_c^{\rm HF}$, the coefficients of the effective action
(which are related to HF quantities) can be evaluated in the zero
temperature limit. This means that we neglect the exponentially
small number of thermally excited quasi-particles which give rise to
non-analytic contributions (Landau damping terms) to the effective
action. \cite{Aitchison00,Shaparov01} As shown 
below, $T_X\sim J(1-x^2)/2 \ll T_c^{\rm HF}$ in the strong-coupling
limit ($J=4t^2/U$), but $T_X\sim T_c^{\rm HF}$ in the weak-coupling 
limit. 

Fluctuations can be parametrized by 
\begin{eqnarray}
\delta m_{\bf r} &=& m_{\bf r}-m_0, \nonumber \\ 
\delta m^{\rm HS}_{\bf r} &=& m^{\rm HS}_{\bf r}-m^{\rm HS}_0,
\nonumber \\  
p_{\bf r} &=& \frac{\theta_{\bf r}-\theta_0}{2} ,   \nonumber \\
q_{\bf r} &=& \varphi_{\bf r}- {\bf Q}\cdot {\bf r} . \label{qdef} 
\end{eqnarray}
In the HF state, $\delta m_{\bf r}=\delta m^{\rm HS}_{\bf r}=p_{\bf
  r}=q_{\bf r}=0$.
The effective action $S[p,q,m^{\rm HS},m]$ is obtained by integrating
out the fermions, and assuming $p_{\bf r},\delta m_{\bf r},\delta
  m^{\rm HS}_{\bf r}$ and $\partial_\mu p_{\bf r},\partial_\mu \delta
  m_{\bf r},\partial_\mu \delta m^{\rm HS}_{\bf r}, \partial_\mu q_{\bf
  r}$ to be small ($\mu=0,x,y$ and $\partial_0\equiv \partial_\tau$). 
We do not assume
  $q_{\bf r}$ to be small so that our approach is valid even in the
  absence of AF long-range order. It is convenient
to introduce a new fermionic variable $\phi_{\bf r}=(\phi_{{\bf
    r}\uparrow}, \phi_{{\bf r}\downarrow})^T$ defined by
$c_{\bf r}=R_{\bf r}\phi_{\bf r}$ where $R_{\bf r}$ is a time- and
site-dependent SU(2)/U(1) matrix satisfying
\begin{equation}
R_{\bf r} \boldsymbol{\sigma}\cdot {\bf\Omega}^{\rm cl}_{\bf r} 
R^\dagger_{\bf r} = \boldsymbol{\sigma}\cdot {\bf\Omega_r} .
\end{equation}
The above definition means that ${\cal R}_{\bf r}$, the SO(3) element
associated to $R_{\bf r}$, maps ${\bf\Omega}^{\rm cl}_{\bf r}$ onto
${\bf\Omega_r}$. The U(1) gauge freedom is due to rotations around
${\bf\Omega}^{\rm cl}_{\bf r}$, which do not change the physical state
of the system. The pseudofermion $\phi_{\bf r}$ has its spin quantized along
${\cal R}_{\bf r}\hat {\bf z}$. The action
(\ref{action1a}-\ref{action1b}) can then be expressed as 
\begin{eqnarray}
S &=& S_{\rm HF}+S_1+S_2 \nonumber \\ && 
+\int_0^\beta d\tau  \sum_{\bf r} \biggl( -\frac{U}{4}
\delta m_{\bf r}^2 +i\delta m^{\rm HS}_{\bf r} \delta m_{\bf r}
\nonumber \\ && -\frac{2}{U}
m_0^{\rm HS}\delta m^{\rm HS}_{\bf r} \biggr) ,
\label{action3} 
\end{eqnarray}
where $S_{\rm HF}$ is the HF action (\ref{HFaction}) and
\begin{eqnarray}
S_1 &=& \int_0^\beta d\tau \sum_{\bf r}\phi^\dagger_{\bf r} A_{0{\bf r}}  
\phi_{\bf r} , \nonumber \\ 
S_2 &=& - t \int_0^\beta d\tau \sum_{\langle {\bf r},{\bf r}'\rangle} 
(\phi^\dagger_{\bf r} A_{{\bf r},{\bf r}'} \phi_{{\bf r}'} + {\rm c.c.}).
\label{S1_2} 
\end{eqnarray}
$\langle {\bf r},{\bf r}'\rangle$ denotes nearest neighbors and we
have introduced 
\begin{eqnarray}
A_{0{\bf r}} &=& R^\dagger_{\bf r} \dot R_{\bf r} - h_0(R^\dagger_{\bf
  r} \sigma^z R_{\bf r}-\sigma^z) -i\delta m^{\rm HS}_{\bf r} \boldsymbol{\sigma}
\cdot {\bf\Omega}^{\rm cl}_{\bf r} \nonumber \\ 
&=& \sum_{\nu=x,y,z} A^\nu_{0{\bf r}} \sigma^\nu , \nonumber \\
A_{{\bf r},{\bf r}'} &=& R^\dagger_{\bf r}R_{{\bf r}'} -1 \nonumber \\
&=& \sum_{\nu=0,x,y,z} A^\nu_{{\bf r},{\bf r}'} \sigma^\nu .
\end{eqnarray}
We use the notation $\dot R_{\bf r}=\partial_\tau R_{\bf
  r}$. $\sigma^0$ is the $2\times 2$ unit matrix. 

The effective action $S[p,q,m]$ is obtained by integrating out the
fermion field $\phi_{\bf r}$ and the Hubbard Stratonovich field $m^{\rm
  HS}_{\bf r}$. The final result reads 
\begin{eqnarray}
S[p,q,m] &=& 
\frac{1}{2} \sum_{\tilde q} \bigl[
p_{-\tilde q} \tilde\Pi_{pp}(\tilde q) p_{\tilde q} + q_{-\tilde
  q} \tilde\Pi_{qq}(\tilde q) q_{\tilde q}   \nonumber \\ && +
\delta m_{-\tilde q} \tilde\Pi_{mm}(\tilde q)\delta m_{\tilde q}
+ 2 p_{-\tilde q} \tilde\Pi_{pq}(\tilde q) q_{\tilde q}
\nonumber \\ &&  + 2 p_{-\tilde q}\tilde\Pi_{pm}(\tilde q)  \delta
m_{\tilde q} 
+ 2 q_{-\tilde q} \tilde\Pi_{qm}(\tilde q) \delta m_{\tilde q} \bigr]
\nonumber \\ &&  + \delta S_B ,
\label{Spqm}
\end{eqnarray}
where $\tilde q=({\bf q},i\omega_\nu)$, with $\omega_\nu$ ($\nu$
integer) a bosonic Matsubara frequency. 
The coefficients $\tilde\Pi$ are given by (\ref{tildePi}) and
$\delta S_B$ by (\ref{dSB}). 
Details of this rather long calculation are given in Appendix
\ref{app:leea}, to which we refer readers interested in results for
the repulsive Hubbard model [Eq.~(\ref{hamtr})]. In the bulk of
the manuscript, we shall focus on the attractive Hubbard model as
defined in (\ref{ham}).

\subsection{Effective action $S[\rho,\Delta]$} 
\label{subsec:leea1}

Results of Sec.~\ref{subsec:leea} can be easily translated to the attractive
model. Below $T_{\rm pair}\equiv T_c^{\rm HF}$, the finite amplitude
$\Delta_0$ of the superconducting order parameter can be interpreted
as the appearance of  incoherent pairs. At weak coupling, the size
$\xi_0\sim t/\Delta_0^{\rm HS}$ of 
these (Cooper) pairs is much larger than the lattice
spacing ($\xi_0\gg 1$). At strong coupling, the preformed pairs are
local and are expected to behave as hard-core bosons (the hard-core
constraint comes from the Pauli principle which prevents double
occupancy of a lattice site). The crossover temperature $T_X$ marks
the onset of strong superconducting fluctuations (i.e. $\xi\gg 1$ with
$\xi$ the superconducting correlation length).

As already pointed out, in general $\frac{m_{\bf r}}{2}{\bf\Omega_r}$
cannot be directly identified with the spin density ${\bf S_r}=c^\dagger_{\bf
  r}\frac{\boldsymbol{\sigma}}{2} c_{\bf r}$. In order to find the
relation between ${\bf S_r}$ and $\frac{m_{\bf r}}{2}{\bf\Omega_r}$,
we add to the action the source term 
\begin{equation}
S_J = \int_0^\beta d\tau \sum_{\bf r} c^\dagger_{\bf r} {\bf J_r}
\cdot \boldsymbol{\sigma} c_{\bf r} ,
\label{SJ}
\end{equation}
with ${\bf J_r}=(J^x_{\bf r},J^y_{\bf r},J^z_{\bf r})$. 
The charge and pairing fields in the attractive model are then
obtained from [see Eq.~(\ref{transformation})]
\begin{eqnarray}
\frac{\delta S}{\delta J^x_{\bf r}} \biggl|_{J=0} &\equiv& 2 (-1)^{\bf r}
|\Delta_{\bf r}| \cos\Theta_{\bf r} , \nonumber \\ 
\frac{\delta S}{\delta J^y_{\bf r}} \biggl|_{J=0} &\equiv& -2 (-1)^{\bf r}
|\Delta_{\bf r}| \sin\Theta_{\bf r} , \nonumber \\ 
\frac{\delta S}{\delta J^z_{\bf r}} \biggl|_{J=0} &\equiv& \rho_{\bf r}-1.
\label{transformation2}
\end{eqnarray}
We then proceed as in the preceding section (see also Appendix
\ref{app:leea}). We integrate out the fermions 
to obtain the effective action to second order in $p_{\bf r}, \delta
m_{\bf r},\partial_\mu p_{\bf r}, \partial_\mu \delta m_{\bf r},
\partial_\mu q_{\bf r}$ and first order in ${\bf J_r}$. Using then
Eqs.~(\ref{transformation2}), we obtain the relation between $m_{\bf
  r},{\bf\Omega_r}$ and $\Delta_{\bf r},\rho_{\bf r}$: see
Eq.~(\ref{transformation3}) in Appendix
\ref{app:rho_delta}. Eq.~(\ref{transformation3}) takes a simple form
in the BCS and Bose limits (see Secs.~\ref{subsubsec:wc1} and
\ref{subsubsec:sc3}).  

\begin{figure}
\epsfxsize 6.cm
\epsffile[50 460 245 590]{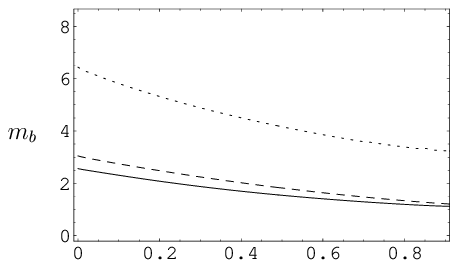}
\epsfxsize 6.cm
\epsffile[50 460 245 590]{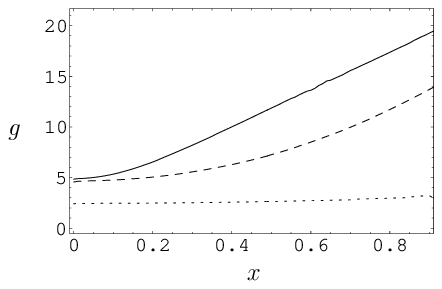}
\caption{Boson mass $m_b$ and interaction strength $g$  {\it vs} doping
  $x=1-\rho_0$ for 
  $U=2t$, $4t$  and $12t$ (solid, dashed, and dotted
  lines, respectively).  } 
\label{fig:masse_U_fixe}
\end{figure}

\begin{figure}
\epsfxsize 6.cm
\epsffile[50 460 245 590]{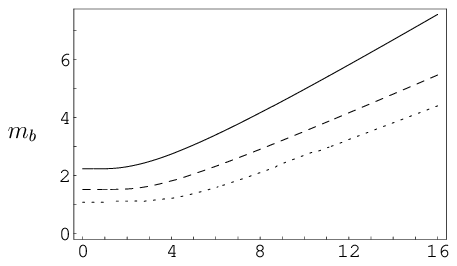}
\epsfxsize 6.cm
\epsffile[50 460 245 590]{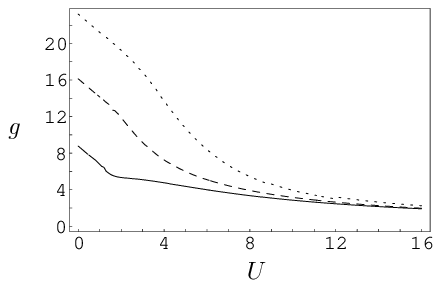}
\caption{Boson mass $m_b$ and  interaction strength $g$ {\it vs} $U$
  for $x=0.1$, $0.5$ and $0.9$ 
  (solid, dashed, and dotted lines, respectively). } 
\label{fig:masse_dopage_fixe}
\end{figure}

To proceed further in the general case, we note that there is no
coupling between $\boldsymbol{\nabla}q$ and $p,\dot q,m$ in the
action $S[p,q,m]$ (Appendix \ref{app:leea}). 
Since $\boldsymbol{\nabla}\Theta_{\bf r}=-\boldsymbol{\nabla}q_{\bf
  r}$, and $\dot\Theta,\delta\rho,\delta |\Delta|$ are functions of
  $p,\dot q,m$ (Appendix \ref{app:rho_delta}), $S[\rho,\Delta]$ takes
the general form
\begin{eqnarray}
S[\rho,\Delta] &=& \frac{1}{2} \int_0^\beta d\tau \int d^2r \biggl[
  i\rho_{\bf r}\dot\Theta_{\bf r} + 
\frac{\langle -K\rangle}{8} (\boldsymbol{\nabla}\Theta_{\bf r})^2
  \nonumber \\ && +
  \Pi_{\rho\rho}(\delta\rho_{\bf 
  r})^2 + \Pi_{|\Delta||\Delta|} (\delta|\Delta_{\bf r}|)^2 \nonumber \\ && 
+ 2 \Pi_{\rho|\Delta|} \delta\rho_{\bf r} \delta |\Delta_{\bf r}| \biggr] . 
\end{eqnarray}
Here we use the fact that half the fermion density is the
conjugate variable of the phase of the superconducting order
parameter, as required by gauge invariance. \cite{note4}
The action $S[\rho,\Delta]$ was previously obtained in
Ref.~\onlinecite{DePalo99}. For our purpose, it is not necessary to
determine  the expression of the coefficients $\Pi_{\rho\rho}$,
$\Pi_{|\Delta||\Delta|}$ and $\Pi_{\rho|\Delta|}$ which can be found
in Ref.~\onlinecite{DePalo99}. \cite{note7} 
To make contact with the usual description of a
Bose superfluid, we integrate out the
amplitude field $|\Delta|$. We thus obtain the action
\begin{eqnarray}
S[\rho,\Theta] &=& \frac{1}{2} \int_0^\beta d\tau \int d^2r \biggl\lbrace
  i\rho_{\bf r}\dot\Theta_{\bf r}
+ \frac{\langle -K\rangle}{8} (\boldsymbol{\nabla}\Theta_{\bf r})^2
  \nonumber \\ && +
  (\delta\rho_{\bf r})^2  \biggl[  \Pi_{\rho\rho} -
  \frac{\Pi_{\rho|\Delta|}^2}{\Pi_{|\Delta||\Delta|}} \biggr]
  \biggr\rbrace . 
\label{action_rT} 
\end{eqnarray}
This action is similar to the superfluid action
\cite{Nagaosa99}    
\begin{equation}
S_b = \int_0^\beta d\tau \int d^2r \biggl[ i\rho_{b{\bf r}}
  \dot\Theta_{\bf r} + \frac{\rho_{b0}}{2m_b}
  (\boldsymbol{\nabla}\Theta_{\bf r})^2 + \frac{g}{2}
  (\delta\rho_{b{\bf r}})^2 \biggr]
\label{action_b} 
\end{equation}
for bosons of mass $m_b$ and density $\rho_{b{\bf r}}$ ($\rho_{b0}$ is
the mean density and $\delta\rho_{b{\bf r}}=\rho_{b{\bf r}}-\rho_{b0}$). 
$g$ is the amplitude of the repulsive interaction between
bosons. Comparing Eqs.~(\ref{action_rT}) and (\ref{action_b}), we obtain
\begin{eqnarray}
\rho_{b{\bf r}} &=& \frac{\rho_{\bf r}}{2} , \nonumber \\ 
m_b &=& \frac{4\rho_0}{\langle -K\rangle} , \nonumber \\ 
g &=& 4 \biggl(  \Pi_{\rho\rho} -
  \frac{\Pi_{\rho|\Delta|}^2}{\Pi_{|\Delta||\Delta|}} \biggr) .
\label{rho_m_g} 
\end{eqnarray}
In order to calculate $g$ without computing  $\Pi_{\rho\rho}$,
$\Pi_{|\Delta||\Delta|}$ and $\Pi_{\rho|\Delta|}$, we integrate out
charge fluctuations. This yields the phase-only action
\begin{eqnarray}
S[\Theta] &=& \int_0^\beta d\tau \int d^2r \biggl[ \frac{\rho_0}{4m_b}
  (\boldsymbol{\nabla}\Theta_{\bf r})^2 + \frac{\dot\Theta_{\bf
  r}^2}{2g} \biggr] \nonumber \\ 
&& + \frac{i}{2} \rho_0 \int_0^\beta d\tau \int d^2r \dot\Theta_{\bf r} .
\end{eqnarray}
$S[\Theta]$ corresponds to an O(2) sigma model with an additional term
proportional to the first-order time derivative of $\Theta$. 
$m_b$ and $g$ are related to the phase stiffness $\rho_s^0$ and
the velocity $c$ of the O(2) sigma model:
\begin{eqnarray}
m_b &=& \frac{\rho_0}{2\rho_s^0} , \nonumber \\
g &=& \frac{c^2}{\rho_s^0} . 
\label{mbg}
\end{eqnarray}
$S[\Theta]$ can be directly obtained from the action
$S[p,q,m]$ (Sec.~\ref{subsec:phase-only}), which allows to
determine $\rho_s^0$ and $c$, and therefore 
$m_b$ and $g$. Anticipating on the results of
Sec.~\ref{subsec:phase-only}, we show in
Figs.~\ref{fig:masse_U_fixe}-\ref{fig:masse_dopage_fixe} $m_b$ and $g$
as a function of doping and interaction strength. Because both
$\rho_s^0$ and $c$ vanish in the low-density limit, the numerical
determination of $m_b$ and $g$ from (\ref{mbg}) is difficult when
$x\to 1$. We have therefore only considered $x\leq 0.9$. 

The low-energy effective action (\ref{action_rT}-\ref{action_b})
ensures that the Fermi superfluid will behave similarly to a Bose
superfluid. Indeed, from the classical equation of motion, we obtain
the two basic equations (in imaginary time) of a superfluid:\cite{Nagaosa99}
\begin{eqnarray}
i\dot\rho_{b{\bf r}}+\boldsymbol{\nabla} \cdot \left( \frac{\rho_{b0}}{m_b}
\boldsymbol{\nabla}\Theta_{\bf r} \right) &=& 0 , \nonumber \\
i\dot\Theta_{\bf r}+ g \delta\rho_{b{\bf r}} &=& 0 .
\label{eqsf}
\end{eqnarray}
For a Bose superfluid, the coefficient in front of
$(\boldsymbol{\nabla}\Theta)^2$ and $\boldsymbol{\nabla}\Theta$ in 
Eqs.~(\ref{action_b}) and (\ref{eqsf}) is the superfluid density
(divided by the boson mass). Here, because the
coefficients of the effective action $S[\rho,\Theta]$ are calculated
from the HF action, we expect the superfluid density to be given by
the full density $\rho_{b0}=\rho_0/2$. 
In Sec.~\ref{sec:scr}, we derive the effective action
$S[\rho,\Theta]$ in the strong-coupling low-density limit [including
the terms proportional to $(\boldsymbol{\nabla}\rho_{\bf 
  r})^2$ that are omitted in (\ref{action_b})], without
relying on the HF theory. We obtain a boson mass $m_b=1/J$ in
agreement with (\ref{rho_m_g}) when $U\gg 4t$ (see
Eq.~(\ref{mass_g_sc}) below for $x\simeq 1$). 

There is nevertheless an important difference with the Bose superfluid
case. $S[\rho,\Theta]$ has been obtained 
by integrating out amplitude fluctuations ($|\Delta|$) and therefore does
not describe the full dynamics of the superconducting order parameter
$\Delta_{\bf r}=|\Delta_{\bf r}|e^{i\Theta_{\bf r}}$ (except in the
strong coupling limit, see Secs.~\ref{subsubsec:sc3} and
\ref{sec:scr}).  

Effective actions similar to (\ref{action_rT}) have been derived
previously for continuum or lattice models. 
\cite{Loktev01,Stone95,Aitchison95,DePalo99} 
In this section, we have obtained
an effective action for the lattice case which is valid for all values
of the interaction strength. The validity of our approach is quite
obvious both at weak and strong couplings. In the latter limit, it
simply follows from the mapping between the attractive model and the
Heisenberg model under the particle-hole transformation
(\ref{phtr}) (see Sec.~\ref{sec:scr}). We therefore expect our
approach to provide a good description of the BCS-Bose crossover as
the interaction strength increases. 

We discuss below in more detail the BCS and Bose limits.

\subsubsection{BCS limit}
\label{subsubsec:wc1}

Using the results of Appendix \ref{subsubsec:wc}, we deduce\cite{note2}
from (\ref{transformation3})  
\begin{eqnarray}
\delta\rho_{\bf r} &=& - \delta m_{\bf r}, \nonumber \\
\delta |\Delta_{\bf r}| &=& -p_{\bf r}x.
\label{transformation4}
\end{eqnarray}
Eqs.~(\ref{transformation4}) can be directly obtained by assuming
${\bf S_r}=\frac{m_{\bf r}}{2}{\bf\Omega_r}$ and considering the limit
$\Delta_0\to 0$. This shows that the identification between ${\bf
  S_r}$ and $\frac{m_{\bf r}}{2}{\bf\Omega_r}$ holds in the
weak-coupling limit. 
When inverting the particle-hole transformation (\ref{phtr}), the 
kinetic energy remains unchanged and the HF spin susceptibility
$\Pi^{zz}_{00}(0,0)$ becomes the HF compressibility $\kappa$ of the
attractive model. The effective action $S[q,m]$ [Eq.~(\ref{action7})]
then becomes  
\begin{eqnarray}
S[\rho,\Theta] &=& \frac{1}{2} \int_0^\beta d\tau \int d^2 r 
\biggl[ i \rho_{\bf r} \dot \Theta_{\bf r} + \frac{\langle -K \rangle
  }{8} (\boldsymbol{\nabla} \Theta_{\bf 
  r})^2 \nonumber \\ && 
+ \left( \frac{1}{\kappa}-\frac{U}{2} \right) (\delta \rho_{\bf r})^2
\biggr] .
\label{action7bis}
\end{eqnarray}
Amplitude fluctuations ($|\Delta|$) do not couple to density and phase
fluctuations in the weak-coupling limit. We also verify that half the
fermion density is the conjugate variable of the phase $\Theta$ of
the pairing field. The gauge 
choice $\psi_{\bf r}=\varphi_{\bf r}/m_0$ [Eq.~(\ref{dSB})] is crucial to
obtain this result. For instance, with $\psi_{\bf r}=0$, one would
obtain $(\rho_{\bf r}-1)/2$ as the conjugate variable of $\Theta_{\bf r}$. 

From (\ref{action7bis}), we conclude that in the BCS limit the system
behaves as a Bose superfluid  with $\rho_{b{\bf r}}=\rho_{\bf r}/2$ and
\begin{eqnarray}
m_b &=& 4\frac{\rho_0}{\langle-K\rangle}, \nonumber \\ 
g &=& 4 \left(\frac{1}{\kappa}-\frac{U}{2} \right) .
\end{eqnarray}

\subsubsection{Bose limit}
\label{subsubsec:sc3}

In the strong-coupling limit, the auxiliary field $\frac{m_{\bf
    r}}{2}{\bf\Omega_r}$ can be identified with the spin density ${\bf
    S_r}$ (see Appendix \ref{subapp:scl}) so that
\begin{eqnarray}
\rho_{\bf r}-1 &=& \Omega^z_{\bf r} , \nonumber \\ 
\Delta_{\bf r} &=& \frac{(-1)^{\bf r}}{2} \Omega^-_{\bf r} , 
\label{rho_delta} 
\end{eqnarray} 
where $\Omega^\pm_{\bf r}=\Omega^x_{\bf r}\pm i \Omega^y_{\bf r}$. 
The condition ${\bf\Omega}_{\bf r}^2=1$ implies that charge and amplitude
fluctuations are not independent but tied by the relation  
\begin{equation}
|\Delta_{\bf r}| = \frac{1}{2} \sqrt{\rho_{\bf r}(2-\rho_{\bf r})} .
\label{mfixe}
\end{equation}
There is therefore a one-to-one correspondence between the bosonic
field $\Psi_{\bf r}=\sqrt{\rho_{\bf r}/2}e^{i\Theta_{\bf 
r}}$ and the pairing field $\Delta_{\bf 
  r}=|\Delta_{\bf r}|e^{i\Theta_{\bf r}}$. In the low-density limit,
both fields coincide, since $|\Delta_{\bf r}|\simeq\sqrt{\rho_{\bf r}/2}$.  

From (\ref{rho_delta}), we deduce $\delta\rho_{\bf r}=-2\sin\theta_0
p_{\bf r}$ and $\delta|\Delta_{\bf r}|=\cos\theta_0 p_{\bf r}$ (with
$\cos\theta_0=-x$).  
These equations can also be obtained from (\ref{transformation3}). 
The effective action $S[\rho,\Theta]$ then reads [see Eq.~(\ref{action_sc})] 
\begin{eqnarray}
S[\rho,\Theta] &=& \frac{1}{2} \int_0^\beta d\tau \int d^2 r 
\biggl[ i \rho_{\bf r} \dot \Theta_{\bf r} + \frac{J}{4}(1-x^2)
  (\boldsymbol{\nabla} \Theta_{\bf 
  r})^2 \nonumber \\ && + 2J (\delta\rho_{\bf r})^2
 \biggr] . 
\label{action_scbis} 
\end{eqnarray}
Again we verify that half the
fermion density is the conjugate variable of the phase $\Theta$ of
the pairing field. In the Bose limit, we therefore have 
\begin{eqnarray}
m_b &=& \frac{2}{J(1+x)} ,
\nonumber \\ 
g &=& 8J .
\label{mass_g_sc}
\end{eqnarray}

The BCS and Bose limits differ in the role of the Berry phase term
\begin{equation}
\langle \phi^\dagger_{\bf r} R^\dagger_{\bf r}\dot R_{\bf r}\phi_{\bf
  r} \rangle = i \Delta_0 (p_{\bf r} \dot q_{\bf r}- \dot p_{\bf r}
  q_{\bf r}) - \frac{i}{2} \rho_0 \dot q_{\bf r} ,
\end{equation}
where we have included the gauge-dependent contribution
(\ref{gauge}). For $U\gg 4t$, the Berry phase term becomes
$\frac{i}{2}\rho_{\bf r}\dot\Theta_{\bf r}$ and therefore determines
the entire dynamics of the phase $\Theta$. This is expected since the
dynamics of the Heisenberg model entirely comes from the Berry phase
term. \cite{Auerbach94}  For $U\ll 4t$, the Berry phase term only
gives $\frac{i}{2}\rho_0\dot\Theta_{\bf r}$ and does not contribute to
the $T=0$ phase collective mode. The missing term,
$\frac{i}{2} \delta\rho_{\bf r}\dot\Theta_{\bf r}$, comes from the
second-order cumulant $\langle (S_1+S_2)^2\rangle_c$. For all values of the
interaction strength, the contribution $\frac{i}{2}\rho_0 \int_0^\beta
d\tau \sum_{\bf r}\dot\Theta_{\bf r}$ to the action, which is
responsible for the Magnus force acting a vortex, \cite{Stone95} is
given by the Berry phase term of the effective action
$S[m,{\bf\Omega}]$. \cite{Ao93}

\subsection{Phase-only action $S[\Theta]$}
\label{subsec:phase-only}

Integrating out $p$, $\delta m^{\rm HS}$ and $\delta m$ in the action
$S[p,q,m^{\rm HS},m]$, and using $\Theta_{\bf r}=-q_{\bf r}$, we obtain
the phase-only action
\begin{eqnarray}
S[\Theta] &=& \frac{\rho^0_s}{2} \int_0^\beta d\tau \int d^2r 
\left[ (\boldsymbol{\nabla} \Theta_{\bf r})^2 
+ \frac{\dot\Theta_{\bf r}^2}{c^2} \right]  \nonumber \\ 
&& +  \frac{i}{2} \rho_0 \int_0^\beta d\tau
\int d^2 r  \dot \Theta_{\bf r}  ,
\label{phaseonly}
\end{eqnarray}
where we have taken the continuum limit in real
space. Eq.~(\ref{phaseonly}) is valid in the hydrodynamic regime
defined by the momentum-space cutoff $\Lambda \sim {\rm
  min}(1,2\Delta_0^{\rm HS}/c)$. Its validity also requires the
fluctuations of $\theta$ to be small, a condition which is not
fulfilled in the vicinity of half-filling. This case is discussed in
detail in Sec.~\ref{sec:O3}. The (bare) phase stiffness
$\rho^0_s$ and the velocity $c$ of the O(2) sigma model in
(\ref{phaseonly}) are determined by 
\begin{eqnarray}
\rho^0_s &=& \frac{\langle -K\rangle}{8} , \nonumber  \\ 
\frac{\rho^0_s}{c^2} &=& \frac{\Pi^{zz}_{00}(0,0)}{4}
+ \lim_{\tilde q\to 0} \frac{1}{\omega_\nu^2} \Biggl[ 
\frac{\Pi_{qm^{\rm HS}}^2(\tilde q)}{\Pi'_{m^{\rm HS}m^{\rm HS}}(\tilde
  q)} \nonumber \\ && +
\frac{\left(\Pi_{qp}(\tilde q)-\frac{\Pi_{qm^{\rm HS}}(\tilde
    q)\Pi_{m^{\rm HS}p}(\tilde
    q)}{\Pi'_{m^{\rm HS}m^{\rm HS}}(\tilde q)} 
  \right)^2}{\Pi_{pp}(\tilde q)- \frac{\Pi^2_{pm^{\rm HS}}(\tilde
    q)}{\Pi'_{m^{\rm HS}m^{\rm HS}}(\tilde q)}} 
\Biggr] .   
\label{velocity} 
\end{eqnarray}
At zero temperature, there is superconducting long-range order (in the
attractive model). From
(\ref{phaseonly}), we deduce the existence of a gapless (Goldstone)
mode with dispersion $\omega =c|{\bf q}|$. This collective mode
smoothly evolves from the Anderson-Bogoliubov mode\cite{Anderson58} at
weak coupling to the Bogoliubov mode\cite{Bogoliubov58}  of a Bose
superfluid at strong coupling (see
Secs.~\ref{subsubsec:po_bcs}-\ref{subsubsec:po_bose}).   

\begin{figure}
\epsfxsize 6.cm
\epsffile[50 460 245 590]{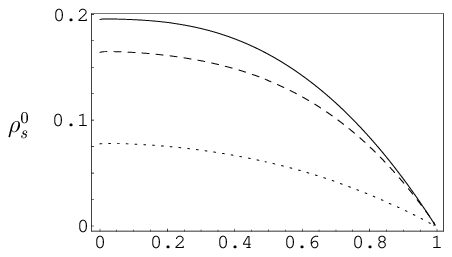}
\epsfxsize 6.cm
\epsffile[50 460 245 590]{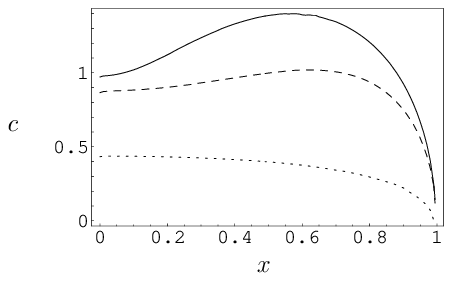}
\caption{Phase stiffness $\rho_s^0$ and velocity $c$  {\it vs} doping
  $x=1-\rho_0$ for 
  $U=2t$, $4t$  and $12t$ (solid, dashed, and dotted
  lines, respectively).  } 
\label{fig:vitesse_U_fixe}
\end{figure}

\begin{figure}
\epsfxsize 6.cm
\epsffile[50 460 245 590]{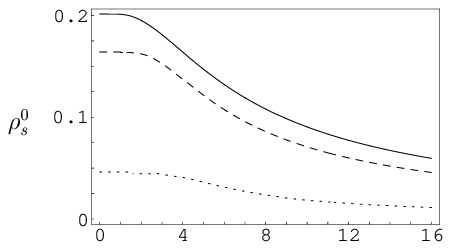}
\epsfxsize 6.cm
\epsffile[50 460 245 590]{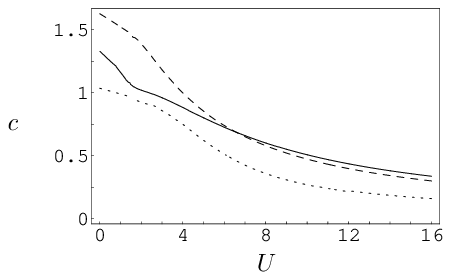}
\caption{Phase stiffness $\rho_s^0$ and  velocity $c$ {\it vs} $U$ for
  $x=0.1$, $0.5$ and $0.9$ (solid, dashed, and dotted
  lines, respectively). } 
\label{fig:vitesse_dopage_fixe}
\end{figure}

Figs.~\ref{fig:vitesse_U_fixe}-\ref{fig:vitesse_dopage_fixe} show $c$
and $\rho_s^0$ {\it vs} $U$ and $x$. Our results reproduce the
collective mode velocity obtained from a random-phase-approximation (RPA)
calculation about the zero-temperature HF superconducting
state. \cite{Belkhir92,Belkhir94,Kostyrko92}  

\subsubsection{BCS limit}
\label{subsubsec:po_bcs}

In the weak-coupling limit, using the results of
Appendix \ref{subsubsec:wc} and Sec.~\ref{subsubsec:wc1}, we obtain
\begin{equation}
c^2 = \frac{4\rho^0_s}{\kappa} \left(1 - \frac{U}{2} \kappa \right) .
\end{equation}
We further simplify this result by noting that
${\Delta_0^{\rm HS}}^2/[(\epsilon-\epsilon_F)^2+{\Delta_0^{\rm
    HS}}^2]^{3/2} \equiv 2 
\delta(\epsilon- \epsilon_F)$ in the limit $\Delta_0^{\rm HS}\to 0$. This
yields [see Eqs.~(\ref{crtcrt16}) in Appendix \ref{app:hfcorr}]
\begin{equation}
\kappa =2{\cal N}_0(\epsilon_F) .
\end{equation}
Using also (Appendix \ref{app:kehf})
\begin{equation}
\langle -K\rangle = 2 \int_{\bf k} v^2_{\bf k}
\delta(\epsilon_{\bf k}-\epsilon_F) ,
\label{kealt} 
\end{equation}
where ${\bf v}_{\bf k}=\boldsymbol{\nabla}_{\bf k} \epsilon_{\bf k}$,
we finally obtain
\begin{eqnarray}
c^2 &=& \frac{\overline{v^2_{\bf k}}}{2} \left[1- U {\cal N}_0(\epsilon_F)
    \right], \label{velocity_wc}  \nonumber \\   
\overline{v^2_{\bf k}} &=& \frac{\int_{\bf k} v^2_{\bf k} \delta(\epsilon_{\bf
  k}-\epsilon_F)}{\int_{\bf k} \delta(\epsilon_{\bf
  k}-\epsilon_F)} .
\label{c_bcs} 
\end{eqnarray}
$\overline{v^2_{\bf k}}$ is the mean square velocity on the
(non-interacting) Fermi surface. Eqs.~(\ref{c_bcs}) reproduce the
expression of the velocity of the Anderson-Bogoliubov mode
\cite{Anderson58} in a 2D continuum model if we identify
$\overline{v^2_{\bf k}}$ with $v_F^2=(k_F/m)^2$ (with $k_F$ the Fermi
momentum and $m$ the fermion mass). They were also obtained in
Ref.~\onlinecite{Belkhir94} from an RPA
calculation about the zero-temperature HF superconducting state.

\subsubsection{Bose limit}
\label{subsubsec:po_bose}

In the strong-coupling limit, using the results of Appendix
\ref{subsubsec:sc2}, we obtain
\begin{eqnarray}
\rho^0_s &=& \frac{J}{4} (1-x^2) , \nonumber \\
c &=& \sqrt{2} J \sqrt{1-x^2} 
\label{rho_c_sc}. 
\end{eqnarray}
We recover the velocity $c=\sqrt{\rho_{b0}g/m_b}$ of the Bogoliubov
mode in a Bose superfluid, \cite{Nagaosa99} where the mass $m_b$
and the interaction amplitude $g$ of the bosons are defined in
(\ref{mass_g_sc}) ($\rho_{b0}=\rho_0/2$).
At half-filling, $x=0$, Eqs.~(\ref{rho_c_sc}) agree with
results obtained directly from the Heisenberg model.

\subsection{Berezinskii-Kosterlitz-Thouless transition}
\label{subsec:kt}

The effective action $S[\Theta]$ derived in
Sec.~\ref{subsec:phase-only} is valid for temperatures below the
crossover temperature $T_X$, i.e. $T\ll T_X$ (see
Sec.~\ref{subsec:leea}). Since $T_{\rm BKT}\sim T_X \sim T_{\rm
  pair}$ in the weak-coupling limit (as will be shown below), this
effective action is not sufficient if one is interested in the BKT
phase transition. In this section, we derive the phase-only action
$S[\Theta]$ for temperatures $T\leq T_{\rm pair}$ and obtain the BKT phase
transition temperature $T_{\rm BKT}$. We consider the
static (i.e. classical) limit, where non-analytic contributions due to
the Landau damping terms \cite{Aitchison00,Shaparov01} are not
present. In this limit, the $q$ 
field decouple from other fluctuations ($p$, $\delta m^{\rm HS}$ and
$\delta m$), and we obtain (see Appendix \ref{app:hfcorr}) 
\begin{eqnarray}
S[\Theta] &=& \frac{\rho_s^0}{2 T} \int d^2 r
(\boldsymbol{\nabla} \Theta_{\bf r})^2 , \label{Sxy} \\ 
\rho_s^0 &=& \frac{\langle -K \rangle }{8} -
\frac{\Pi^{zz}_{xx}(0,0)}{4} \nonumber \\ 
&=& \int_{\bf k} \left[ \frac{\epsilon_{\bf k}\epsilon_{{\bf
      k}\uparrow}}{8 E_{{\bf k}\uparrow}} \tanh \frac{E_{{\bf
      k}\uparrow}}{2 T} -\frac{t^2\sin^2k_x}{2T \cosh^2 \left(\frac{E_{{\bf 
      k}\uparrow}}{2T}\right)} \right].
\label{rho_s0}  
\end{eqnarray}
The action (\ref{Sxy}) is valid in the
hydrodynamic regime and must therefore be supplemented with a momentum-space
cutoff $\Lambda \sim {\rm min}(1,2\Delta_0^{\rm HS}(T)/c)$. 
$\Pi^{zz}_{xx}(0,0)$ is a current-current correlation function in the
attractive model. It vanishes at zero temperature, and the phase
stiffness reduces to the mean kinetic energy as obtained in 
Sec.~\ref{subsec:phase-only}. The expression (\ref{rho_s0}) of the
finite-temperature phase stiffness has also been obtained in
Refs.~\onlinecite{Denteneer91,Denteneer93,Singer98}.

In order to extract the BKT phase transition
temperature, we assume that the system is described by an XY model
whose continuum limit is given by (\ref{Sxy}).   
\cite{Denteneer91,Denteneer93,Alvarez96,Singer98,Paramekanti00} 
Although this
assumption can be justified at strong-coupling (except in the
low-density limit) (see Sec.~\ref{sec:scr}), it is in general not
correct. \cite{Benfatto03} Nevertheless, it should give a reasonable
estimate of the BKT phase transition temperature. This leads us to the 2D XY
Hamiltonian 
\begin{equation}
H_{XY} = - \rho_s^0 \sum_{\langle {\bf r},{\bf r}'\rangle }
\cos(\Theta_{\bf r}-\Theta_{{\bf r}'}) ,
\label{Hxy}
\end{equation}
defined on a square lattice with spacing $\sim \Lambda^{-1}$.
The XY model (\ref{Hxy}) is known to have a BKT phase 
transition at the temperature $T_{\rm BKT}$ defined by
\begin{equation}
T_{\rm BKT} = Q \rho_s^0 (T_{\rm BKT}),
\label{Tkt}
\end{equation}
where $Q\simeq 0.898$ is estimated from Monte-Carlo simulations.
\cite{Tobochnik79,Fernandez86,Gupta88} In (\ref{Tkt}), we have written
explicitly the temperature dependence of $\rho_s^0$.  
$T_{\rm BKT}$ is obtained by solving simultaneously
Eqs.~(\ref{gapeq1},\ref{gapeq2},\ref{Tkt}).

From the XY model, we can extract another characteristic temperature,
\begin{equation}
T_X = 2\rho_s^0(T_X), 
\label{Txdef} 
\end{equation}
defined as the mean-field transition temperature of $H_{XY}$. $T_X$
marks the onset of a RC regime of superconducting
fluctuations. In this regime, superconducting fluctuations become
quasi-static, the amplitude  $|\Delta_{\bf r}|$ of the order parameter
takes a finite value, and strong phase fluctuations develop
(i.e.  $\xi\gg 1$, with $\xi$ the correlation length). \cite{Kyung01}   

In the strong coupling limit, $T_{\rm
  BKT}\simeq Q \frac{J}{4}(1-x^2)$ and $T_X\simeq
\frac{J}{2}(1-x^2)\ll T_{\rm pair}$. In the weak-coupling
limit, $T_{\rm BKT}\sim T_X \sim T_{\rm pair} \propto e^{-1/U{\cal
    N}_0(\epsilon_F)}$ is exponentially small.

\subsection{Phase diagram}
\label{subsec:diag}

Near half-filling, ${\bf q}={\bf Q}$ charge fluctuations become
important and are suppressed only at very small temperature. [This
corresponds to a weak magnetic field $h_0$ in the repulsive model.]
We call $T_{\rm SO(3\to 2)}$ the crossover temperature above which
charge fluctuations restore the SO(3) symmetry of the order
parameter. Postponing the determination of $T_{\rm SO(3\to 2)}$ to 
Sec.~\ref{sec:O3}, we discuss in this section the phase diagram of
the 2D attractive Hubbard model. 

\begin{figure}
\epsfxsize 6.cm
\epsffile[50 455 275 605]{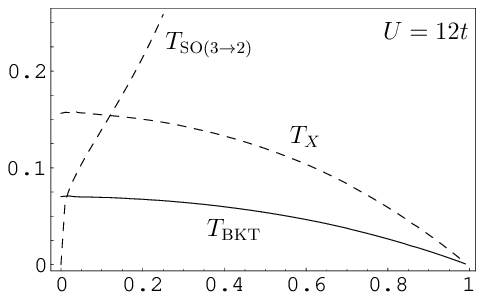}
\epsfxsize 6.cm
\epsffile[50 455 275 605]{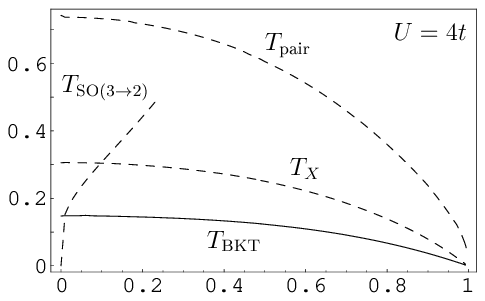}
\epsfxsize 6.cm
\epsffile[50 455 275 605]{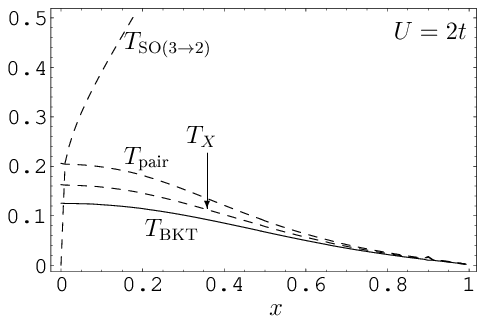}
\caption{$T_{\rm BKT}$,  $T_{\rm SO(3\to 2)}$, $T_X$ and $T_{\rm pair}$
  {\it vs} doping $x$ for $U=12t$, $4t$ and $2t$.
  For $U=12t$, $T_{\rm pair}$, which is nearly a vertical line
  around $x=1$ on the scale of the figure, is not shown. } 
\label{fig:temp_U_fixe}
\end{figure}
\begin{figure}
\epsfxsize 6.cm
\epsffile[50 455 275 620]{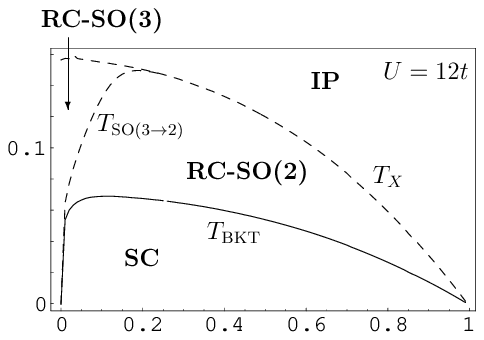}
\epsfxsize 6.cm
\epsffile[50 455 275 620]{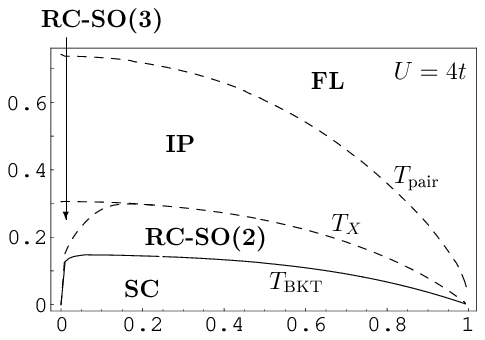}
\epsfxsize 6.cm
\epsffile[50 455 275 620]{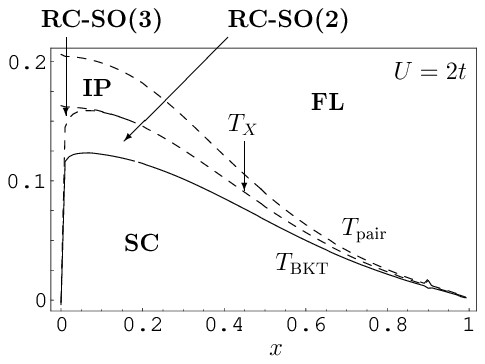}
\caption{Phase diagram of the 2D attractive Hubbard model {\it vs}
  doping $x$ for $U=12t$, $4t$ and $2t$. SC:
  superconducting phase (algebraic order). RC-SO(2): renormalized
  classical regime with fluctuations of SO(2) symmetry. RC-SO(3):
  RC regime where the fluctuations exhibit an
  effective SO(3) symmetry due to the presence of strong ${\bf q}={\bf
  Q}$ charge fluctuations. FL: Fermi liquid. The incoherent-pair
  regime (IP) corresponds to the formation of Cooper (local) pairs at weak
  (strong) coupling, without short-range SC order. Superconducting long-range
  order sets in at $T=0$. 
} 
\label{fig:diag_U_fixe}
\end{figure}
\begin{figure}
\epsfxsize 6.cm
\epsffile[50 465 275 620]{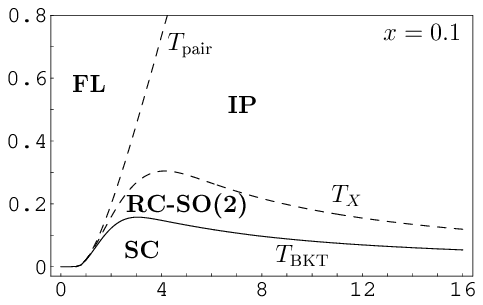}
\epsfxsize 6.cm
\epsffile[50 465 275 620]{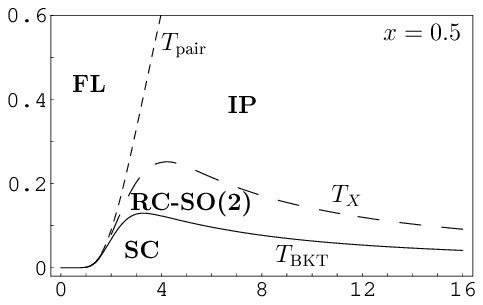}
\epsfxsize 6.cm
\epsffile[50 455 275 620]{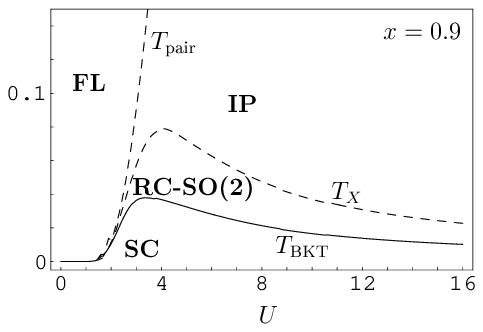}
\caption{Phase diagram of the 2D attractive Hubbard model {\it vs }
  $U$ for $x=0.1$, $0.5$ and $0.9$.    } 
\label{fig:diag_dopage_fixe}
\end{figure}

Let us first discuss the phase diagram as a function of doping 
(Figs.~\ref{fig:temp_U_fixe}-\ref{fig:diag_U_fixe}). Below 
$T_{\rm pair}$ (not shown in the figure for $U=12t$), the amplitude of
the superconducting order parameter takes a finite value, but with no
short-range order of the phase (i.e. $\xi \sim 1$ with $\xi$ the
superconducting correlation length). This corresponds to the appearance of
incoherent pairs (Cooper pairs at weak-coupling and local pairs at
strong coupling). Below $T_X$, superconducting correlations start to grow. 
The system enters a RC regime of 
fluctuations with SO(2) symmetry: the phase correlation length $\xi$
increases and rapidly becomes much larger than the lattice spacing
($\xi \gg 1$). At $T_{\rm BKT}$, a BKT phase
transition occurs and the system 
becomes superconducting. Superconducting long-range order sets in at
$T=0$. The situation is slightly different near
half-filling where $T_{\rm SO(3\to 2)}<T_X$. Below $T_X$, both
superconducting and ${\bf q}={\bf Q}$ charge fluctuations
start to grow, and the order parameter has an effective SO(3)
symmetry. For $T<T_{\rm SO(3\to 2)}$, charge fluctuations are
suppressed and only superconducting fluctuations continue to grow. 
This corresponds to a regime of fluctuations with SO(2) symmetry which
eventually drive a BKT phase transition at the temperature $T_{\rm
  BKT}$. Once the SO(2) regime is reached, the
BKT phase transition occurs rapidly, so that $T_{\rm BKT}$ can 
be estimated by $T_{\rm SO(3\to 2)}$ (see Sec.~\ref{subsec:O3_ht}).  
Therefore, the true transition temperature is estimated as
${\rm min}(T_{\rm BKT},T_{\rm SO(3\to 2)})$, where $T_{\rm BKT}$ is
the transition temperature obtained in Sec.~\ref{subsec:kt},
i.e. by neglecting the ${\bf q}={\bf Q}$ charge fluctuations. The
resulting phase diagram is shown in Fig.~\ref{fig:diag_U_fixe}. 
In the intermediate and strong-coupling regimes (i.e. for $U\gtrsim
4t$), we clearly identify a RC regime above the BKT superconducting
phase and an incoherent-pair regime at higher temperatures. We believe
the persistence of the incoherent-pair regime at weak-coupling (see
Fig.~\ref{fig:diag_U_fixe} for $U=2t$) to be likely an artifact of our
simple estimate of $T_{\rm pair}$. Due to many-body effects not taken
into account in our approach, $T_{\rm pair}$ is reduced with respect
to the HF transition temperature. \cite{Vilk97} 
We expect $T_{\rm pair}\simeq T_X$ at
weak coupling and thus the disappearance of the incoherent-pair
regime. We also observe that the RC regime with SO(3) fluctuations
extends to higher doping at strong coupling. 
Fig.~\ref{fig:diag_dopage_fixe} shows the phase diagram as a function
of $U$ for $x=0.1$, 0.5 and 0.9.  

Away from half-filling, the BKT phase transition
temperature derived from the criterion $T_{\rm BKT}\propto
\rho_s^0 (T_{\rm BKT})$, with $\rho_s^0$ given by
(\ref{rho_s0}), was also discussed in
Refs.~\onlinecite{Denteneer91,Denteneer93,Alvarez96,Singer98}. Good
quantitative agreement with Quantum Monte-Carlo simulations was
obtained. \cite{Singer98} Our estimation of the crossover temperature
$T_X$ for $U=4t$ also agrees with the Quantum Monte-Carlo simulations of
Ref.~\onlinecite{Kyung01}. 

In the weak-coupling limit, we expect a pseudogap to appear in the
density of states and the spectral function $A({\bf k},\omega) =
-\pi^{-1} {\rm Im}G({\bf k},\omega)$ below the crossover temperature
$T_X$ [$G({\bf k},\omega)$ is the single-particle
Green's function]. \cite{Singer99,Vilk97,Kyung01} This pseudogap results from
strong scattering of particles on collective fluctuations in the RC
regime. \cite{Vilk97,Kyung01} In the BCS limit, the pseudogap region is
small, except in the vicinity of half-filling where an SO(3) RC
regime is observed over a wide temperature range ($0\leq T \lesssim
T_X$) [see Fig.~\ref{fig:diag_U_fixe}]. \cite{Allen99} As the
interaction strength increases, the weak-coupling pseudogap should
progressively evolves into a strong-coupling gap due to the formation
of local singlet pairs when $T\lesssim T_{\rm pair}$. 
Pseudogaps in the 2D attractive Hubbard model have been seen both in
Quantum Monte Carlo simulations \cite{Singer99,Kyung01,Allen99} and
analytical approaches. 
\cite{Vilk97,Kyung01,Loktev01,Gusynin00,Gusynin02,note6} 
At half-filling, the transformation of the weak-coupling pseudogap
into the strong-coupling gap has been studied in
Refs.~\onlinecite{KB1,KB2}.

\section{Near half-filling}
\label{sec:O3}

In this section, we calculate the crossover temperature $T_{\rm
SO(3\to 2)}$ introduced in Sec.~\ref{subsec:diag}. The approach of
Sec.~\ref{sec:O2} holds only when ${\bf q}={\bf Q}$ charge
fluctuations (in the attractive model) are weak for all temperatures
below $T_X$. In the repulsive model, this corresponds to the limit
where the magnetic field $h_0$ is strong enough to suppress
fluctuations of the $S^z$ component of the spin density. Near
half-filling, ${\bf q}={\bf Q}$ charge fluctuations are suppressed
only at very low temperature. We therefore expect a RC regime with SO(3)
symmetry in the temperature range $T_{\rm SO(3\to 2)}\leq T\leq T_X$
followed by a RC regime with SO(2) symmetry at lower temperature
($T_{\rm BKT}\leq T\leq T_{\rm SO(3\to 2)}$). 

In the repulsive model, a weak magnetic field ${\bf h}_0 =
h_0\hat{\bf z}$ can be treated perturbatively. When it vanishes, 
the low-energy effective action for spin
fluctuations is a NL$\sigma$M. \cite{Schulz95,KB1,KB2} From general
arguments, \cite{Sachdev94} we expect the magnetic field to modify the
kinetic term of the NL$\sigma$M, $\dot {\bf n}_{\bf r}^2
\to (\dot {\bf n}_{\bf r}-2i {\bf h}_0\times {\bf n_r})^2$, as
obtained in the Heisenberg limit. \cite{Sachdev99} [Here
  ${\bf n}$  (${\bf n}^2=1$) is the N\'eel field describing low-energy
  AF fluctuations.] We show
below that this is indeed correct. We consider the low-temperature
limit $T\ll T_X$ where the coefficients of the effective action
can be evaluated in the zero-temperature limit. 

Following Ref.~\onlinecite{KB2}, we introduce only one auxiliary field
($m_{\bf r}$) to decouple the spin term $(c^\dagger_{\bf
  r}\boldsymbol{\sigma} \cdot {\bf\Omega_r}c_{\bf r})^2$ in
(\ref{decomp}). \cite{note3} The action then reads 
\begin{equation}
 S=\int_0^\beta d\tau \sum_{\bf r} \Bigl[ c^\dagger_{\bf r}(
 \partial_\tau - \hat t -h_0 \sigma^z - m_{\bf r} \boldsymbol{\sigma}
 \cdot {\bf\Omega_r}) c_{\bf r} + \frac{m^2_{\bf r}}{U} \Bigr].
\end{equation}
At the HF level,
neglecting the magnetic field, we have $m_{\bf
r}=m_0$ and ${\bf\Omega}^{\rm cl}_{\bf r}=(-1)^{\bf r}\hat {\bf
  z}$. We choose the AF magnetization along the $z$ axis as in
Ref.~\onlinecite{KB2}.
The saddle-point equation reads $2m_0/U = (-1)^{\bf r} \langle
c^\dagger_{\bf r} \sigma^z c_{\bf r} \rangle$. 
Following Haldane, \cite{Haldane83,Auerbach94} in the presence of AF
short-range order ($T\ll T_X$), we write 
\begin{equation}
{\bf\Omega_r} = (-1)^{\bf r} {\bf n}_{\bf r} \sqrt{1-{\bf L}_{\bf r}^2} +
{\bf L}_{\bf r} .
\label{Haldec}
\end{equation}
${\bf n_r}$ is the slowly varying N\'eel field, whereas ${\bf L_r}$ is a
canting vector, orthogonal to ${\bf n_r}$, taking account of local
ferromagnetic fluctuations. We assume ${\bf L_r}$ to be small, which 
clearly restricts the validity of this approach to weak magnetic
fields. As in Sec.~\ref{sec:O2}, we introduce a
new field $\phi$ defined by $c_{\bf r}=R_{\bf r}\phi_{\bf r}$ where $R_{\bf
  r}$ is a time- and site-dependent SU(2)/U(1) matrix satisfying 
\begin{equation}
R_{\bf r} \sigma^z R^\dagger_{\bf r} = \boldsymbol{\sigma} \cdot
{\bf n_r} . 
\end{equation}
${\cal R}_{\bf r}$, the SO(3) element associated to $R_{\bf r}$, maps
$\hat {\bf z}$ onto ${\bf n_r}$. We also define the rotated canting
field ${\bf l_r}={\cal R}^{-1}_{\bf r} {\bf L_r}$. Given that ${\cal
  R}^{-1}_{\bf r} {\bf n_r}=\hat {\bf z}$ and ${\bf L_r} \perp {\bf
  n_r}$, the ${\bf l_r}$ vector lies in the $(x-y)$ plane. 

In order to express the action in terms of the $\phi$ field, it is
convenient to make use of the SU(2) gauge field $A_{\mu{\bf
    r}}=\sum_{\nu=x,y,z}A^\nu_{\mu{\bf r}} \sigma^\nu$ defined as
\begin{eqnarray}
A_{0{\bf r}} &=& - R^\dagger_{\bf r} \partial_\tau R_{\bf r} ,
\nonumber \\
A_{\mu{\bf r}} &=& i R^\dagger_{\bf r} \partial_\mu R_{\bf r}
\,\,\,\, (\mu=x,y) .
\end{eqnarray}
Since the gauge field is of order $O(\partial_\mu)$, we can expand the
action with respect to ${\bf l}$, $A_\mu$, $h_0$ and $\delta m$. To
second order, we obtain
\begin{eqnarray}
S &=& S_{\rm HF} + S_p +S_l + S_d+S_{l^2} +S_{h_0} +S_{\delta m}
\nonumber \\ 
&& + \int_0^\beta d\tau \sum_{\bf r} \frac{\delta m^2_{\bf
    r}+2m_0 \delta m_{\bf r}}{U} ,  \nonumber \\
S_{ \mathrm{p} }  &=&   - \int_0^{\beta} d \tau 
                \sum_{\mu=0,x,y \atop {\nu=x,y,z \atop \mathbf{r}} }
                j_{\mu \mathbf{r}}^{\nu} A_{\mu \mathbf{r}}^{\nu}
        \label{eq0act0para} , \nonumber \\
                S_{ \mathrm{d} } 
        &=&     \frac{t}{2} \int_0^{\beta} d \tau 
                \sum_{\mu=x,y \atop {\nu=x,y,z \atop \mathbf{r}} }
                  {A^\nu_{\mu \mathbf{r}}}^2
                \phi^{\dagger}_{\mathbf{r}} 
                        \cos(-i \partial_{\mu}) 
                \phi_{\mathbf{r} }
         + \mathrm{c.c.}
        \label{eq0act0dia} , \nonumber \\
                S_{l}   &=&
                - m_0 \int_0^{\beta} d \tau 
                \sum_{\nu=x,y \atop \mathbf{r}}
                l_{\mathbf{r} }^{\nu} j_{0 \mathbf{r}}^{\nu}
        \label{eq0act0l} , \nonumber \\
                S_{l^2}
        &=&      \frac{m_0}{2} \int_0^{\beta} d \tau
        \sum_{\mathbf{r}} (-1)^{\mathbf{r}} \mathbf{l}_{\mathbf{r} }^2
  j_{0 \mathbf{r}}^z , \nonumber \\ 
S_{h_0} &=& -h_0 \int_0^\beta d\tau \sum_{\mathbf{r}} \phi^\dagger_{\bf r}
        R^\dagger_{\bf r} \sigma^z R_{\bf r} \phi_{\bf r} , \nonumber \\
 S_{\delta m} &=& -\int_0^\beta d\tau \sum_{\bf r} (-1)^{\bf r} \delta
        m_{\bf r} j^z_{0{\bf r}} , 
\end{eqnarray} 
where the spin-density current $j^\nu_{\mu{\bf r}}$ is defined in
(\ref{spincrt}). For $h_0=0$, $S$ reduces to the action derived in
Ref.~\onlinecite{KB2}. $S_{\rm HF}$ is the HF action.
$S_p$ and $S_d$ are paramagnetic and
diamagnetic terms, respectively. $S_l$ and $S_{l^2}$ are first-order
and second-order corrections in ${\bf l}$. $S_{\delta m}$ is the
contribution due to amplitude fluctuations.  \cite{note1} 

The effective action $S[{\bf n},{\bf L},\delta m]$ is obtained by integrating
out the fermions. To second order in $A_\mu$, ${\bf l}$, $\delta m$
and $h_0$, one finds 
\begin{eqnarray}
S[{\bf n},{\bf L},\delta m] &=& S[{\bf n},{\bf L};h_0=\delta m=0] 
+\langle S_p + S_{h_0} + S_{\delta m} \rangle\nonumber \\ &&  -
\frac{1}{2} \langle (S_{h_0}+S_{\delta m})^2 \nonumber \\ 
&& +2(S_{h_0}+S_{\delta m})(S_p+S_l) \rangle_c \nonumber \\ && 
 + \int_0^\beta d\tau \sum_{\bf r} \frac{\delta m^2_{\bf
    r}+2m_0\delta m_{\bf r}}{U} ,
\label{SnLdm}
\end{eqnarray}
where the averages $\langle \cdots\rangle$ are taken with the HF
action. $S[{\bf n},{\bf L};h_0=\delta m=0]$ is the action with no magnetic
field (and no amplitude fluctuations) and was derived in
Ref.~\onlinecite{KB2}. $\langle S_p\rangle$ is a Berry phase term. It
was ignored in Ref.~\onlinecite{KB2}, since it does not play any role
in the RC regime of the $h_0=0$ NL$\sigma$M. In order to calculate the
HF averages, we write  
\begin{equation}
S_{h_0} = -h_0 \int_0^\beta d\tau \sum_{{\bf r} \atop {\nu=x,y,z}}
B^\nu_{0{\bf r}} j^\nu_{0{\bf r}} ,
\end{equation} 
where $B^\nu_{0{\bf r}}$ is defined by $R^\dagger_{\bf r}\sigma^z R_{\bf
r}=\sum_{\nu=x,y,z} B^\nu_{0{\bf r}}\sigma^\nu$. $A^\nu_{\mu{\bf r}}$ and
$B^\nu_{0{\bf r}}$ are calculated using
\begin{equation}
R_{\bf r}= 
\left (
\begin{array}{lr}
\cos\left(\frac{\theta_{\bf r}}{2}\right) e^{-\frac{i}{2}(\varphi_{\bf
    r}+\psi_{\bf r})} & 
- \sin\left(\frac{\theta_{\bf r}}{2}\right)
    e^{-\frac{i}{2}(\varphi_{\bf r}-\psi_{\bf r})}
\\ 
\sin\left(\frac{\theta_{\bf r}}{2}\right) e^{\frac{i}{2}(\varphi_{\bf
    r}-\psi_{\bf r})} & 
\cos\left(\frac{\theta_{\bf r}}{2}\right) e^{\frac{i}{2}(\varphi_{\bf
    r}+\psi_{\bf r})} \\
\end{array}
\right ) .
\label{Rr} 
\end{equation}
Here we define ${\bf n_r}$ by its polar and azimuthal angles
$\theta_{\bf r}$ and $\varphi_{\bf r}$. The angle $\psi_{\bf r}$ comes
from the U(1) gauge freedom in the definition of $R_{\bf r}$. We obtain
\begin{eqnarray}
A^x_{0{\bf r}} &=& \frac{i}{2} \dot\theta_{\bf r} \sin\psi_{\bf r} 
-\frac{i}{2} \dot \varphi_{\bf r} \sin\theta_{\bf
  r} \cos\psi_{\bf r}, \nonumber \\
A^y_{0{\bf r}} &=& \frac{i}{2} \dot \theta_{\bf r} \cos\psi_{\bf r}
+ \frac{i}{2} \dot\varphi_{\bf r} \sin\theta_{\bf r}\sin\psi_{\bf r} ,
\nonumber \\ 
A^z_{0{\bf r}} &=& \frac{i}{2} \dot \varphi_{\bf r} \cos\theta_{\bf
  r} + \frac{i}{2} \dot\psi_{\bf r} , \nonumber \\ 
B^x_{0{\bf r}} &=& -\sin\theta_{\bf r} \cos\psi_{\bf r} , \nonumber \\
B^y_{0{\bf r}} &=& \sin\theta_{\bf r} \sin\psi_{\bf r} , \nonumber \\
B^z_{0{\bf r}} &=& \cos\theta_{\bf r} . 
\end{eqnarray}
$A^\nu_{\mu{\bf r}}$ ($\mu\neq 0$) is obtained from $A^\nu_{0{\bf r}}$ with the
replacement $\partial_\tau \to -i\partial_\mu$. 

 We find (see Appendix \ref{app:O3}) 
\begin{eqnarray}
\langle S_p \rangle &=&  - i \frac{m_0}{U} \int_0^\beta d\tau
    \sum_{\bf r} (-1)^{\bf r} ( 
    \dot\varphi_{\bf r} \cos\theta_{\bf r} + \dot\psi_{\bf r} ) ,
\nonumber \\ 
\langle S_{h_0} \rangle &=& 0, \nonumber \\
\langle S^2_{h_0}\rangle  &=& \Pi^{xx}_{00} \int_0^\beta d\tau
\sum_{\bf r}  ({\bf  h}_0\times {\bf n_r})^2 , \nonumber \\
\langle S_{h_0} S_p \rangle &=& \frac{i}{2} \Pi^{xx}_{00}
\int_0^\beta d\tau \sum_{\bf r} {\bf h}_0 \cdot ({\bf n_r} \times
    \dot {\bf n}_{\bf r}) , \nonumber \\
\langle S_{h_0} S_l \rangle &=& m _0 \Pi^{xx}_{00}
\int_0^\beta d\tau \sum_{\bf r} {\bf h}_0 \cdot {\bf L_r} ,
\end{eqnarray}
and $\langle S_{h_0} S_{\delta m} \rangle = \langle S_{\delta m} S_p
\rangle = \langle S_{\delta m} S_l \rangle =0$. There is therefore no
coupling between amplitude and direction fluctuations in the limit of
a small magnetic field.
The HF correlation function $\Pi^{xx}_{00}\equiv\Pi^{xx}_{00}(0,0) $
is given in Appendix \ref{app:hfcorr2}. Using the result of
Ref.~\onlinecite{KB2} for $S[{\bf n},{\bf L};h_0=\delta m=0]$, one has
\begin{eqnarray}
S[{\bf n},{\bf L}] &=& \int_0^\beta d\tau \sum_{\bf r} \biggl[
\frac{\langle -K\rangle}{16} (\boldsymbol{\nabla} {\bf n_r})^2
\nonumber \\ && 
+\frac{\Pi^{xx}_{00}}{8} (\dot {\bf n}_{\bf r}-2i {\bf h}_0 \times {\bf
  n_r})^2 \nonumber \\ &&
+ m_0^2 \left(\frac{1}{U}-\frac{\Pi^{xx}_{00}}{2} \right) {\bf
  L}^2_{\bf r} \nonumber \\ && 
- \frac{i}{2} m_0 \Pi^{xx}_{00} {\bf L_r} \cdot
({\bf n_r} \times \dot {\bf n}_{\bf r}-2i {\bf h}_0) \biggr] \nonumber
\\ && + S_B[{\bf n}] , 
\label{SnL}
\end{eqnarray}
where we have taken the continuum limit in real space. $\langle
K\rangle$ is the mean kinetic energy in the HF state 
(for $h_0=0$). We denote the Berry phase term $\langle S_p \rangle$ by
$S_B[{\bf n}]$. Integrating out the ${\bf L}$ field with the
constraint ${\bf L_r} \perp {\bf n_r}$, we finally obtain 
\begin{eqnarray}
S[{\bf n}] &=& \frac{\bar \rho^0_s}{2} \int_0^\beta d\tau \int d^2r \left[
  (\boldsymbol{\nabla} {\bf n_r})^2 +\frac{ (\dot {\bf n}_{\bf r}-2i {\bf
  h}_0 \times {\bf n_r})^2}{\bar c^2} \right]   \nonumber \\ &&
+  S_B[{\bf n}] , \label{nlsmO3} 
\end{eqnarray}
where
\begin{eqnarray}
\bar\rho^0_s &=& \frac{\langle -K\rangle}{8} , \nonumber \\
\bar c^2 &=&  \frac{\langle -K\rangle}{2} \left(
  \frac{1}{\Pi^{xx}_{00}} -\frac{U}{2} \right) .
\end{eqnarray}
$\bar\rho^0_s$ and
$\bar c$ are the spin stiffness and the spin-wave velocity in the absence
of field ($h_0=0$, i.e. at half-filling in the attractive model). It
can be checked 
analytically that the expression of $\bar c$ agrees with
(\ref{velocity}) evaluated at $h_0=0$. The action (\ref{nlsmO3}) is
valid in the hydrodynamic regime defined by the momentum-space cutoff
$\Lambda\sim{\rm min}(1,2m_0/\bar c)$. In the strong-coupling limit, $\bar
c=\sqrt{2}J$ and $\bar\rho^0_s=J/4$, we recover the NL$\sigma$M
obtained from the Heisenberg model in a magnetic
field. \cite{Sachdev99} The crossover temperature $T_X$ can
be obtained from the criterion $\xi\sim 1$, where $\xi$ is the AF
correlation length deduced from the NL$\sigma$M. \cite{KB2} However,
since $T_X$ is weakly doping dependent near half-filling, we can also
consider the estimate obtained in Sec.~\ref{subsec:diag}
[Eq.~(\ref{Txdef})].

\subsection{Low-temperature limit}
\label{subsec:O3_lt} 

\begin{figure}
\epsfxsize 6.cm
\epsffile[50 460 245 590]{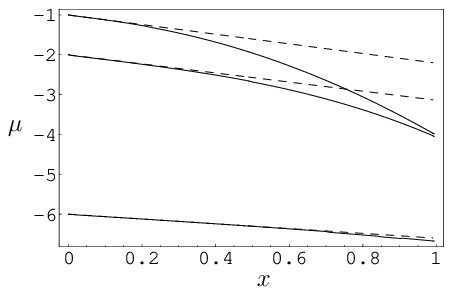}
\caption{Zero-temperature chemical potential $\mu=h_0-U/2$  {\it vs}
   doping $x$ for 
  $U=2t$, $4t$ and $12t$. The dashed lines show the result obtained  from
  Eq.~(\ref{h0O3}).  The solid lines reproduce the results obtained
   by solving Eqs.~(\ref{gapeq1}-\ref{gapeq2}) [see
   Fig.~\ref{fig:mu_delta_theta_U_fixe}]. 
} 
\label{fig:mu_so2_U=4}
\end{figure}

In the low-temperature limit, one expects to recover the results
obtained in Sec.~\ref{sec:O2}. Let us
first consider the NL$\sigma$M (\ref{nlsmO3}) within a static
saddle-point approximation where the N\'eel field ${\bf n}_{\bf
  r}^{\rm cl}$ lies in the $(x,y)$ plane. The classical action reads
\begin{equation}
S_{\rm cl} = - N\beta \frac{2\bar\rho^0_s h_0^2}{\bar c^2} .
\label{Scl}
\end{equation}
The magnetic field $h_0$ is determined by the condition $\langle
c^\dagger_{\bf r} \sigma^z c_{\bf r}\rangle=(N\beta)^{-1}
\partial_{h_0} \ln Z =-x$. From (\ref{Scl}), we deduce
\begin{equation}
h_0 = - x \frac{\bar c^2}{4\bar\rho^0_s} .
\label{h0O3}
\end{equation}
The chemical potential $\mu\equiv h_0-U/2$ obtained from (\ref{h0O3})
is in very good agreement with the result of Sec.~\ref{sec:O2} for
$x\lesssim 0.2$ (Fig.~\ref{fig:mu_so2_U=4}). In the strong-coupling limit, 
$\bar c=\sqrt{2}J$ and $\bar\rho^0_s=J/4$, we find $h_0=-2Jx$ as in
Sec.~\ref{sec:O2}.  

Let us now consider fluctuations about the N\'eel state defined by
${\bf n}^{\rm cl}_{\bf r}$. \cite{note5} We have shown in
Sec.~\ref{subsec:leea1} that the phase of the superconducting order
parameter is obtained from $\Theta_{\bf r}=-{\rm arctan}(\Omega^y_{\bf
  r}/\Omega^x_{\bf r})+{\bf Q}\cdot {\bf r}$ [Eq.~(\ref{transformation3})]. 
From (\ref{Haldec}), we then deduce $\Theta_{\bf r}=-\varphi_{\bf
  r}$. It is also clear that the $\theta$ field is related to the
${\bf q}={\bf Q}$ charge fluctuations. In the low-temperature limit
$|n^z_{\bf r}|\ll 1$. The low-energy effective action
$S[\Theta,\theta]$ is obtained by assuming
$|\boldsymbol{\nabla}\Theta_{\bf r}|,|\dot \Theta_{\bf
  r}|,|\boldsymbol{\nabla}\theta_{\bf r}|,|\dot\theta_{\bf r}|,
|\theta_{\bf r}-\pi/2|\ll 1$. For the 
calculation of the Berry phase term $S_B[{\bf n}]$, we choose
$\psi_{\bf r}=(-1)^{\bf r}\varphi_{\bf r}U/(2m_0)$. As in
Sec.~\ref{sec:O2},  this gauge choice ensures that electromagnetic
gauge invariance is satisfied. \cite{note4} This gives
$S[\Theta,\theta] =S_{\rm cl}+S[\Theta]+S[\theta]$ where 
\begin{eqnarray}
S[\Theta] &=& \frac{\bar\rho^0_s}{2} \int_0^\beta d\tau \int d^2 r 
\left[ (\boldsymbol{\nabla} \Theta_{\bf r})^2 + \frac{\dot
    \Theta_{\bf r}^2}{\bar c^2} \right] \nonumber \\ &&
+ \frac{i}{2} \rho_0 \int_0^\beta d\tau \int d^2 \dot \Theta_{\bf r}, 
 \label{Sphi} \\ 
S[\theta] &=& \frac{\bar\rho^0_s}{2} \int_0^\beta d\tau \int d^2 r \biggl[
  (\boldsymbol{\nabla} \theta_{\bf r})^2  + \frac{\dot
    \theta_{\bf r}^2}{\bar c^2} \nonumber \\ && + \frac{(2h_0)^2}{\bar
    c^2} \left(\theta_{\bf r}-   \frac{\pi}{2} \right)^2 \biggr] .
\label{Stheta}  
\end{eqnarray}
$S[\Theta]$ corresponds to the $h_0=0$ limit of the phase-only action
(\ref{phaseonly}) derived in Sec.~\ref{sec:O2}. $S[\theta]$ describes
${\bf q}={\bf Q}$ charge fluctuations with a gap $2|h_0|$. These
fluctuations were not considered in Sec.~\ref{sec:O2}. We show in the
next section that they play an important role when $|h_0|$ is small,
i.e. near half-filling in the attractive model.

\subsection{High-temperature limit: SO(3)$\to$SO(2) crossover and BKT
  transition} 
\label{subsec:O3_ht} 

The preceding results are valid at low temperatures when
fluctuations of $\theta_{\bf r}$ are small. In this regime, the SO(3)
rotation symmetry is broken by the magnetic field. At higher temperatures,
the gapped mode will be thermally excited thus restoring the SO(3)
spin-rotation symmetry. In this section, we determine the crossover
temperature $T_{\rm SO(3\to 2)}$ separating these two regimes and the
BKT transition temperature $T_{\rm BKT}\leq T_{\rm SO(3\to 2)}$. 

The SO(3)$\to$SO(2) crossover can be understood from 
renormalization group (RG) arguments. Following the standard
procedure, \cite{Chakravarty89} we write the N\'eel field ${\bf n_r}$
as $((1-{{\bf\Pi}_{\bf r}}^2)^{1/2},{\bf\Pi_r})$,
${\bf\Pi_r}=(\Pi^y_{\bf r},\Pi^z_{\bf r})$, with
$|{\bf\Pi_r}|\ll 1$. This yields the spin-wave action
\begin{equation}
S = \frac{\bar\rho_s^0}{2} \int_0^\beta d\tau \int d^2 r \left[
  (\boldsymbol{\nabla} {\bf\Pi_r})^2 + \frac{\dot {\bf \Pi}_{\bf r}^2}{\bar
  c^2} + \frac{4h_0^2}{\bar c^2} {\Pi^z_{\bf r}}^2 \right] + S_1 ,
\end{equation} 
with momenta bounded by the NL$\sigma$M cutoff $\Lambda$. 
In agreement with Sec.~\ref{subsec:O3_lt}, we find a gap in the (bare)
propagator 
of the $\Pi^z$ field. The interaction part $S_1$ of the action is
evaluated at $h_0=0$. RG equations are obtained by integrating out
degrees of freedom with momenta between $\Lambda$ and $\Lambda
e^{-dl}$ and rescaling momenta, energies and
fields. \cite{Chakravarty89} At the beginning of the RG procedure, we
can ignore the gap in the fluctuations of $\Pi^z$ and treat the magnetic
field perturbatively. We then obtain the usual RG equations of the SO(3)
NL$\sigma$M ($h_0=0$) for the dimensionless coupling constants $g=\bar
c\Lambda/\bar\rho_s^0$ and 
$t=T/\bar\rho_s^0$, together with the flow equation $dh_0/dl=h_0$. The RG flow
should be stopped when we reach the strong-coupling regime of the
SO(3) NL$\sigma$M ($t(l)\sim 1$) or when the magnetic field cannot be
treated perturbatively anymore ($|2h_0(l)|\sim \bar c
\Lambda$). Introducing the characteristic length $\Lambda^{-1}e^l$, the  
former condition defines the SO(3) AF correlation length
$\xi$ while the latter defines the magnetic length
$\xi_{h_0}=c/|2h_0|$. We are therefore left to consider two different
cases. If $\xi\lesssim \xi_{h_0}$, $t(l)\sim 1$ occurs before $|2h_0(l)|\sim
\bar c\Lambda$. The system is then disordered by SO(3) fluctuations before
reaching the SO(2) regime. This will occur above the crossover
temperature $T_{\rm SO(3\to 2)}$ defined by $\xi(T_{\rm SO(3\to 2)}) \sim
\xi_{h_0}$.\cite{Fisher89}  
If $\xi_{h_0}\lesssim \xi$, i.e. $T\lesssim T_{\rm SO(3\to 2)}$, an
SO(3)$\to$SO(2) crossover takes place. For $|2h_0(l)|>\bar c\Lambda$, 
fluctuations of $\Pi^z$ are suppressed, and only fluctuations of
$\Pi^y$ survive. This regime is described by an SO(2) NL$\sigma$M with
coupling constants $t(l_{h_0})$ and $g(l_{h_0})$ where $l_{h_0}$ is
defined by $|2h_0(l_{h_0})|\sim \bar c\Lambda$. The very existence of the SO(2)
regime implies $t(l_{h_0})\lesssim 1$. The latter inequality is also
approximately the condition
for the SO(2) NL$\sigma$M to be in the low-temperature BKT phase. This
implies that the temperature range where the system is disordered
by SO(2) fluctuations is very narrow, so that we can identify the BKT
transition temperature with the SO(3)$\to$SO(2) crossover temperature:
$T_{\rm BKT}\sim T_{\rm SO(3\to 2)}$. 

In order to determine $T_{\rm SO(3\to 2)}$, we use the following
expression for the $h_0=0$ correlation length $\xi$:\cite{Chakravarty89}  
\begin{equation}
\xi(T)=K \frac{\bar c}{\bar\rho_s} \exp\left(\frac{2\pi\bar\rho_s}{T}\right) ,
\end{equation}
where $K\simeq 0.05$ and $\bar\rho_s$ is the zero-temperature spin
stiffness in the N\'eel state. To estimate $\bar\rho_s$, we use the
one-loop RG result,  \cite{Chakravarty89} 
\begin{equation}
\bar\rho_s = \bar\rho^0_s \left( 1- \frac{\bar c\Lambda}{4\pi
  \bar\rho^0_s} \right) .
\end{equation}
$T_{\rm  SO(3\to 2)}$ is therefore given by 
\begin{equation}
T_{\rm SO(3\to 2)} = \frac{2\pi \bar\rho_s}{\ln \left( \frac{
    \bar\rho_s}{2K |h_0|} \right) } 
\simeq  \frac{2\pi \bar\rho_s}{\ln \left( \frac{10 \bar\rho_s}{|h_0|}
  \right) } ,
\label{Tso3} 
\end{equation}
where we have taken $K=0.05$. $T_{\rm SO(3\to 2)}$,
obtained from (\ref{Tso3}), is shown in
Fig.~\ref{fig:temp_U_fixe}. 

The definition of  $T_{\rm SO(3\to
2)}$ is meaningful only  below the crossover temperature $T_X$ which
marks the onset of AF short-range order and defines the temperature
range where the NL$\sigma$M holds.  
Once $T_{\rm SO(3\to 2)}$ becomes of the order of $T_X$, as $|h_0|$
increases, ${\bf q}={\bf Q}$ charge fluctuations (in the attractive
model) can be ignored. In this regime, the analysis of
Sec.~\ref{sec:O2} holds.

\section{Strong-coupling limit}
\label{sec:scr} 

In the strong-coupling limit $U\gg 4t$, we can directly integrate out
the fermions to recover the action $S[{\bf\Omega}]$ of the Heisenberg
model in a magnetic field (Sec.~\ref{subsec:Heis}). We can then go
beyond the hydrodynamic limit (${\bf q}\to 0$) considered in
Sec.~\ref{sec:O2} and obtain the collective mode dispersion over the
entire Brillouin zone. When $|h_0|$ is weak enough
(i.e. away from the low-density limit in the attractive model), the
Heisenberg model reduces to the quantum XY model
(Sec.~\ref{subsec:QXY}). In the low-density 
limit, the Heisenberg model allows to recover the usual action of a
Bose superfluid, including the terms proportional to
$(\boldsymbol{\nabla}\rho_{\bf r})^2$ that were omitted in
Sec.~\ref{sec:O2}, and in turn the Gross-Pitaevskii equation
(Sec.~\ref{subsec:GP}).  

\subsection{Heisenberg model}
\label{subsec:Heis}

We write the action (\ref{action1a}-\ref{action1b}) in terms of the
$\phi$ field 
defined by $c_{\bf r}=R_{\bf r}\phi_{\bf r}$ and $R_{\bf r} \sigma^z
R^\dagger_{\bf r} = \boldsymbol{\sigma} \cdot {\bf\Omega_r}$:
\begin{eqnarray}
S &=& S_{\rm at} + \int_0^\beta d\tau \Bigl[ \sum_{\bf r}
  \phi^\dagger_{\bf r} (R^\dagger_{\bf r} \dot R_{\bf r} - h_0
  R^\dagger_{\bf r} 
  \sigma^z R_{\bf r}) \phi_{\bf r} \nonumber \\ &&
- t \sum_{\langle {\bf r},{\bf r}' \rangle} (\phi^\dagger_{\bf r}
  R^\dagger_{\bf r} R_{{\bf r}'} \phi_{{\bf r}'} + {\rm c.c.}) \Bigr] ,
\\ 
S_{\rm at} &=& \int_0^\beta d\tau \sum_{\bf r} [
  \phi^\dagger_{\bf r} (\partial_\tau -im_0^{\rm HS} \sigma^z) \phi_{\bf r} ] ,
\label{Sat} 
\end{eqnarray}
where we neglect spin amplitude fluctuations and $im_0^{\rm HS}=U/2$
to leading order in $1/U$ (see
Sec.~\ref{sec:O2}). $S_{\rm at}$ is the action in the atomic limit
($t=0$). The effective action of the angular variable ${\bf\Omega}$ is
obtained by integrating out the fermions. To lowest order in $t/U$ and
$h_0$, we obtain
\begin{eqnarray}
S &=& \int_0^\beta d\tau \sum_{\bf r} \bigl[ \langle \phi^\dagger_{\bf
    r} R^\dagger_{\bf r} \dot R_{\bf r} \phi_{\bf r} \rangle -h_0  \langle
    \phi^\dagger_{\bf r} R^\dagger_{\bf r} \sigma^z R_{\bf r} \phi_{\bf r}
    \rangle \bigr] \nonumber \\ && 
-\frac{1}{2} \int_0^\beta d\tau d\tau' \sum_{{\bf r}_1,{\bf r}_2 \atop
    {{\bf r}_1',{\bf r}_2'}} t_{{\bf r}_1,{\bf r}_2}  t_{{\bf
    r}_1',{\bf r}_2'} \bigl\langle (\phi^\dagger_{{\bf r}_1}
    R^\dagger_{{\bf r}_1} R_{{\bf r}_2} \phi_{{\bf r}_2})_\tau
      \nonumber \\ && \times (\phi^\dagger_{{\bf r}_1'}
    R^\dagger_{{\bf r}_1'} R_{{\bf r}_2'} \phi_{{\bf r}_2'})_{\tau'}
    \bigr\rangle ,
\label{action4}
\end{eqnarray}
where $t_{{\bf r},{\bf r}'}$ equals $t$ for ${\bf r},{\bf r}'$ nearest
neighbors and vanishes otherwise. The averages in (\ref{action4}) are
taken with the atomic action. They can be easily calculated using the
parameterization (\ref{Rr}) of $R_{\bf r}$. One finds
\begin{eqnarray}
 && \langle \phi^\dagger_{\bf r} R^\dagger_{\bf r} \dot R_{\bf r}
 \phi_{\bf r} \rangle = \langle {\bf\Omega_r} | \dot {\bf\Omega}_{\bf
 r} \rangle , \nonumber \\ && 
 \langle \phi^\dagger_{\bf r} R^\dagger_{\bf r} \sigma^z R_{\bf r}
 \phi_{\bf r} \rangle  = \Omega^z_{\bf r} , \nonumber \\ &&
\bigl\langle (\phi^\dagger_{{\bf r}_1}
    R^\dagger_{{\bf r}_1} R_{{\bf r}_2} \phi_{{\bf r}_2})_\tau
    (\phi^\dagger_{{\bf r}_1'}
    R^\dagger_{{\bf r}_1'} R_{{\bf r}_2'} \phi_{{\bf r}_2'})_{\tau'}
    \bigr\rangle \nonumber \\ && \simeq \delta_{{\bf r}_1',{\bf r}_2}
 \delta_{{\bf 
 r}_2',{\bf r}_1} \delta(\tau-\tau') 
\frac{1}{U} (1-{\bf\Omega}_{{\bf r}_1} \cdot {\bf\Omega}_{{\bf r}_2}) ,
\end{eqnarray}
where we have used $(R^\dagger_{\bf r}R_{{\bf r}'})_{\sigma\bar\sigma} 
(R^\dagger_{{\bf r}'}R_{\bf r})_{\bar\sigma\sigma}=(1-{\bf\Omega_r} \cdot
{\bf\Omega}_{{\bf r}'})/2$. We have introduced the spin$-\frac{1}{2}$
coherent state $|{\bf\Omega_r}\rangle = R_{\bf r} |\uparrow\rangle
=\cos\left(\frac{\theta_{\bf r}}{2} 
\right) e^{-\frac{i}{2}(\varphi_{\bf r}+\psi_{\bf r})} |\uparrow\rangle +
\sin \left(\frac{\theta_{\bf r}}{2}\right)
e^{\frac{i}{2}(\varphi_{\bf r} -\psi_{\bf r})}
|\downarrow\rangle$. \cite{Auerbach94} $\psi_{\bf r}$ is arbitrary and
corresponds to the U(1) gauge freedom in the definition of the SU(2)
matrix $R_{\bf r}$. We therefore recover the action
\begin{equation}
S = \int_0^\beta d\tau \biggl\lbrace \sum_{\bf r} \bigl[ \langle
  {\bf\Omega_r} | \dot {\bf\Omega}_{\bf r} \rangle - {\bf h}_0 \cdot
  {\bf\Omega_r} \bigr]
+ J \sum_{\langle {\bf r},{\bf
  r}' \rangle } \frac{{\bf\Omega_r} \cdot {\bf\Omega}_{{\bf r}'}}{4}
\biggr\rbrace 
\label{action5} 
\end{equation}
of the Heisenberg model in a magnetic field ${\bf h}_0$. 

Consider first the classical ground-state defined by 
${\bf\Omega}^{\rm cl}_{\bf r}=(-1)^{\bf r}\sin\theta_0 \hat {\bf x}+
\cos\theta_0 \hat {\bf z}$. Minimizing the classical action 
\begin{equation}
S_{\rm cl} = N\beta \left[ -h_0 \cos\theta_0 + \frac{J}{2} 
\cos(2\theta_0) \right] 
\end{equation}
with respect to $\theta_0$, we find
\begin{equation}
\cos\theta_0= \frac{h_0}{2J} 
\end{equation}
if $|h_0|\leq 2J$ and $\theta=\pi$ otherwise (for $h_0\leq 0$). The
condition $\langle c^\dagger_{\bf r} \sigma^z c_{\bf r}\rangle=-x$ translates
into $\langle \Omega^z_{\bf r}\rangle=-x$ (see Appendix
\ref{subapp:scl}), i.e. $\cos\theta_0=-x$ within the classical
approximation. We thus obtain $h_0=-2Jx$ as in
Sec.~\ref{subsubsec:sc1}.  

The Heisenberg model allows to obtain the collective excitations of
the attractive Hubbard model in the strong-coupling limit without
taking the continuum limit.  We introduce the variables $p_{\bf
  r}=(\theta_{\bf r}-\theta_0)/2$ and $q_{\bf r}=\varphi_{\bf r}-{\bf
  Q}\cdot {\bf r}$ and derive the effective action to quadratic order
in $p_{\bf r}$ and $q_{\bf r}-q_{{\bf r}'}$ (for ${\bf r},{\bf r}'$ first
neighbors): 
\begin{eqnarray}
S &=& \int_0^\beta d\tau \biggl\lbrace \sum_{\bf r} \biggl[ \frac{i}{2}
  \sin\theta_0 (p_{\bf 
  r}\dot q_{\bf r}-\dot p_{\bf r}q_{\bf r}) +2h_0 \cos\theta_0
  p^2_{\bf r} \nonumber \\ && -\frac{i}{2} \rho_0 \dot q_{\bf r} \biggr]
+ \frac{J}{4} \sum_{\langle {\bf r},{\bf r}'\rangle} \biggl[
  -2\cos(2\theta_0) (p_{\bf r}+p_{{\bf r}'})^2 \nonumber \\ && +
  \frac{\sin^2\theta_0}{2} (q_{\bf r}-q_{{\bf r}'})^2 \biggr]
  \biggr\rbrace .  \label{Heis1} 
\end{eqnarray}
The Berry phase term $\langle {\bf\Omega_r} | \dot {\bf\Omega}_{\bf
 r} \rangle$ has been evaluated with $\psi_{\bf r}=\varphi_{\bf
 r}$. This gauge choice is similar to the one made in
 Sec.~\ref{sec:O2}. We thus obtain
\begin{equation}
S = -\frac{i}{2}\rho_0 \int_0^\beta d\tau \sum_{\bf r} \dot q_{\bf r} 
+ \frac{1}{2} \sum_{\tilde q} (p_{-\tilde q},q_{-\tilde q})
{\cal D}^{-1}(\tilde q) 
\left(
\begin{array}{l}
p_{\tilde q} \\ q_{\tilde q} 
\end{array}
\right) ,  \label{Heis2} 
\end{equation}
\begin{widetext}
\begin{equation}
{\cal D}^{-1}(\tilde q) = \left(
\begin{array}{lr} 
-4J\cos(2\theta_0)(1+\gamma_{\bf q})+4h_0\cos\theta_0 &
 \omega_\nu\sin\theta_0 \\
-\omega_\nu\sin\theta_0 & J \sin^2\theta_0 (1-\gamma_{\bf q}) 
\end{array}
\right) ,
\label{Heis3} 
\end{equation}
where $\gamma_{\bf q}=(\cos q_x+\cos q_y)/2$. 
Eqs.~(\ref{Heis1}-\ref{Heis3}) assume the fluctuations of $p$ to be
small and are therefore valid only for $h_0\neq 0$,
i.e. away from half-filling in the attractive model.  
At zero-temperature, there is AF long-range order
(i.e. superconducting order in the attractive model). 
Collective modes are then obtained from ${\rm det}{\cal D}^{-1}(\tilde
q)=0$ with the analytic continuation to real frequencies
$i\omega_\nu\to \omega_{\bf q}$. This gives \cite{Kostyrko92a}
\begin{eqnarray}
\omega^2_{\bf q} &=& 2h_0^2 (1-\gamma_{\bf q}) -(2h_0^2-4J^2)
(1-\gamma^2_{\bf q}) \nonumber \\ &=& 
8J^2x^2 (1-\gamma_{\bf q}) -4J^2 (2x^2-1) (1-\gamma^2_{\bf q}) .
\label{dispersion_sc} 
\end{eqnarray}
Fig.~\ref{fig:dispersion_sc} shows $\omega_{\bf q}$ for different
values of the doping $x$. 
For ${\bf q}\to 0$, we obtain a spin-wave mode (satisfying $p\simeq
0$) with dispersion $\omega_{\bf q}=c|{\bf q}|$,
$c=\sqrt{2}J\sqrt{1-x^2}$.  This mode corresponds to the Bogoliubov
mode obtained in Sec.~\ref{subsubsec:po_bose}. In
the vicinity of ${\bf Q}=(\pi,\pi)$, we find a mode with the dispersion
\begin{equation}
\omega^2_{\bf q} = (2h_0)^2 \pm \sqrt{2}J|1-3 x^2|^{1/2} {\bf q}^2 ,
\end{equation}
where the $+$ ($-$) sign refers to the case $x<1/\sqrt{3}$
($x>1/\sqrt{3}$). When $x<1/\sqrt{3}$, this mode
corresponds to a local minimum of the energy with a gap $|2h_0|$. It
involves fluctuations along the magnetic field axis which correspond
in the attractive model to ${\bf q}={\bf Q}$ charge-density
fluctuations. At the critical value of the doping $x_c=1/\sqrt{3}$, this
local minimum becomes a local maximum. Note that the value $|2h_0|$ of
the gap was also found in Sec.~\ref{subsec:O3_lt} for a weak magnetic
field but for all values of the interaction $U$. \cite{Benfatto02} 

\begin{figure}
\epsfxsize 6.cm
\epsffile[30 450 245 590]{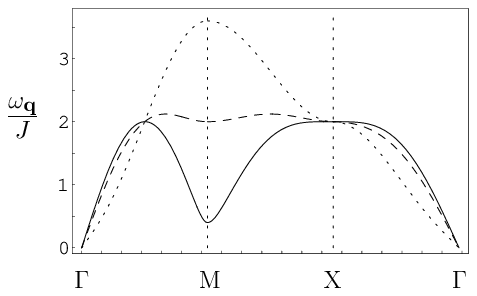}
\caption{Collective mode dispersion $\omega_{\bf q}$ in the
  strong-coupling limit as obtained from the Heisenberg model
  [Eq.~(\ref{dispersion_sc})] for $x=0.1$ (solid line), 0.5 (dashed
  line) and 0.9 (dotted line). [$\Gamma=(0,0)$, ${\rm M}=(\pi,\pi)$
  and ${\rm X}=(\pi,0)$.]
} 
\label{fig:dispersion_sc}
\end{figure}

\subsection{Quantum XY model} 
\label{subsec:QXY}

In this section, we show that the attractive Hubbard
model in the strong-coupling limit reduces to the quantum XY model
(except in the low-density limit). 
The effective action of the phase $\Theta_{\bf r}=-q_{\bf r}=-\varphi_{\bf
  r}+{\bf Q}\cdot {\bf r}$ of the superconducting order parameter is
obtained by 
integrating out the $p$ field. To quadratic order in $p$, the action reads
\begin{eqnarray}
S[p,\Theta] &=& \int_0^\beta d\tau \biggl\lbrace \sum_{\bf r} \biggl[
  \frac{i}{2} \rho_0 \dot\Theta_{\bf r} - i \sin\theta_0
  p_{\bf r}\dot\Theta_{\bf r} + 2h_0 \cos\theta_0 p^2_{\bf r} \biggr] 
  - \frac{J}{2} \cos(2\theta_0)
  \sum_{\langle {\bf r},{\bf r}'\rangle} (p_{\bf r}+p_{{\bf r}'})^2
  \nonumber \\ && 
+ \frac{J}{4} \sum_{\langle {\bf r},{\bf r}'\rangle} \bigl[
  1-\cos(\Theta_{\bf r}-\Theta_{{\bf r}'}) \bigr] 
\bigl[ \sin^2\theta_0 + \sin(2\theta_0)(p_{\bf r}+p_{{\bf r}'}) +
  4\cos^2\theta_0 p_{\bf r}p_{{\bf r}'} -2 \sin^2\theta_0(p^2_{\bf
  r}+p^2_{{\bf r}'}) \bigr] \biggr\rbrace .
\end{eqnarray}
\end{widetext}
Since the fluctuations of $p$ are small, we can neglect all the terms
but $\sin^2\theta_0$ in the coefficient of $1-\cos(\Theta_{\bf
  r}-\Theta_{{\bf r}'})$. This approximation breaks down when
$\sin\theta_0\to 0$ (i.e. $\theta_0\to\pi$), which corresponds to the
low-density limit of the attractive Hubbard model. Integrating out the
$p$ field, we obtain the action of the quantum XY model
\begin{eqnarray}
S[\Theta] &=&  \int_0^\beta d\tau \biggl\lbrace \sum_{\bf r}
\biggl[\frac{i}{2} \rho_0 \dot\Theta_{\bf r} + 
\frac{\dot\Theta^2_{\bf r}}{16J} \biggr] \nonumber \\ && + \frac{J}{4} (1-x^2) 
\sum_{\langle {\bf r},{\bf r}'\rangle} \bigl[1-\cos(\Theta_{\bf
    r}-\Theta_{{\bf r}'}) \bigr] \biggr\rbrace . 
\end{eqnarray}
Taking the continuum limit, we recover the phase-only action derived
in Sec.~\ref{sec:O2} for $U\gg 4t$
[Eqs.~(\ref{phaseonly},\ref{rho_c_sc})].

\subsection{Low-density limit: Gross-Pitaevskii equation}
\label{subsec:GP}

In the strong-coupling limit, fermions form tightly bound pairs which
behave as hard-core bosons (the hard-core
constraint comes from the Pauli principle which prevents double
occupancy of a lattice site). In the low-density limit, the hard-core
constraint does not matter anymore and we expect to recover the usual
action of a Bose superfluid, and in turn the Gross-Pitaevskii equation. 

Using Eqs.~(\ref{rho_delta},\ref{mfixe}),
\begin{eqnarray}
\Omega^\pm_{\bf r} &=& (-1)^{\bf r} [\rho_{\bf r}(2-\rho_{\bf
  r})]^{1/2} e^{\mp i \Theta_{\bf r}} , \nonumber \\ 
\Omega^z_{\bf r} &=& \rho_{\bf r}-1 ,
\label{omega_delta}
\end{eqnarray}
we deduce from the Heisenberg model [Eq.~(\ref{action5})]
\begin{eqnarray}
S[\rho,\Theta] &=& \int_0^\beta d\tau \int d^2 r \biggl\lbrace
\frac{i}{2} \rho_{\bf r} \dot \Theta_{\bf r} + J \rho^2_{\bf r}
-(h_0+2J) \rho_{\bf r} \nonumber \\ && + \frac{J}{4} \biggl[
\frac{(\boldsymbol{\nabla} \rho_{\bf r})^2}{4\rho_{\bf r}} + \rho_{\bf r}
(\boldsymbol{\nabla} \Theta_{\bf r})^2 \biggr] \biggr\rbrace 
\end{eqnarray}
in the low-density limit ($\rho_{\bf r}\ll 1$) and to second order in
gradient. We have taken the continuum limit in real space. 
Introducing the boson density $\rho_{b{\bf r}}=\rho_{\bf
  r}/2$, we recover the action of a Bose superfluid, 
\begin{eqnarray}
S[\rho_b,\Theta] &=& \int_0^\beta d\tau \int d^2 r \biggl\lbrace
i \rho_{b{\bf r}} \dot \Theta_{\bf r} + \frac{g}{2} \rho^2_{b{\bf r}}
-\mu_b \rho_{b{\bf r}} \nonumber \\ && + \frac{1}{2m_b} \biggl[
\frac{(\boldsymbol{\nabla} \rho_{b{\bf r}})^2}{4\rho_{b{\bf r}}} +
\rho_{b{\bf r}} 
(\boldsymbol{\nabla} \Theta_{\bf r})^2 \biggr] \biggr\rbrace 
\nonumber \\ 
&=& \int_0^\beta d\tau \int d^2 r \biggl[ \Psi^*_{\bf r}
  \biggl(\partial_\tau - \mu - \frac{\boldsymbol{\nabla}^2}{2m_b} \biggr)
  \Psi_{\bf r} \nonumber \\ && 
+\frac{g}{2} (\Psi^*_{\bf r}\Psi_{\bf r})^2 \biggr] , 
\label{action6} 
\end{eqnarray}
where $\Psi_{\bf r}=\sqrt{\rho_{b{\bf r}}}e^{i\Theta_{\bf r}}$. 
\begin{eqnarray}
m_b &=& \frac{1}{J}, \nonumber \\  
\mu_b &=& 2h_0+4J, \nonumber \\ 
g &=& 8J , 
\end{eqnarray}
are the mass, chemical potential and interaction constant of the
bosons, respectively. Note that if we neglect the
$(\boldsymbol{\nabla}\rho_{b{\bf r}})^2$ term and replace $\rho_{b{\bf
    r}}$ by $\rho_{b0}$ in the $ (\boldsymbol{\nabla} \Theta_{\bf
  r})^2$ term, we recover the action $S[\rho,\Theta]$ (for $x\to 1$)
derived in Sec.~\ref{subsubsec:sc3}. 

The classical equations of motion derived from the action
(\ref{action6}) yield the Gross-Pitaevskii equation
\begin{equation}
\left( \partial_\tau -\mu_b - \frac{\boldsymbol{\nabla}^2}{2m_b}
\right) \Psi_{\bf r} + g |\Psi_{\bf r}|^2 \Psi_{\bf r} = 0 
\label{GPeq}
\end{equation}
for the complex order parameter $\Psi_{\bf r} = \sqrt{\rho_{b{\bf r}}}
e^{i\Theta_{\bf r}}$. As pointed out in Sec.~\ref{subsubsec:sc3},
$\Psi_{\bf r}$ equals the superconducting order parameter $\Delta_{\bf
  r}=|\Delta_{\bf r}|e^{i\Theta_{\bf r}}$ in the strong-coupling
low-density limit of the attractive Hubbard model. 
Alternatively, the Gross-Pitaevskii equation can be obtained from the
semiclassical spin dynamics,
\begin{equation}
i \dot {\bf\Omega}_{\bf r} = \frac{J}{2} \sum_{\boldsymbol{\delta}}
     {\bf\Omega}_{{\bf r}+{\boldsymbol{\delta}}} \times {\bf\Omega_r} -2 {\bf
     h}_0 \times {\bf\Omega_r} ,
\label{eqsc1}
\end{equation}
within a second-order gradient expansion and in the low-density
limit $\rho_{\bf r}=\Omega^z_{\bf r}+1 \ll 1$. The sum over
${\boldsymbol{\delta}}$ in (\ref{eqsc1}) denotes a sum over nearest
neighbors.  

For continuum models, the time-independent Gross-Pitaevskii equation
has been obtained as the strong-coupling limit of the Bogoliubov-de
Gennes equations. \cite{Pieri03} It has also been shown that the
results obtained in the strong-coupling limit of an RPA calculation
about the BCS state can be reproduced from the linearized version of
the time-dependent Gross-Pitaevskii equation. \cite{Strinati03} 
In this section, we have directly shown the equivalence, in the
low-density limit of the lattice case, between the
low-energy effective action of the superfluid order parameter
$\Delta_{\bf r}\equiv \sqrt{\rho_{\bf r}/2}e^{i\Theta_{\bf r}}$  and
the action of a Bose superfluid. Our approach is not limited to the
low-density limit. Eq.~(\ref{eqsc1}), together with (\ref{omega_delta}),
holds for any density and can be considered as a generalization of the
Gross-Pitaevskii equation to the lattice case.

\section{Concluding remarks}

In this paper, we have studied the 2D attractive Hubbard model using
the mapping onto the half-filled repulsive model in a uniform magnetic
field coupled to the fermion spins. Our approach reproduces, in a
unique framework, a number of previously known results.

One of our main new results is the derivation of a low-energy effective 
action $S[\rho,\Theta]$ which is valid for all values of doping
$x=1-\rho_0$ and interaction strength $U$. $S[\rho,\Theta]$ has been obtained 
by integrating out amplitude fluctuations ($|\Delta|$) and therefore does
not describe the full dynamics of the superconducting order parameter
$\Delta_{\bf r}=|\Delta_{\bf r}|e^{i\Theta_{\bf r}}$. Nevertheless, it
is similar to the action of a Bose superfluid with order parameter
$\Psi_{\bf r}=\sqrt{\rho_{\bf r}/2}e^{i\Theta_{\bf r}}$ where the mass
of the ``bosons'' and their 
mutual interaction depend on $x$ and $U$. This ensures that a Fermi
superfluid, as described by the 2D attractive Hubbard model, will
behave similarly to a Bose superfluid and exhibit the same macroscopic
quantum phenomena. The effective action $S[\rho,\Theta]$ also
describes the smooth crossover between the weak-coupling BCS limit and
the Bose limit of preformed (local) pairs. 

Another important result obtained by our approach is a complete
description of the phase diagram of the 2D attractive Hubbard
model. From the phase-only action $S[\Theta]$, we are able to extract
an effective XY model and in turn the BKT phase transition
temperature $T_{\rm BKT}$. We identify a RC regime of superconducting
fluctuations in the temperature range $T_{\rm BKT}\leq T\leq T_X$ and
an incoherent-pair regime (with no superconducting short-range order)
for $T_X\leq T\leq T_{\rm pair}$. The values 
obtained for $T_{\rm pair}$ and $T_X$ are in good agreement with
numerical results (when available). Near half-filling, we find that
$T_{\rm BKT}$ is suppressed due to the strong ${\bf q}=(\pi,\pi)$
charge fluctuations which enlarge the symmetry of the order parameter
to SO(3). 

In the strong-coupling limit, the attractive Hubbard model maps onto
the Heisenberg model in a uniform field. The latter reduces to the
quantum XY model (except for a weak field, i.e. in the low-density
limit of the attractive model). In the low-density limit, we recover
the usual action of a Bose superfluid (including the terms
proportional to $(\boldsymbol{\nabla}\rho_{\bf r})^2$) and in turn the
Gross-Pitaevskii equation.

\acknowledgments

We thank A.M.-S. Tremblay and B. Delamotte for discussions and
K. Borejsza for advice regarding the numerical calculations.

\begin{widetext}

\appendix

\section{HF current-current correlation function}
\label{app:hfcorr}

In this appendix we calculate the HF current-current
correlation function
\begin{equation}
\Pi^{\nu\nu'}_{\mu\mu'}(\tilde q,\tilde q') = \langle j^\nu_\mu(\tilde q)
j^{\nu'}_{\mu'}(-\tilde q') \rangle_c ,
\label{jjHF}
\end{equation}
for $\tilde q,\tilde q'=0,{\bf Q}$. The calculation is performed at
finite temperature in the classical limit: $\lim_{\tilde q \to 0}\equiv
\lim_{{\bf q} \to 0} \lim_{\omega_\nu\to 0}$ and $\lim_{\tilde q \to
  {\bf Q}}\equiv \lim_{{\bf q} \to {\bf Q}}\lim_{\omega_\nu \to 0}$.
$j^\nu_\mu(\tilde q)$ is the Fourier transformed field of
\begin{eqnarray}
j^\nu_{0{\bf r}} &=& \phi^\dagger_{\bf r} \sigma^\nu \phi_{\bf r} ,
\nonumber \\
j^\nu_{\mu{\bf r}} &=& -it \phi^\dagger_{\bf r} \sigma^\nu \phi_{{\bf
    r}+\hat {\boldsymbol{\mu}}} + {\rm c.c.} \,\,\, (\mu=x,y) .
\label{spincrt}
\end{eqnarray}
We have 
\begin{eqnarray}
&& j^\nu_\mu(\tilde q) = \frac{1}{\sqrt{\beta N}} \sum_{\tilde k} v_\mu({\bf
  k},{\bf q}) \phi^\dagger_{\tilde k} \sigma^\nu \phi_{\tilde k+\tilde
  q} , \nonumber \\
&& v_0({\bf k},{\bf q}) = 1, \nonumber \\
&& v_{\mu \neq 0}({\bf k},{\bf q}) = -it (e^{i(k_\mu+q_\mu)}-e^{-ik_\mu}) . 
\end{eqnarray}
$\tilde k=({\bf k},i\omega)$ where $\omega=\pi T(2n+1)$ ($n$ integer)
is a fermionic Matsubara frequency. 
In (\ref{jjHF}), one must have $\tilde q=\tilde q'$ or $\tilde
q=\tilde q'+{\bf Q}$. The correlation functions of interest are 
\begin{eqnarray}
\Pi^{\nu\nu'}_{\mu\mu'}(0,0) &=& -\frac{1}{\beta N} \sum_{\tilde
  k,\sigma_1,\sigma_2} \bigl[ v_\mu({\bf k},0) v_{\mu'}({\bf k},0)
  \sigma^\nu_{\sigma_1 \sigma_2}  \sigma^{\nu'}_{\sigma_2 \sigma_1}
G_{\sigma_1}(\tilde k) G_{\sigma_2}(\tilde k) \nonumber \\ && + 
v_\mu({\bf k},0) v_{\mu'}({\bf k}+{\bf Q},0)
  \sigma^\nu_{\sigma_1 \sigma_2}  \sigma^{\nu'}_{\bar\sigma_2 \bar\sigma_1}
F_{\sigma_1}(\tilde k) F_{\sigma_2}(\tilde k) \bigr] , \nonumber \\
\Pi^{\nu\nu'}_{\mu\mu'}({\bf Q},{\bf Q}) &=& -\frac{1}{\beta N} \sum_{\tilde
  k,\sigma_1,\sigma_2} \bigl[ v_\mu({\bf k},{\bf Q}) v_{\mu'}({\bf
  k}+{\bf Q},-{\bf Q})
  \sigma^\nu_{\sigma_1 \sigma_2}  \sigma^{\nu'}_{\sigma_2 \sigma_1}
G_{\sigma_1}(\tilde k) G_{\sigma_2}(\tilde k+{\bf Q}) \nonumber
  \nonumber \\ && +  
v_\mu({\bf k},{\bf Q}) v_{\mu'}({\bf k},-{\bf Q})
  \sigma^\nu_{\sigma_1 \sigma_2}  \sigma^{\nu'}_{\bar\sigma_2 \bar\sigma_1}
F_{\sigma_1}(\tilde k) F_{\sigma_2}(\tilde k+{\bf Q}) \bigr] , \nonumber \\
\Pi^{\nu\nu'}_{\mu\mu'}(0,{\bf Q}) &=& -\frac{1}{\beta N} \sum_{\tilde
  k,\sigma_1,\sigma_2} \bigl[ v_\mu({\bf k},0) v_{\mu'}({\bf k},-{\bf Q})
  \sigma^\nu_{\sigma_1 \sigma_2}  \sigma^{\nu'}_{\sigma_2 \bar\sigma_1}
F_{\sigma_1}(\tilde k) G_{\sigma_2}(\tilde k)  \nonumber \nonumber \\ && + 
v_\mu({\bf k},0) v_{\mu'}({\bf k}+{\bf Q},-{\bf Q})
  \sigma^\nu_{\sigma_1 \sigma_2}  \sigma^{\nu'}_{\bar\sigma_2 \sigma_1}
G_{\sigma_1}(\tilde k) F_{\sigma_2}(\tilde k) \bigr] , 
\label{A3}
\end{eqnarray}
where 
\begin{eqnarray}
v_\mu({\bf k},0) &=& \delta_{\mu,0} + \delta_{\mu\neq 0} 2t \sin
k_\mu, \nonumber \\ 
v_\mu({\bf k}+{\bf Q},0) &=& \gamma_\mu v_\mu({\bf k},0) , \nonumber \\
v_\mu({\bf k},{\bf Q}) &=& \delta_{\mu,0} + \delta_{\mu\neq 0} 2it
\cos k_\mu, \nonumber \\ 
v_\mu({\bf k}+{\bf Q},-{\bf Q}) &=& \gamma_\mu v_\mu({\bf k},{\bf Q}) ,
\end{eqnarray}
and $\gamma_\mu = \delta_{\mu,0}-\delta_{\mu\neq 0}$. We use the
notation $\delta_{\mu\neq 0}=1-\delta_{\mu,0}$ and
$\bar\sigma=-\sigma$. The HF propagators $G$ 
and $F$ are defined in (\ref{GF}). We have 
\begin{eqnarray}
\frac{1}{\beta} \sum_\omega G_\sigma({\bf k},i\omega) G_{\sigma'}({\bf
  k}',i\omega) 
&=& \frac{-1}{2(E^2_{{\bf k}\sigma}-E^2_{{\bf k}'\sigma'})} \left[
  T_{{\bf k}\sigma} \left( E_{{\bf k}\sigma} +
  \frac{\epsilon_{{\bf k}\sigma}\epsilon_{{\bf k}'\sigma'}}{E_{{\bf
        k}\sigma}} \right) -
T_{{\bf k}'\sigma'} \left( E_{{\bf k}'\sigma'} +
  \frac{\epsilon_{{\bf k}\sigma}\epsilon_{{\bf k}'\sigma'}}{E_{{\bf
        k}'\sigma'}} \right) \right] , 
\nonumber \\ 
\frac{1}{\beta} \sum_\omega G_\sigma({\bf k},i\omega) G_{\sigma'}({\bf
k}'+{\bf Q},i\omega)
&=&   \frac{-1}{2(E^2_{{\bf k}\sigma}-E^2_{{\bf k}'\bar\sigma'})} \left[
  T_{{\bf k}\sigma} \left( E_{{\bf k}\sigma} -
  \frac{\epsilon_{{\bf k}\sigma}\epsilon_{{\bf k}'\bar\sigma'}}{E_{{\bf
        k}\sigma}} \right) -
T_{{\bf k}'\bar\sigma'} \left( E_{{\bf k}'\bar\sigma'} -
  \frac{\epsilon_{{\bf k}\sigma}\epsilon_{{\bf k}'\bar\sigma'}}{E_{{\bf
        k}'\bar\sigma'}} \right) \right] ,
 \nonumber \\ 
\frac{1}{\beta} \sum_\omega F_\sigma({\bf k},i\omega) F_{\sigma'}({\bf
  k}',i\omega) 
&=&  \frac{-{\Delta^{\rm HS}_0}^2}{2(E^2_{{\bf k}\sigma}-E^2_{{\bf
  k}'\sigma'})} \left(
  \frac{T_{{\bf k}\sigma}}{E_{{\bf k}\sigma}} - \frac{T_{{\bf
  k}'\sigma'}}{E_{{\bf k}'\sigma'}} \right) ,  
\nonumber  \\
 \frac{1}{\beta} \sum_\omega F_\sigma({\bf k},i\omega)
  G_{\sigma'}({\bf k}',i\omega) 
&=&  \frac{\Delta^{\rm HS}_0 \epsilon_{{\bf k}'\sigma'}}{2(E^2_{{\bf
  k}\sigma}-E^2_{{\bf k}'\sigma'})} \left(
  \frac{T_{{\bf k}\sigma}}{E_{{\bf k}\sigma}} - \frac{T_{{\bf
  k}'\sigma'}}{E_{{\bf k}'\sigma'}} \right) .
\end{eqnarray}
where $T_{{\bf k}\sigma}=\tanh(\beta E_{{\bf k}\sigma}/2)$. 
Performing the sum over $\sigma_1,\sigma_2$ in (\ref{A3}), we find that the
only non-vanishing correlation functions are [using
$\Pi^{\nu\nu'}_{\mu\mu'}(\tilde q,\tilde q')=
\Pi^{\nu'\nu}_{\mu'\mu}(\tilde q',\tilde q)$]
\label{crtcrt}
\begin{eqnarray}
\Pi^{zz}_{\mu\mu}(0,0) &=& \frac{1}{2} \int_{\bf k} v_\mu^2({\bf k},0) 
\left[ \frac{T_{{\bf k}\uparrow}}{E^3_{{\bf k}\uparrow}}
  {\Delta_0^{\rm HS}}^2(1+\gamma_\mu) + \frac{U_{{\bf
        k}\uparrow}}{E^2_{{\bf k}\uparrow}} (E^2_{{\bf k}\uparrow}+
  \epsilon_{{\bf k}\uparrow}^2-\gamma_\mu {\Delta_0^{\rm HS}}^2)
  \right] ,  
\nonumber \\ 
\Pi^{xx}_{\mu\mu}(0,0) &=&  \int_{\bf k} \frac{v_\mu^2({\bf
    k},0)}{E^2_{{\bf k}\uparrow}-E^2_{{\bf k}\downarrow}}  
\left[ \frac{T_{{\bf k}\uparrow}}{E_{{\bf k}\uparrow}}
(E^2_{{\bf k}\uparrow}+\epsilon_{{\bf k}\uparrow} \epsilon_{{\bf
      k}\downarrow} +\gamma_\mu 
  {\Delta_0^{\rm HS}}^2 ) - \frac{T_{{\bf k}\downarrow}}{E_{{\bf k}\downarrow}}
(E^2_{{\bf k}\downarrow}+\epsilon_{{\bf k}\uparrow} \epsilon_{{\bf
      k}\downarrow} +\gamma_\mu {\Delta_0^{\rm HS}}^2 ) \right] , 
\nonumber \\
\Pi^{yy}_{\mu\mu}(0,0) &=&  \int_{\bf k} \frac{v_\mu^2({\bf
    k},0)}{E^2_{{\bf k}\uparrow}-E^2_{{\bf k}\downarrow}}  
\left[ \frac{T_{{\bf k}\uparrow}}{E_{{\bf k}\uparrow}}
(E^2_{{\bf k}\uparrow}+\epsilon_{{\bf k}\uparrow} \epsilon_{{\bf
      k}\downarrow} -\gamma_\mu 
  {\Delta_0^{\rm HS}}^2 ) - \frac{T_{{\bf k}\downarrow}}{E_{{\bf k}\downarrow}}
(E^2_{{\bf k}\downarrow}+\epsilon_{{\bf k}\uparrow} \epsilon_{{\bf
      k}\downarrow} -\gamma_\mu {\Delta_0^{\rm HS}}^2 ) \right] , 
\nonumber \\
\Pi^{zz}_{\mu\mu'}({\bf Q},{\bf Q}) &=& \bar\delta_{\mu,\mu'}
 \int_{\bf k} \frac{v_\mu({\bf k},{\bf Q}) v_{\mu'}({\bf k},{\bf
     Q})}{E^2_{{\bf k}\uparrow}-E^2_{{\bf k}\downarrow}}   
\left[ \frac{T_{{\bf k}\uparrow}}{E_{{\bf k}\uparrow}} \left( 
\gamma_\mu(E^2_{{\bf k}\uparrow}-\epsilon_{{\bf k}\uparrow} \epsilon_{{\bf
      k}\downarrow}) - {\Delta_0^{\rm HS}}^2 \right) - 
\frac{T_{{\bf k}\downarrow}}{E_{{\bf k}\downarrow}} \left( 
\gamma_\mu(E^2_{{\bf k}\downarrow}-\epsilon_{{\bf k}\uparrow} \epsilon_{{\bf
      k}\downarrow}) - {\Delta_0^{\rm HS}}^2 \right) \right] ,
 \nonumber \\
\Pi^{xx}_{\mu\mu'}({\bf Q},{\bf Q}) &=& - \bar\delta_{\mu,\mu'} \gamma_\mu
\frac{1}{2} 
 \int_{\bf k} v_\mu({\bf k},{\bf Q}) v_{\mu'}({\bf k},{\bf Q})  
\left[ \frac{T_{{\bf k}\uparrow}}{E_{{\bf k}\uparrow}} \left( -2 
+ \frac{{\Delta_0^{\rm HS}}^2}{E^2_{{\bf k}\uparrow}}(1+\gamma_\mu)
\right) - U_{{\bf k}\uparrow} \frac{{\Delta_0^{\rm HS}}^2}{E^2_{{\bf
      k}\uparrow}}(1+\gamma_\mu) \right]  ,
\nonumber \\
\Pi^{yy}_{\mu\mu'}({\bf Q},{\bf Q}) &=& - \bar\delta_{\mu,\mu'} \gamma_\mu
\frac{1}{2} 
 \int_{\bf k} v_\mu({\bf k},{\bf Q}) v_{\mu'}({\bf k},{\bf Q})  
\left[ \frac{T_{{\bf k}\uparrow}}{E_{{\bf k}\uparrow}} \left( -2 
+ \frac{{\Delta_0^{\rm HS}}^2}{E^2_{{\bf k}\uparrow}}(1-\gamma_\mu)
\right) - U_{{\bf k}\uparrow} \frac{{\Delta_0^{\rm HS}}^2}{E^2_{{\bf
      k}\uparrow}}(1-\gamma_\mu) \right]  ,
\nonumber \\
\Pi^{xy}_{0,\mu\neq 0}({\bf Q},{\bf Q}) &=& -\frac{1}{2} 
      \int_{\bf k}  \epsilon_{\bf k} \left( T_{{\bf k}\uparrow}
      \frac{\epsilon^2_{{\bf k}\uparrow}}{E^3_{{\bf k}\uparrow}} +
      U_{{\bf k}\uparrow}  \frac{{\Delta_0^{\rm HS}}^2}{E^2_{{\bf
      k}\uparrow}} \right) ,
\nonumber \\
\Pi^{xy}_{\mu\neq 0,0}({\bf Q},{\bf Q}) &=& \frac{h_0}{U} , 
\nonumber \\ 
\Pi^{zx}_{00}(0,{\bf Q}) &=& \Delta_0^{\rm HS} \int_{\bf k}
      \epsilon_{{\bf k}\uparrow} \left( \frac{T_{{\bf
            k}\uparrow}}{E^3_{{\bf k}\uparrow}} -  \frac{U_{{\bf
            k}\uparrow}}{E^2_{{\bf k}\uparrow}} \right)  ,
\nonumber \\ 
\Pi^{zy}_{0,\mu\neq 0}(0,{\bf Q}) &=& -\frac{\Delta_0^{\rm HS}}{2} \int_{\bf k}
 \epsilon_{\bf k} \epsilon_{{\bf k}\uparrow} \left( \frac{T_{{\bf
            k}\uparrow}}{E^3_{{\bf k}\uparrow}} -  \frac{U_{{\bf
            k}\uparrow}}{E^2_{{\bf k}\uparrow}} \right)  ,
\nonumber \\ 
\Pi^{xz}_{00}(0,{\bf Q}) &=& 2\Delta_0^{\rm HS} h \int_{\bf k}  
\frac{1}{E^2_{{\bf k}\uparrow}-E^2_{{\bf k}\downarrow}} \left( 
\frac{T_{{\bf k}\uparrow}}{E_{{\bf k}\uparrow}} - 
\frac{T_{{\bf k}\downarrow}}{E_{{\bf k}\downarrow}} \right) , 
\nonumber \\ 
\Pi^{yz}_{0,\mu\neq 0}(0,{\bf Q}) &=& - \Delta_0^{\rm HS} \int_{\bf k}
      \frac{\epsilon^2_{\bf k}} {E^2_{{\bf k}\uparrow}-E^2_{{\bf
            k}\downarrow}} \left(  
\frac{T_{{\bf k}\uparrow}}{E_{{\bf k}\uparrow}} - 
\frac{T_{{\bf k}\downarrow}}{E_{{\bf k}\downarrow}} \right) , 
\label{crtcrt16} 
\end{eqnarray}
where $U_{{\bf k}\sigma}= \beta/2\cosh^2\Bigl(\frac{\beta E_{{\bf
      k}\sigma}}{2}\Bigr)$ and
$\bar\delta_{\mu,\mu'}= \delta_{\mu,0}\delta_{\mu',0}+
\delta_{\mu\neq 0}\delta_{\mu'\neq 0}$. In order to obtain
Eqs.~(\ref{crtcrt16}), we have used the gap equations
(\ref{gapeq1}-\ref{gapeq2}) and the 
symmetry relations $\epsilon_{{\bf k}\sigma}=-\epsilon_{{\bf k}+{\bf
    Q}\bar\sigma}$, $E_{{\bf k}\sigma}=E_{{\bf k}+{\bf
    Q}\bar\sigma}$, $v_{\mu\neq 0}({\bf k},0)=-v_{\mu\neq 0}(-{\bf
  k},0) =-v_{\mu\neq 0}({\bf k}+{\bf Q},0)$, etc. 

\subsection{Strong-coupling limit $U\gg 4t$ ($T=0$)} 
\label{app:hfcorr1}

Expanding Eqs.~(\ref{crtcrt16}) to leading order in
$1/U$, we obtain ($T=0$)
\begin{eqnarray}
\Pi^{zz}_{00}(0,0) &=& \frac{2}{U} \sin^2\theta_0 , \nonumber \\
 \Pi^{xx}_{00}({\bf Q},{\bf Q}) &=& \frac{2}{U} \cos^2\theta_0 , \nonumber \\
 \sum_{\mu,\mu'=x,y} \Pi^{yy}_{\mu\mu'}({\bf Q},{\bf Q}) &=& 2J
  \cos^2\theta_0 , \nonumber \\ 
 \Pi^{xy}_{0x}({\bf Q},{\bf Q}) &=& \frac{8t^2}{U^2}
  \cos\theta_0(2-3\cos^2\theta_0) , \nonumber \\
 \Pi^{zx}_{00}(0,{\bf Q}) &=& - \frac{1}{U} \sin(2\theta_0) , \nonumber \\
 \Pi^{zy}_{0x}(0,{\bf Q}) &=& - \frac{8t^2}{U^2}
  \sin\theta_0(1-3\cos^2\theta_0) . \label{A6sc}
\end{eqnarray} 
Here we consider only the correlation functions that are useful for
the derivation of the effective action $S[\rho,\Delta]$ and the
calculation of the velocity $c$ of the phase collective mode.  

\subsection{Correlation functions for $h_0=0$ and ${\bf\Omega}_{\bf r}^{\rm
    cl}=(-1)^{\bf r} \hat {\bf z}$ ($T=0$)}  
\label{app:hfcorr2}

In Sec.~\ref{sec:O3}, we need the HF current-current correlation
function at half filling and for a magnetization parallel to the $z$
axis. They can be deduced from Eqs.~(\ref{crtcrt16})
with $h_0=h=0$ and 
by making the rotation in spin space $\hat {\bf x}\to \hat {\bf z}$,
$\hat {\bf z}\to \hat {\bf y}$ and $\hat {\bf y}\to \hat {\bf x}$. The
only non-vanishing correlation functions are then (the notations are
those of Sec.~\ref{sec:O3}) 
\begin{eqnarray}
\Pi^{xx}_{00}(0,0) &=& \Pi^{yy}_{00}(0,0) = m_0^2 \int_{\bf k}
  \frac{1}{E^3_{\bf k}} , \nonumber \\
\Pi^{zz}_{xx}(0,0) &=& \Pi^{zz}_{yy}(0,0) = 4t^2 m_0^2 \int_{\bf k}
  \frac{\sin^2 k_x}{E^3_{\bf k}} , \nonumber \\
\Pi^{xx}_{\mu\mu}({\bf Q},{\bf Q}) &=&  \Pi^{yy}_{\mu\mu}({\bf Q},{\bf Q})
= - \int_{\bf k} \frac{\epsilon^2_{\bf k}}{2E^3_{\bf k}}
  (\delta_{\mu,0} E^2_{\bf k}-\delta_{\mu \neq 0} \epsilon^2_{\bf k})
  , \nonumber \\
\Pi^{zz}_{\mu\mu}({\bf Q},{\bf Q}) &=& \int_{\bf k}
  \frac{\epsilon^2_{\bf k}}{2E^3_{\bf k}} 
  (\delta_{\mu\neq 0} E^2_{\bf k}-\delta_{\mu,0} \epsilon^2_{\bf k}) ,
  \nonumber \\
\Pi^{yx}_{0,\mu\neq 0}(0,{\bf Q}) &=& -\frac{m_0}{2} \int_{\bf k}
  \frac{\epsilon^2_{\bf k}}{E^3_{\bf k}} , \nonumber \\
\Pi^{xy}_{0,\mu\neq 0}(0,{\bf Q}) &=&  \frac{m_0}{2} \int_{\bf k}
  \frac{\epsilon^2_{\bf k}}{E^3_{\bf k}} ,
\end{eqnarray} 
where $E_{\bf k}=\sqrt{\epsilon_{\bf k}^2+m_0^2}$.

\section{Effective action $S[m,{\bf\Omega}]$} 
\label{app:leea}

In this appendix, we derive the effective action $S[m,{\bf\Omega}]=S[p,q,m]$
[Eq.~(\ref{Spqm})] of spin fluctuations in the repulsive Hubbard model. 

It is convenient to express the rotation matrix
$R_{\bf r}$ as $R_{\bf 
  r}=M({\bf\Omega_r}) M^\dagger({\bf\Omega}^{\rm cl}_{\bf r})$ where
  $M({\bf\Omega_r})$ is defined by $M({\bf\Omega_r}) \sigma^z
  M^\dagger ({\bf\Omega_r})= \boldsymbol{\sigma}\cdot
  {\bf\Omega_r}$:
\begin{equation}
M({\bf\Omega_r})= 
\left (
\begin{array}{lr}
\cos\left(\frac{\theta_{\bf r}}{2}\right) e^{-\frac{i}{2}\varphi_{\bf r}} & 
- \sin\left(\frac{\theta_{\bf r}}{2}\right) e^{-\frac{i}{2}\varphi_{\bf r}}
\\ 
\sin\left(\frac{\theta_{\bf r}}{2}\right) e^{\frac{i}{2}\varphi_{\bf r}} & 
\cos\left(\frac{\theta_{\bf r}}{2}\right) e^{\frac{i}{2}\varphi_{\bf r}} \\
\end{array}
\right ) .
\end{equation}
We then obtain
\begin{eqnarray}
A^x_{0{\bf r}} &=& (-1)^{\bf r} \left[ \left(\frac{i}{2} \dot q_{\bf
  r}+h_0 \right) \sin(2p_{\bf r}) 
  -\sin\theta_0 i\delta m^{\rm HS}_{\bf r} \right] ,
  \nonumber \\
A^y_{0{\bf r}} &=& -i (-1)^{\bf r} \dot p_{\bf r} ,
  \nonumber \\
A^z_{0{\bf r}} &=& -\left( \frac{i}{2} \dot q_{\bf r}+h_0\right)
  \cos(2p_{\bf r}) +h_0 -\cos\theta_0 i\delta m^{\rm HS}_{\bf r} ,
  \nonumber \\
A^0_{{\bf r},{\bf r}'} &=&  \cos\left(\frac{q_{\bf r}-q_{{\bf
  r}'}}{2}\right) \cos\left( p_{\bf r}+p_{{\bf
  r}'} \right) -1 , \nonumber \\
A^x_{{\bf r},{\bf r}'} &=&  -i (-1)^{\bf r}
  \sin\left(\frac{q_{\bf r}- q_{{\bf
  r}'}}{2}\right) \sin\left( p_{\bf r}-p_{{\bf
  r}'} \right) , \nonumber \\
A^y_{{\bf r},{\bf r}'} &=&  i (-1)^{\bf r} \cos\left(\frac{q_{\bf
  r}- q_{{\bf
  r}'}}{2}\right) \sin\left( p_{\bf r}+p_{{\bf
  r}'} \right) ,  \nonumber \\
A^z_{{\bf r},{\bf r}'} &=&  i \sin\left(\frac{q_{\bf r}-q_{{\bf
  r}'}}{2}\right) \cos\left( p_{\bf r}-p_{{\bf
  r}'} \right) . \label{Arrz}
\end{eqnarray}
Eqs.~(\ref{Arrz}) can be rewritten as
\begin{eqnarray}
A^x_{0{\bf r}} &=& (-1)^{\bf r} \biggl[ \frac{i}{2}(p_{\bf r}\dot
q_{\bf r}-\dot p_{\bf r} q_{\bf r})+2h_0p_{\bf r}
-\sin\theta_0 i\delta m^{\rm HS}_{\bf r} \biggr] , \nonumber \\
A^y_{0{\bf r}} &=& - i (-1)^{\bf r} \dot p_{\bf r} , \nonumber \\
A^z_{0{\bf r}} &=& -\frac{i}{2} \dot q_{\bf r} + 2h_0 p^2_{\bf r} -
\cos\theta_0 i\delta m^{\rm HS}_{\bf r} , \nonumber \\
A^0_{{\bf r},{\bf r}'} &=& -\frac{(p_{\bf r}+p_{{\bf r}'})^2}{2}-
\frac{(q_{\bf r}-q_{{\bf r}'})^2}{8} , \nonumber \\
A^x_{{\bf r},{\bf r}'} &=& -\frac{i}{2} (-1)^{\bf r} (q_{\bf r}-q_{{\bf
    r}'}) (p_{\bf r}-p_{{\bf r}'}) , \nonumber \\
A^y_{{\bf r},{\bf r}'} &=& i (-1)^{\bf r} (p_{\bf r}+p_{{\bf
    r}'}) , \nonumber \\
A^z_{{\bf r},{\bf r}'} &=& \frac{i}{2}  (q_{\bf r}-q_{{\bf r}'}) ,
\end{eqnarray}
to quadratic order in $p_{\bf r}$, $\delta m^{\rm HS}_{\bf r}$, $\dot p_{\bf
  r}$, $\dot q_{\bf r}$ and $q_{\bf r}-q_{{\bf r}'}$.  

Keeping terms up to second-order in
$p_{\bf r}$, $\delta m^{\rm HS}_{\bf r}$, $\delta m_{\bf r}$, $\dot
p_{\bf r}$, $\dot q_{\bf r}$ and $q_{\bf r}-q_{{\bf 
    r}'}$ (${\bf r},{\bf r}'$ nearest 
neighbors), the effective action is given by first- and second-order
cumulants of $S_1$ and $S_2$ with respect to the HF action: 
\begin{equation}
S[p,q,m^{\rm HS},m] = \langle S_1+S_2 \rangle -\frac{1}{2} \langle
(S_1+S_2)^2 \rangle_c   + 
\int_0^\beta d\tau  \sum_{\bf r} \biggl( -\frac{U}{4}
\delta m_{\bf r}^2 +i\delta m^{\rm HS}_{\bf r} \delta m_{\bf r}
-\frac{2}{U} m_0^{\rm HS}\delta m^{\rm HS}_{\bf r} \biggr) .
\end{equation}

Using
\begin{eqnarray}
\langle \phi^\dagger_{\bf r} \sigma^0 \phi_{\bf r}\rangle &=& 1 , \nonumber \\
 \langle \phi^\dagger_{\bf r} \sigma^x \phi_{\bf r}\rangle &=&
(-1)^{\bf r}  2 \Delta_0 , \nonumber \\
 \langle \phi^\dagger_{\bf r} \sigma^y \phi_{\bf r}\rangle &=& 0, \nonumber \\
 \langle \phi^\dagger_{\bf r} \sigma^z \phi_{\bf r}\rangle &=& -x, 
\end{eqnarray}
where the averages are taken with the HF action, we obtain 
\begin{equation}
\langle S_1 \rangle = \int_0^\beta d\tau  \sum_{\bf r} \biggl[ 
i\Delta_0 (p_{\bf r}\dot q_{\bf r}- \dot p_{\bf r} q_{\bf
    r}) + 4 \Delta_0 h_0 p_{\bf r}
- 2 \Delta_0 \sin\theta_0 i \delta m^{\rm HS}_{\bf r}
+ \frac{i}{2}x \dot q_{\bf r}  -2x h_0
  p^2_{\bf r} +x \cos\theta_0 i \delta m^{\rm HS}_{\bf r} \biggr] .
\label{appS1}
\end{equation}
Similarly, from  (${\bf r},{\bf r}'$ are nearest neighbors)
\begin{eqnarray}
\langle \phi^\dagger_{\bf r} \sigma^0 \phi_{{\bf r}'} \rangle &=&
\frac{\langle -K\rangle}{4t}  
= \frac{1}{4t} \int_{\bf k} \frac{\epsilon_{\bf k}\epsilon_{{\bf
      k}\uparrow}}{E_{{\bf k}\uparrow}} \tanh \frac{\beta E_{{\bf
      k}\uparrow}}{2} , \nonumber \\ 
\langle \phi^\dagger_{\bf r} \sigma^x \phi_{{\bf r}'} \rangle &= & 0,
\nonumber \\ 
\langle \phi^\dagger_{\bf r} \sigma^y \phi_{{\bf r}'} \rangle &=&
- i (-1)^{\bf r} \frac{\Delta_0 h_0}{2t} , \nonumber \\
\langle \phi^\dagger_{\bf r} \sigma^z \phi_{{\bf r}'} \rangle &=& 0 , 
\end{eqnarray}
we deduce 
\begin{equation}
\langle S_2 \rangle = \frac{\langle -K\rangle}{4}  \int_0^\beta
  d\tau \sum_{\langle {\bf r},{\bf r}'\rangle} \left[ (p_{\bf
  r}+p_{{\bf r}'})^2 +  \frac{(q_{\bf r}-q_{{\bf r}'})^2}{4} \right]
-4 \Delta_0 h_0\int_0^\beta d\tau \sum_{\bf r} p_{\bf r} . 
\label{appS2}
\end{equation}
We have introduced the mean value $\langle K\rangle$ of the kinetic energy
per site in the HF state. 

When calculating the second-order cumulant, it is sufficient to consider
$A_{0{\bf r}}$ and $A_{{\bf r},{\bf r}'}$ to linear order in $p_{\bf r}$,
$\delta m^{\rm HS}_{\bf r}$, $\dot p_{\bf r}$, $\dot q_{\bf r}$, and $q_{\bf
  r}-q_{{\bf r}'}$: 
\begin{eqnarray}
A^x_{0{\bf r}} &=& (-1)^{\bf r} (2h_0 p_{\bf r}-
\sin\theta_0 i\delta m^{\rm HS}_{\bf r}) , \nonumber \\
A^y_{0{\bf r}} &=& - i (-1)^{\bf r} \dot p_{\bf r} ,   \nonumber \\
A^z_{0{\bf r}} &=& -\frac{i}{2} \dot q_{\bf r} - \cos\theta_0 i \delta
m^{\rm HS}_{\bf r},  \nonumber \\
A^0_{{\bf r},{\bf r}'} &=& 0 , \nonumber \\
A^x_{{\bf r},{\bf r}'} &=& 0 ,  \nonumber \\
A^y_{{\bf r},{\bf r}'} &=& i (-1)^{\bf r} (p_{\bf r}+p_{{\bf r}'}) ,
\nonumber \\ 
A^z_{{\bf r},{\bf r}'} &=& \frac{i}{2} (q_{\bf r}-q_{{\bf r}'}) .
\label{Azmu} 
\end{eqnarray} 
In order to evaluate $\langle (S_1+S_2)^2 \rangle_c$, we
write the action $S_1+S_2$ as
\begin{eqnarray}
S_1+S_2 &=& \int_0^\beta d\tau \sum_{{\bf r} \atop {{\mu=0,x,y} \atop
    {\nu=x,y,z}}} A^{\nu{\rm (tot)}}_{\mu{\bf r}} j^\nu_{\mu{\bf r}} ,
\\
A^{\nu{\rm (tot)}}_{\mu{\bf r}} &=& \delta_{\mu,0} A^\nu_{0{\bf r}} 
-i (1-\delta_{\mu,0}) A^\nu_{\mu{\bf r}} , 
\label{Atot}
\end{eqnarray}
where $A^\nu_{\mu{\bf r}}=A^\nu_{{\bf r},{\bf r}+\hat {\boldsymbol{\mu}}}$  for
$\mu=x,y$. The spin-density current $j^\nu_{\mu{\bf r}}$ is defined by
(\ref{spincrt}).  We then have
\begin{equation}
\langle (S_1+S_2)^2 \rangle_c = \int_0^\beta d\tau d\tau' 
\sum_{{\bf r},{\bf r}' \atop  
{\mu,\mu'=0,x,y \atop {\nu,\nu'=x,y,z}}}
A^{\nu{\rm (tot)}}_{\mu{\bf r}}(\tau) 
\Pi^{\nu\nu'}_{\mu\mu'}({\bf
  r},\tau; {\bf r}',\tau') A^{\nu'{\rm (tot)}}_{\mu'{\bf r}'}(\tau') ,
\label{S2} 
\end{equation}
where 
\begin{equation}
\Pi^{\nu\nu'}_{\mu\mu'}({\bf r},\tau; {\bf r}',\tau') =
\langle j^\nu_{\mu{\bf r}}(\tau) j^{\nu'}_{\mu'{\bf r}'}(\tau')\rangle_c 
\end{equation}
is the HF current-current correlation
function. $\Pi^{\nu\nu'}_{\mu\mu'}$ is calculated in Appendix
\ref{app:hfcorr}.

In order to calculate the second-order cumulant, we write the actions
$S_1$ and $S_2$ [Eqs.~(\ref{S1_2})] as
\begin{eqnarray}
S_1 &=& \int_0^\beta  d\tau\sum_{{\bf r} \atop {\nu=x,y,z}}
A^\nu_{0{\bf r}} j^\nu_{0{\bf r}} , \nonumber \\
S_2 &=& -i  \int_0^\beta d\tau \sum_{{\bf r} \atop {\mu=x,y \atop
    {\nu=x,y,z}}} A^\nu_{\mu{\bf r}} j^\nu_{\mu{\bf r}} ,
\end{eqnarray}
where $A^\nu_{\mu{\bf r}}=A^\nu_{{\bf r},{\bf r}+\hat
{\boldsymbol{\mu}}}$  for $\mu=x,y$. $A^\nu_{0{\bf r}}$ and
$A^\nu_{\mu{\bf r}}$ are given by Eqs.~(\ref{Azmu}). The
spin-density current $j^\nu_{\mu{\bf r}}$ is defined in
(\ref{spincrt}).  In Fourier space, we obtain
\begin{eqnarray}
A^\nu_0(\tilde q) &=& z_0(\tilde q) \left[\delta_{\nu,y}p_{\tilde
    q+{\bf Q}} + \delta_{\nu,z} \frac{q_{\tilde q}}{2} \right] 
+ \delta_{\nu,x} [2h_0 p_{\tilde q+{\bf Q}}-\sin\theta_0
i\delta m^{\rm HS}_{\tilde q+{\bf Q}}]
-\delta_{\nu,z} \cos\theta_0 i\delta m^{\rm HS}_{\tilde q} , 
\nonumber \\  
-i A^\nu_{\mu}(\tilde q) &=& z_\mu(\tilde q) \left[
    \delta_{\nu,y}p_{\tilde q+{\bf Q}} + \delta_{\nu,z} \frac{q_{\tilde
    q}}{2} \right]   \,\,\,\,\, (\mu=x,y) , 
\end{eqnarray}
where 
\begin{eqnarray}
z_0(\tilde q) &=& - \omega_\nu , \nonumber \\
z_{\mu=x,y}(\tilde q) &=& 1-e^{iq_\mu} . 
\end{eqnarray}
We deduce
\begin{eqnarray}
\langle (S_1+S_2)^2 \rangle_c &=& \sum_{\tilde q,\tilde q'} \Biggl\lbrace 
\sum_{\mu,\mu'=0,x,y} \biggl[ 
z_\mu(-\tilde q-{\bf Q}) z_{\mu'}(\tilde q'+{\bf Q}) 
p_{-\tilde q} p_{\tilde q'} 
\Pi^{yy}_{\mu\mu'}(\tilde q+{\bf Q},\tilde q'+{\bf Q}) 
+\frac{1}{4} z_\mu(-\tilde q) z_{\mu'}(\tilde q')
q_{-\tilde q} q_{\tilde q'} 
\Pi^{zz}_{\mu\mu'}(\tilde q,\tilde q') \nonumber \\ &&  
+\frac{1}{2} z_\mu(-\tilde q-{\bf Q}) z_{\mu'}(\tilde q') 
p_{-\tilde q} q_{\tilde q'} 
\Pi^{yz}_{\mu\mu'}(\tilde q+{\bf Q},\tilde q') 
+\frac{1}{2} z_\mu(-\tilde q) z_{\mu'}(\tilde q'+{\bf Q}) 
q_{-\tilde q} p_{\tilde q'} 
\Pi^{zy}_{\mu\mu'}(\tilde q,\tilde q'+{\bf Q}) \biggr] \nonumber \\ &&
+ \sum_{\mu=0,x,y} \biggl[
z_\mu(\tilde q'+{\bf Q}) r_{-\tilde q} p_{\tilde q'} 
\Pi^{xy}_{0\mu}(\tilde q+{\bf Q},\tilde q'+{\bf Q}) 
+ \frac{1}{2} z_\mu(\tilde q') r_{-\tilde q}q_{\tilde q'}  
\Pi^{xz}_{0\mu}(\tilde q+{\bf Q},\tilde q') \nonumber \\ &&
 + z_\mu(-\tilde q-{\bf Q}) p_{-\tilde q} r_{\tilde q'} 
\Pi^{yx}_{\mu0}(\tilde q+{\bf Q},\tilde q'+{\bf Q})
+ \frac{1}{2} z_\mu(-\tilde q) q_{-\tilde q}  r_{\tilde q'}
\Pi^{zx}_{\mu0}(\tilde q,\tilde q'+{\bf Q}) \nonumber \\ &&
-\cos\theta_0 \Bigl(
z_\mu(\tilde q'+{\bf Q}) i\delta {m^{\rm HS}_{-\tilde q}} p_{\tilde q'} 
\Pi^{zy}_{0\mu}(\tilde q,\tilde q'+{\bf Q}) 
+ \frac{1}{2} z_\mu(\tilde q') i\delta {m^{\rm HS}_{-\tilde q}} q_{\tilde q'}  
\Pi^{zz}_{0\mu}(\tilde q,\tilde q') \nonumber \\ &&
+ z_\mu(-\tilde q-{\bf Q}) p_{-\tilde q} i\delta m^{\rm HS}_{\tilde q'} 
\Pi^{yz}_{\mu0}(\tilde q+{\bf Q},\tilde q')
+ \frac{1}{2} z_\mu(-\tilde q) q_{-\tilde q}  i\delta m^{\rm HS}_{\tilde q'}
\Pi^{zz}_{\mu0}(\tilde q,\tilde q') \Bigr)
\biggr] \nonumber \\ &&
+r_{-\tilde q} r_{\tilde q'} 
\Pi^{xx}_{00}(\tilde q+{\bf Q},\tilde q'+{\bf Q}) 
- \cos^2\theta_0 \delta {m^{\rm HS}_{-\tilde q}} \delta m^{\rm
    HS}_{\tilde q'}   
\Pi^{zz}_{00}(\tilde q,\tilde q') \nonumber \\ &&
-\cos\theta_0 \bigl[ r_{-\tilde q} i\delta m^{\rm HS}_{\tilde q'} 
\Pi^{xz}_{00}(\tilde q+{\bf Q},\tilde q')
+ i\delta {m^{\rm HS}_{-\tilde q}}  r_{\tilde q'}
\Pi^{zx}_{00}(\tilde q,\tilde q'+{\bf Q}) \bigr] 
\Biggr\rbrace ,
\label{S12}
\end{eqnarray}
where $r_{\tilde q}=2h_0p_{\tilde q}-\sin\theta_0 i\delta m^{\rm HS}_{\tilde
q}$. To proceed further, we use the expression of the correlation
function $\Pi^{\nu\nu'}_{\mu\mu'}(\tilde q,\tilde q')$ obtained in Appendix
\ref{app:hfcorr}. We thus have $\tilde q=\tilde q'$ in (\ref{S12}). In
order to obtain the effective action of the $q$ field to lowest order
in $\partial_\mu q$, it is sufficient to retain the first-order
derivative terms 
$\partial_\mu q$. Since $z_\mu(\tilde q)=O(\partial_\mu)$, one can
use $z_0(\tilde q)=-\omega_\nu$, $z_{\mu\neq 0}(\tilde q)=-iq_\mu$,
$z_0(\tilde q+{\bf Q})=0$, $z_{\mu\neq 0}(\tilde q+{\bf Q})=2$, and evaluate
the correlation functions $\Pi^{\nu\nu'}_{\mu\mu'}$ at $\tilde
q=\tilde q'=0$ in (\ref{S12}). This gives
\begin{eqnarray}
\langle (S_1+S_2)^2 \rangle_c &=& \sum_{\tilde q} \biggl\lbrace 
p_{-\tilde q} p_{\tilde q} \biggl( 4 \sum_{\mu,\mu'\neq 0}
\Pi^{yy}_{\mu\mu'}({\bf Q},{\bf Q}) + 4 h_0^2 \Pi^{xx}_{00}({\bf
  Q},{\bf Q}) + 16h_0 \Pi^{xy}_{0x}({\bf Q},{\bf Q}) \biggr) 
\nonumber \\ &&
+ q_{-\tilde q} q_{\tilde q} \Bigl( - \frac{\omega_\nu^2}{4}\Pi^{zz}_{00}(0,0)
+ \frac{{\bf q}^2}{4} \Pi^{zz}_{xx}(0,0) \Bigr) , 
\nonumber \\ &&
- \delta m^{\rm HS}_{-\tilde q} \delta m^{\rm HS}_{\tilde q} 
\left( \sin^2\theta_0
\Pi^{xx}_{00}({\bf Q},{\bf Q})  
+ \cos^2\theta_0 \Pi^{zz}_{00}(0,0) 
+ \sin(2\theta_0) \Pi^{zx}_{00}(0,{\bf Q}) \right) 
\nonumber \\ &&
- \left[ p_{-\tilde q} q_{\tilde q} \omega_\nu \left(  2
  \Pi^{zy}_{0x}(0,{\bf Q}) + h_0 \Pi^{zx}_{00}(0,{\bf Q}) \right) -
     {\rm c.c.} \right] 
\nonumber \\ &&
- i \Bigl[ p_{-\tilde q} \delta m^{\rm HS}_{\tilde q}  \bigl(
 2h_0 \sin\theta_0 \Pi^{xx}_{00}({\bf Q},{\bf Q}) 
+ 4 \sin\theta_0 \Pi^{xy}_{0x}({\bf Q},{\bf Q}) 
\nonumber \\ &&
+ 4 \cos\theta_0 \Pi^{zy}_{0x}(0,{\bf Q}) 
+ 2 h_0 \cos\theta_0 \Pi^{zx}_{00}(0,{\bf Q}) \bigr) + {\rm c.c.} \Bigr] 
\nonumber \\ &&
- i \left[ q_{-\tilde q} \delta m^{\rm HS}_{\tilde q} \frac{\omega_\nu}{2}
  \left(  \sin\theta_0 \Pi^{zx}_{00}(0,{\bf Q}) 
+ \cos\theta_0 \Pi^{zz}_{00}(0,0) \right) - {\rm c.c.} \right] 
\biggr\rbrace   .
\label{S12bis}
\end{eqnarray}

A comment is in order here. The first-order cumulant (\ref{appS1}) gives a term
linear in $q_{\bf r}$:
\begin{equation}
\delta S_B = \frac{i}{2} x \int_0^\beta d\tau \sum_{\bf r} \dot q_{\bf r}.
\end{equation}
This term comes from the Berry phase term $S_B=\int_0^\beta d\tau \sum_{\bf
  r} \langle \phi^\dagger_{\bf r} R^\dagger_{\bf r} \dot R_{\bf r} \phi_{\bf
  r} \rangle$. The SU(2) matrix $R_{\bf r}$ is defined
up to the U(1) gauge transformation $R_{\bf r}\to R_{\bf r} e^{-\frac{i}{2}
  \psi_{\bf r} \boldsymbol{\sigma}\cdot {\bf\Omega}^{\rm cl}_{\bf
  r}}$. The Berry phase term depends on the gauge
choice.\cite{Auerbach94}  The gauge-dependent term is given by
\begin{equation}
-\frac{i}{2} \int_0^\beta d\tau \sum_{\bf r} \langle \phi^\dagger_{\bf
 r} \boldsymbol{\sigma}\cdot {\bf\Omega}^{\rm cl}_{\bf r} \phi_{\bf r}
 \rangle \dot \psi_{\bf r} 
= - \frac{i}{2} m_0 \int_0^\beta d\tau \sum_{\bf r} \dot \psi_{\bf r} .
\label{gauge}
\end{equation}
In the following we take $\psi_{\bf r}=\varphi_{\bf
r}/m_0$. As shown in Sec.~\ref{subsec:leea1}, this choice ensures that
in the attractive model, half the fermion density is identified as the
conjugate variable of the phase $\Theta$ of the superconducting order
parameter, as required by gauge invariance. \cite{note4} Including the
gauge-dependent term of $S_B$ [Eq.~(\ref{gauge})] 
in $\delta S_B$, we obtain 
\begin{equation}
\delta S_B = - \frac{i}{2} \rho_0 \int_0^\beta d\tau
\sum_{\bf r} \dot q_{\bf r} .
\label{dSB}
\end{equation}

From Eqs.~(\ref{appS1},\ref{appS2},\ref{S12bis},\ref{dSB}), we then obtain
\begin{eqnarray}
S[p,q,m^{\rm HS},m] &=& 
\frac{1}{2} \sum_{\tilde q} \bigl[
p_{-\tilde q} \Pi_{pp}(\tilde q) p_{\tilde q} + q_{-\tilde
  q} \Pi_{qq}(\tilde q) q_{\tilde q} 
+ \delta m^{\rm HS}_{-\tilde q} \Pi_{m^{\rm HS}m^{\rm HS}}(\tilde q)
\delta m^{\rm HS}_{\tilde q} 
+2 p_{-\tilde q} \Pi_{pq}(\tilde q) q_{\tilde q} \nonumber \\ &&
+2 p_{-\tilde q} \Pi_{pm^{\rm HS}}(\tilde
q) \delta m^{\rm HS}_{\tilde q} 
+2 q_{-\tilde q} \Pi_{qm^{\rm HS}}(\tilde
q)  \delta m^{\rm HS}_{\tilde q} 
+ 2i \delta m_{-\tilde q} \delta m^{\rm HS}_{\tilde q} - \frac{U}{2}
\delta m_{-\tilde q} \delta m_{\tilde q}
\bigr]  + \delta S_B ,
\label{cumu2}
\end{eqnarray}
where 
\begin{eqnarray}
\Pi_{pp}(\tilde q) &=& -4 \sum_{\mu,\mu'\neq 0}
\Pi^{yy}_{\mu\mu'}({\bf Q},{\bf Q})  -4 h_0^2 \Pi^{xx}_{00}({\bf
  Q},{\bf Q})
-16h_0 \Pi^{xy}_{0x}({\bf Q},{\bf Q}) -4xh_0
+4\langle -K\rangle , \nonumber \\
\Pi_{qq}(\tilde q) &=& \frac{\omega_\nu^2}{4} \Pi^{zz}_{00}(0,0) +
\frac{{\bf q}^2}{8} \langle -K\rangle , \nonumber \\
\Pi_{m^{\rm HS}m^{\rm HS}}(\tilde q) &=& \sin^2\theta_0
\Pi^{xx}_{00}({\bf Q},{\bf Q}) + \cos^2\theta_0 \Pi^{zz}_{00}(0,0)
+ \sin(2\theta_0) \Pi^{zx}_{00}(0,{\bf Q}) , \nonumber \\ 
\Pi_{pq}(\tilde q) &=& \omega_\nu \left[ 2 \Pi^{zy}_{0x}(0,{\bf Q})
+h_0 \Pi^{zx}_{00}(0,{\bf Q}) + 2 \Delta_0 \right] , \nonumber \\
\Pi_{pm^{\rm HS}}(\tilde q) &=& 2i \bigl[ h_0 \sin\theta_0
  \Pi^{xx}_{00}({\bf Q},{\bf Q})  
+2 \sin\theta_0 \Pi^{xy}_{0x}({\bf Q},{\bf Q}) 
+2 \cos\theta_0 \Pi^{zy}_{0x}(0,{\bf Q}) 
+ h_0 \cos\theta_0 \Pi^{zx}_{00}(0,{\bf Q}) \bigr]  , \nonumber \\
\Pi_{qm^{\rm HS}}(\tilde q) &=& \frac{i}{2} \omega_\nu \left[ 
\sin\theta_0 \Pi^{zx}_{00}(0,{\bf Q}) 
+ \cos\theta_0 \Pi^{zz}_{00}(0,0) \right] .
\label{piqm}
\end{eqnarray}
$p_{\tilde q}$, $q_{\tilde q}$, $\delta m^{\rm HS}_{\tilde q}$ and
$\delta m_{\tilde q}$ are the 
Fourier transformed fields of $p_{\bf r}$, $q_{\bf r}$, $\delta
m^{\rm HS}_{\bf r}$ and $\delta
m_{\bf r}$. $\tilde q=({\bf q},i\omega_\nu)$ where $\omega_\nu=\nu
2\pi T$ ($\nu$ integer) is a bosonic Matsubara
frequency. Eqs.~(\ref{piqm}) are valid in the hydrodynamic regime ($\tilde
q\to 0$). They are sufficient to obtain the effective action of the AF
field $q$ to order $O(\partial^2_\mu)$. 

Integrating out the Hubbard-Stratonovich field $m^{\rm HS}$, we obtain
Eq.~(\ref{Spqm}) with 
\begin{eqnarray}
\tilde\Pi_{pp}  &=& \Pi_{pp}  - \frac{\Pi_{pm^{\rm
      HS}}  \Pi_{m^{\rm
      HS}p} }{\Pi_{m^{\rm HS}m^{\rm HS}} } , \nonumber \\ 
\tilde\Pi_{qq}  &=& \Pi_{qq}  
- \frac{\Pi_{qm^{\rm HS}} \Pi_{m^{\rm
      HS}q} }{\Pi_{m^{\rm HS}m^{\rm HS}} } , \nonumber \\ 
\tilde\Pi_{mm}  &=& \frac{1}{\Pi_{m^{\rm HS}m^{\rm
      HS}} } - \frac{U}{2} 
      , \nonumber \\ 
\tilde\Pi_{pq}  &=& \Pi_{pq } 
- \frac{\Pi_{pm^{\rm HS}} \Pi_{m^{\rm
      HS}q} }{\Pi_{m^{\rm HS}m^{\rm HS}} } , \nonumber \\ 
 \tilde\Pi_{pm}  &=& -i \frac{\Pi_{pm^{\rm HS}}}{\Pi_{m^{\rm HS}m^{\rm 
      HS}} } , \nonumber \\ 
 \tilde\Pi_{qm}  &=& -i \frac{\Pi_{qm^{\rm HS}}}{\Pi_{m^{\rm HS}m^{\rm 
      HS}} } .
\label{tildePi} 
\end{eqnarray}

The action $S[p,q,m]$ takes a very simple form in the weak-coupling
(Slater) and strong-coupling (Mott-Heisenberg) limits.

\subsection{Slater limit}
\label{subsubsec:wc}

We assume that we are not too close to half-filling  so that the
zero-temperature order parameter $\Delta_0^{\rm HS}$ is given by
(\ref{deltabcs}). Since 
$\Delta_0^{\rm HS}$ is exponentially small at weak coupling,
$\theta_0\simeq \pi$ and $im_0^{\rm HS}\simeq xU/2$ [see
Eqs.~(\ref{delta_h_def},\ref{spteq2bis})]. We also have
\begin{eqnarray}
\Pi^{zx}_{00}(0,{\bf Q}) &=& \Delta_0^{\rm HS} \int_{\bf k}
\frac{\epsilon_{{\bf k}\uparrow}}{E^3_{{\bf k}\uparrow}} \nonumber \\
&\simeq & \Delta_0^{\rm HS} {\cal N}_0(\epsilon_F) \int_{-4t}^{4t} d\epsilon
\frac{\epsilon-\epsilon_F}{[(\epsilon-\epsilon_F)^2+{\Delta_0^{\rm
        HS}}^2]^{3/2}} \nonumber \\ &\simeq & 0 .
\label{pizx00wc}
\end{eqnarray}
Since the integral in (\ref{pizx00wc}) is peaked around
$\epsilon=\epsilon_F$ for 
$\Delta_0^{\rm HS}\to 0$, we have replaced the density of states
${\cal N}_0(\epsilon)$ by its value at the Fermi energy $\epsilon_F$ [see
  the discussion after Eq.~(\ref{gap1})]. 
For the same reason, we can extend the integration range to
$]-\infty,\infty[$, so that the integral vanishes. This
result is a consequence of the particle-hole symmetry which holds in
the weak-coupling (BCS) limit of the attractive model. Similarly, we find
$\Pi^{zy}_{0x}(0,{\bf Q}) \simeq 0$. We therefore have
\begin{eqnarray}
\Pi_{qq} &=& \frac{\Pi^{zz}_{00}(0,0)}{4} \omega^2_\nu + \frac{\langle
  -K\rangle}{8} {\bf q}^2 , \nonumber \\ 
\Pi_{m^{\rm HS}m^{\rm HS}} &=& \Pi^{zz}_{00}(0,0) , \nonumber \\ 
\Pi_{pq}(\tilde q) &=& 2 \omega_\nu \Delta_0 \simeq 0 , \nonumber \\
\Pi_{pm^{\rm HS}}(\tilde q) &=& 0 , \nonumber \\ 
\Pi_{qm^{\rm HS}}(\tilde q) &=& -\frac{i}{2} \Pi^{zz}_{00}(0,0) \omega_\nu .
\end{eqnarray}
Using Eqs.~(\ref{tildePi}), we deduce
\begin{eqnarray}
\tilde \Pi_{qq} &=&  \frac{\langle -K\rangle}{8} {\bf q}^2 , \nonumber \\ 
\tilde \Pi_{mm} &=& \frac{1}{\Pi^{zz}_{00}(0,0)}-\frac{U}{2} , \nonumber \\
\tilde \Pi_{qm} &=& -\frac{\omega_\nu}{2} 
\end{eqnarray}
and $\tilde\Pi_{pq}=\tilde\Pi_{pm}=0$. $p$ fluctuations do not couple
to $q$ and $m$ fluctuations. We therefore obtain the effective action
\begin{eqnarray}
S[q,m] &=& \frac{1}{2} \int_0^\beta d\tau \int d^2 r 
\biggl[ i \delta m_{\bf r} \dot q_{\bf r} + \frac{\langle -K \rangle
  }{8} (\boldsymbol{\nabla} q_{\bf r})^2 \nonumber \\ && 
+ \left( \frac{1}{\Pi^{zz}_{00}(0,0)}-\frac{U}{2} \right) \delta m^2_{\bf r}
\biggr]  + \delta S_B ,
\label{action7}
\end{eqnarray}
where we have taken the continuum limit in real space. 

\subsection{Mott-Heisenberg limit}
\label{subsubsec:sc2}

In the strong-coupling limit, there are well-defined local moments
with a fixed amplitude so that we can ignore the
fluctuations of $m^{\rm HS}$ and $m$. Low-energy fluctuations
correspond to direction fluctuations of these local moments. This can
be seen explicitly by integrating out the $m$ field in the action
$S[p,q,m^{\rm HS},m]$ [Eq.~(\ref{cumu2})]. This yields the replacement 
\begin{equation}
\Pi_{m^{\rm HS}m^{\rm HS}} \to \Pi_{m^{\rm HS}m^{\rm HS}}' = 
\Pi_{m^{\rm HS}m^{\rm HS}} - \frac{2}{U} .
\end{equation}
To leading order in $1/U$ , we have [see Eqs.~(\ref{A6sc})
in Appendix \ref{app:hfcorr}]
\begin{eqnarray}
\Pi'_{m^{\rm HS}m^{\rm HS}}(\tilde q) &=& - \frac{2}{U} , \nonumber \\
\Pi_{pm^{\rm HS}} &=& O\left( \frac{t^2}{U^2} \right) , \nonumber \\ 
\Pi_{qm^{\rm HS}} &=& O\left( \frac{\omega_\nu t}{U^2} \right) .
\end{eqnarray}
Integrating out $m^{\rm HS}$ yields the correction terms
\begin{eqnarray}
-\frac{\Pi_{pm^{\rm HS}} \Pi_{m^{\rm HS}p}}{\Pi'_{m^{\rm HS}m^{\rm
 HS}}} &=& O \left( \frac{t^4}{U^3} \right) 
, \nonumber \\
-\frac{\Pi_{qm^{\rm HS}} \Pi_{m^{\rm HS}q}}{\Pi'_{m^{\rm HS}m^{\rm
 HS}}}  &=& O \left( \frac{\omega_\nu^2 t^2}{U^3} \right) , \nonumber \\
-\frac{\Pi_{pm^{\rm HS}} \Pi_{m^{\rm HS}q}}{\Pi'_{m^{\rm HS}m^{\rm
 HS}}}  &=& O \left( \frac{\omega_\nu t^3}{U^3} \right) , \nonumber \\
\end{eqnarray}
to $\Pi_{pp}$, $\Pi_{qq}$ and $\Pi_{pq}$, respectively. These terms can
be ignored in the limit $U\gg 4t$. 

Using Eqs.~(\ref{A6sc}) of Appendix \ref{app:hfcorr1}, we therefore have 
\begin{eqnarray}
\Pi_{pp}(\tilde q) &=& 8J \sin^2\theta_0 , \nonumber \\
\Pi_{qq}(\tilde q) &=& {\bf q}^2 \frac{J}{4} \sin^2\theta_0 , \nonumber \\
\Pi_{pq}(\tilde q) &=& \omega_\nu \sin\theta_0 , \label{scpi3}
\end{eqnarray}
and $\Pi_{pm^{\rm HS}}=\Pi_{qm^{\rm HS}}\simeq 0$. Using then
$\tilde\Pi=\Pi$ [Eqs.~(\ref{tildePi})], we obtain the effective action
\begin{eqnarray}
S[p,q] &=& \frac{1}{2} \int_0^\beta d\tau \int d^2 r 
\biggl[ 2i \sqrt{1-x^2} p_{\bf r} \dot q_{\bf r} \nonumber \\ && 
+ \frac{J}{4}(1-x^2) (\boldsymbol{\nabla} q_{\bf 
  r})^2 + 
8J(1-x^2) p^2_{\bf r} \biggr] +\delta S_B . \nonumber \\ && 
\label{action_sc} 
\end{eqnarray}

\section{Kinetic energy in the HF state ($T=0$)}
\label{app:kehf}

In this Appendix, we derive Eq.~(\ref{kealt}). We start from
\begin{eqnarray}
\boldsymbol{\nabla}_{\bf k} \cdot \left( \frac{\epsilon_{{\bf
        k}\uparrow}}{E_{{\bf k}\uparrow}} {\bf v_k} \right) &=&
\frac{\epsilon_{{\bf k}\uparrow}}{E_{{\bf k}\uparrow}}
        \boldsymbol{\nabla}_{\bf k} \cdot {\bf v_k} +
        \boldsymbol{\nabla}_{\bf k}  
\left( \frac{\epsilon_{{\bf k}\uparrow}}{E_{{\bf k}\uparrow}} \right)
        \cdot  {\bf v_k} \nonumber \\ &=&
-  \frac{\epsilon_{{\bf k}\uparrow}\epsilon_{\bf k}}{E_{{\bf
        k}\uparrow}}  + \frac{{\Delta_0^{\rm HS}}^2}{E^3_{{\bf
        k}\uparrow}} v^2_{\bf k} .
\label{div1}
\end{eqnarray}
where $ {\bf v_k}=\boldsymbol{\nabla}_{\bf k} \epsilon_{\bf k}$.
Integrating the left hand side of this equation over the entire
Brillouin zone, we obtain  
\begin{equation}
\int_{\bf k} \boldsymbol{\nabla}_{\bf k} \cdot \left( \frac{\epsilon_{{\bf
        k}\uparrow}}{E_{{\bf k}\uparrow}} {\bf v_k} \right) = \oint 
 \frac{\epsilon_{{\bf k}\uparrow}}{E_{{\bf k}\uparrow}} {\bf v_k}
        \cdot d{\bf l_k} =0 ,
\label{div2}
\end{equation}
where the contour in the last integral is given by the Brillouin zone
(BZ) boundary and $d{\bf l_k}$ is perpendicular to the contour. The
integral vanishes since ${\bf v_k}\cdot d{\bf l_k}=0$ at the BZ
boundary. We deduce from (\ref{div1}-\ref{div2}) 
\begin{equation}
\langle -K\rangle_{T=0} = \int_{\bf k} \frac{\epsilon_{{\bf
        k}\uparrow}\epsilon_{\bf k}}{E_{{\bf k}\uparrow}} =
{\Delta_0^{\rm HS}}^2 \int_{\bf k} \frac{v^2_{\bf
    k}}{E^3_{{\bf k}\uparrow}} .
\end{equation}
In the weak-coupling limit, using ${\Delta_0^{\rm HS}}^2/E^3_{{\bf k}\uparrow}
\equiv 2\delta(\epsilon_{\bf k}-\epsilon_F)$, we obtain
(\ref{kealt}). 

\section{Charge ($\rho_{\bf r}$) and pairing ($\Delta_{\bf r}$)
  fields}
\label{app:rho_delta}

In the section, we relate the charge ($\rho_{\bf r}$) and pairing
($\Delta_{\bf r}$) fluctuations of the attractive model to the fields
$p$, $q$ and $\delta m$ defined in the repulsive model. We rewrite  
$S_J$ [Eq.~(\ref{SJ})] as
\begin{equation}
S_J =  \int_0^\beta d\tau \sum_{\bf r} \phi^\dagger_{\bf r} 
{\bf B}_{0{\bf r}} \phi_{\bf r} ,
\label{SJapp}
\end{equation}
where ${\bf B}_{0{\bf r}}=R^\dagger_{\bf r} {\bf J_r} \cdot
  \boldsymbol{\sigma}R_{\bf 
  r}=\sum_{\nu=x,y,z} B^\nu_{0{\bf r}}\sigma^\nu$ and
\begin{eqnarray}
B^x_{0{\bf r}} &=& (-1)^{\bf r} \cos(\theta_{\bf r}-\theta_0)
\bigl[\cos\varphi_{\bf r} J^x_{\bf r} + \sin\varphi_{\bf r} J^y_{\bf
    r} \bigr] 
-(-1)^{\bf r} \sin(\theta_{\bf r}-\theta_0) J^z_{\bf r} , \nonumber \\
B^y_{0{\bf r}} &=& -(-1)^{\bf r} \sin\varphi_{\bf r} J^x_{\bf r} 
+ (-1)^{\bf r} \cos\varphi_{\bf r} J^y_{\bf r},  \nonumber \\
B^z_{0{\bf r}} &=& \sin(\theta_{\bf r}-\theta_0) \bigl[
\cos\varphi_{\bf r} J^x_{\bf r} + 
\sin\varphi_{\bf r} J^y_{\bf r} \bigr]
+ \cos(\theta_{\bf r}-\theta_0) J^z_{\bf r} . 
\end{eqnarray}

To first order in ${\bf J}$ and second-order in $p,\delta m^{\rm
HS},\delta m$, the effective action is given by $\langle
S_1+S_2+S_J\rangle -\frac{1}{2}\langle (S_1+S_2+S_J)^2\rangle_c$. The
first-order cumulant gives the source-dependent contribution
\begin{equation}
S^{(1)}_J = \int_0^\beta d\tau \sum_{\bf r} \left[ 2 \Delta_0  
  (-1)^{\bf r} B^x_{0{\bf r}} -x B^z_{0{\bf r}} \right] . 
\end{equation}
From the second-order cumulant, we obtain to linear order in the
source 
\begin{eqnarray}
S^{(2)}_J &=& -  \int_0^\beta d\tau d\tau' \sum_{{\bf r},{\bf r}'
  \atop {{\nu,\nu'=x,y,z} \atop {\mu'=0,x,y}}} B^\nu_{0{\bf r}}
  (\tau) \Pi^{\nu\nu'}_{0\mu'}({\bf r},\tau;{\bf r}',\tau')
   A^{\nu(\rm
  tot)}_{\mu '{\bf r}'}(\tau')  \nonumber \\ 
&=&  -  \int_0^\beta d\tau \sum_{\bf r} \Bigl\lbrace (-1)^{\bf r}
  B^x_{0{\bf r}} \Bigl[ 4 \Pi^{xy}_{0x}({\bf Q},{\bf Q}) p_{\bf r}
+\Pi^{xx}_{00}({\bf Q},{\bf Q})(2h_0p_{\bf r}-\sin\theta_0 i\delta
  m^{\rm HS}_{\bf r}) + \Pi^{zx}_{00}(0,{\bf Q})\Bigl(-\frac{i}{2}
  \dot q_{\bf r}-\cos\theta_0 i\delta 
  m^{\rm HS}_{\bf r}\Bigr) \Bigr] \nonumber \\ &&
+ B^z_{0{\bf r}} \Bigl[ 4 \Pi^{zy}_{0x}(0,{\bf Q}) p_{\bf r}
+\Pi^{zx}_{00}(0,{\bf Q})(2h_0p_{\bf r}-\sin\theta_0 i\delta
  m^{\rm HS}_{\bf r}) 
+ \Pi^{zz}_{00}(0,0)\Bigl(-\frac{i}{2} \dot q_{\bf r}-\cos\theta_0 i\delta
  m^{\rm HS}_{\bf r}\Bigr)
\Bigr] \Bigr\rbrace ,
\end{eqnarray}
where the last expression is valid in the hydrodynamic regime. $ A^{\nu(\rm
  tot)}_\mu$ is defined in (\ref{Atot}). In the
presence of the source ${\bf J}$, the effective action $S[p,q,m^{\rm HS},m]$
[Eq.~(\ref{cumu2})] should be supplemented with $S^{(1)}_J+S^{(2)}_J$. The
integration of the $m^{\rm HS}$ field then leads to the
source-dependent action  $S^{(1)}_J+S^{(2)'}_J+S^{(3)}_J$ where
$S^{(2)'}_J=S^{(2)}_J|_{\delta m^{\rm HS}=0}$ and
\begin{eqnarray}
S^{(3)}_J &=& - \frac{i}{\Pi_{m^{\rm HS}m^{\rm HS}}} 
\int_0^\beta d\tau \sum_{\bf r} \Bigl\lbrace (-1)^{\bf r} B^x_{0{\bf
    r}} 
\bigl[ \sin\theta_0 \Pi^{xx}_{00}({\bf Q},{\bf Q}) +
  \cos\theta_0 \Pi^{zx}_{00}(0,{\bf Q}) \bigr] \nonumber \\ && + 
 B^z_{0{\bf r}} \bigl[ \sin\theta_0 \Pi^{zx}_{00}(0,{\bf Q}) +
  \cos\theta_0 \Pi^{zz}_{00}(0,0) \bigr] \Bigr\rbrace  
(\Pi_{m^{\rm HS}p} p_{\bf r} + \Pi^{r}_{m^{\rm HS}q} i \dot
q_{\bf r} + i \delta m_{\bf r} ) .
\label{SJ3}
\end{eqnarray}
We have introduced $\Pi^{r}_{m^{\rm HS}q}= \Pi_{m^{\rm
HS}q}/\omega_\nu$. In Eq.~(\ref{SJ3}), the correlation functions
$\Pi_{m^{\rm HS}m^{\rm HS}}$,$\Pi_{m^{\rm HS}p}$,$\Pi^{r}_{m^{\rm
HS}q}$ are evaluated at $\tilde q=0$. Taking the functional derivative
of $S^{(1)}_J+S^{(2)'}_J+S^{(3)}_J$ with respect to ${\bf J}$, we
finally obtain 
\begin{eqnarray}
\delta\rho_{\bf r} &=& p_{\bf r} \left[ -4\Delta_0 -2h_0
  \Pi^{zx}_{00}(0,{\bf Q}) -4\Pi^{zy}_{0x}(0,{\bf Q}) \right] 
+ \Pi^{zz}_{00}(0,0) \frac{i}{2} \dot q_{\bf r} \nonumber \\ && 
- \frac{i}{\Pi_{m^{\rm HS}m^{\rm HS}}} \left[ \sin\theta_0
  \Pi^{zx}_{00}(0,{\bf Q}) + \cos\theta_0 \Pi^{zz}_{00}(0,0) \right]
  \nonumber (\Pi_{m^{\rm HS}p} p_{\bf r} + \Pi^r_{m^{\rm
  HS}q} i \dot q_{\bf r} +i \delta m_{\bf r} ), \nonumber \\
\delta |\Delta_{\bf r}| &=& p_{\bf r} \left[ -x -h_0
  \Pi^{xx}_{00}({\bf Q},{\bf Q}) -2\Pi^{xy}_{0x}({\bf Q},{\bf Q})
  \right] + \Pi^{zx}_{00}(0,{\bf Q}) \frac{i}{4} \dot
  q_{\bf r} \nonumber \\ &&  
- \frac{i}{2\Pi_{m^{\rm HS}m^{\rm HS}}}
  \left[ \sin\theta_0 
  \Pi^{xx}_{00}({\bf Q},{\bf Q}) + \cos\theta_0 \Pi^{zx}_{00}(0,{\bf
  Q}) \right] 
 (\Pi_{m^{\rm HS}p} p_{\bf r} + \Pi^r_{m^{\rm
  HS}q} i \dot q_{\bf r} +i \delta m_{\bf r} ), \nonumber \\
\Theta_{\bf r} &=& -q_{\bf r} .
\label{transformation3} 
\end{eqnarray}

\subsection{Strong-coupling limit}
\label{subapp:scl}
 
In the strong-coupling limit, we can obtain a simple relation
between $\rho_{\bf r},\Delta_{\bf r}$ and $m_{\bf r}{\bf\Omega_r}$. In
Sec.~\ref{sec:scr}, we show that the Hubbard model reduces to the
Heisenberg model when $U\gg 4t$. We can carry out the same derivation
in the presence of the source term (\ref{SJapp}). Integrating out the
fermions, we obtain the source-dependent term
\begin{equation}
S_J = \int_0^\beta d\tau \sum_{\bf r} \langle \phi^\dagger_{\bf r}
R^\dagger_{\bf r} {\bf J_r} \cdot \boldsymbol{\sigma} R_{\bf r}
\phi_{\bf r} \rangle_{\rm at}
\label{SJapp2}
\end{equation}
to leading order in $1/U$. The average in (\ref{SJapp2}) is taken with
respect to the atomic action (\ref{Sat}). The matrix $R_{\bf r}$
satisfies $R_{\bf r}\sigma^z R^\dagger_{\bf r}= \boldsymbol{\sigma} \cdot
{\bf\Omega_r}$ (see Sec.~\ref{subsec:Heis}). Using $\langle
\phi^\dagger_{\bf r}\sigma^\nu \phi_{\bf r}\rangle_{\rm
  at}=\delta_{\nu,z}$ and $R^\dagger_{\bf r}
{\bf J_r} \cdot \boldsymbol{\sigma} R_{\bf r}= \boldsymbol{\sigma}
\cdot {\cal R}_{\bf r}^{-1} {\bf J_r}$, where 
\begin{equation}
{\cal R}_{\bf r} = 
\left (
\begin{array}{ccc}
 \cos\theta_{\bf r}\cos\varphi_{\bf r} & -\sin\varphi_{\bf r} &
  \sin\theta_{\bf r}  \cos\varphi_{\bf r}
\\ 
 \cos\theta_{\bf r}\sin\varphi_{\bf r} & \cos\varphi_{\bf r} &
  \sin\theta_{\bf r}  \sin\varphi_{\bf r}
\\
-\sin\theta_{\bf r} & 0 & \cos\theta_{\bf r} 
\end{array}
\right ) 
\label{Rcal}
\end{equation}
is the SO(3) rotation matrix
which maps $\hat {\bf z}$ onto ${\bf\Omega_r}$, we obtain  
\begin{equation}
S_J = \int_0^\beta d\tau \sum_{\bf r} {\bf J_r} \cdot {\bf\Omega_r} .
\end{equation}
From Eqs.~(\ref{transformation2}) we then deduce
\begin{eqnarray}
\rho_{\bf r}-1 &=& \Omega^z_{\bf r}, \nonumber \\
\Delta_{\bf r} &=& \frac{(-1)^{\bf r}}{2} \Omega^-_{\bf r} ,
\end{eqnarray}
where $\Omega^\pm_{\bf r}=\Omega^x_{\bf r} \pm i\Omega^y_{\bf r}$.

\section{HF Cumulants $\langle S_p+S_{h_0}+S_{\delta m}\rangle$ and $\langle
  (S_{h_0}+S_{\delta m})^2+2(S_{h_0}+S_{\delta m})(S_p+S_l) \rangle_c$} 
\label{app:O3} 

In this appendix, we calculate $\langle S_p+ S_{h_0}+S_{\delta m}\rangle$ and
$\langle (S_{h_0}+S_{\delta m})^2+2(S_{h_0}+S_{\delta m})(S_p+S_l) \rangle_c$
(Sec.~\ref{sec:O3}). The first-order cumulants read
\begin{eqnarray}
\langle S_p \rangle &=&  -\int_0^\beta d\tau \sum_{\bf r} A^z_{0{\bf
    r}} \langle j^z_{0{\bf r}} \rangle \nonumber \\ 
&=& - i \frac{m_0}{U} \int_0^\beta d\tau \sum_{\bf r} (-1)^{\bf r} (
    \dot\varphi_{\bf r} \cos\theta_{\bf r} + \dot\psi_{\bf r} ) ,
    \nonumber \\  
\langle S_{h_0} \rangle &=& -h_0 \int_0^\beta d\tau \sum_{\bf r}
B^z_{0{\bf r}} \langle j^z_{0{\bf r}} \rangle \nonumber \\ 
&=& - \frac{2h_0 m_0}{U} \int_0^\beta d\tau \sum_{\bf r} (-1)^{\bf r}
n^z_{\bf r},
\end{eqnarray}
and 
\begin{eqnarray}
\langle S_{\delta m} \rangle &=& - \int_0^\beta d\tau \sum_{\bf r}
(-1)^{\bf r} \delta m_{\bf r}  \langle j^z_{0{\bf r}} \rangle
\nonumber \\  &=& - \frac{2m_0}{U} \int_0^\beta d\tau \sum_{\bf r}
\delta m_{\bf r} ,
\label{Sdm}
\end{eqnarray}
where we have used $n^z_{\bf r}=\cos\theta_{\bf r}$ and the saddle-point
equation $2m_0/U=(-1)^{\bf r} \langle c^\dagger_{\bf r}\sigma^z c_{\bf
  r}\rangle$. Since ${\bf n_r}$ is slowly varying, $\langle
S_{h_0}\rangle$ vanishes. From (\ref{Sdm},\ref{SnLdm}), we conclude that there
is no linear contribution in $\delta m$. 

Let us now consider the second-order cumulant
\begin{eqnarray}
\langle S_{h_0} S_{\delta m} \rangle &=& h_0 \int_0^\beta d\tau d\tau'
\sum_{{\bf r},{\bf r}' \atop {\nu=x,y,z}} B^\nu_{0{\bf r}}(\tau)
\Pi^{\nu z}_{00}({\bf r},\tau;{\bf r}',\tau') 
(-1)^{{\bf r}'} \delta m_{{\bf r}'} \nonumber \\ &=& h_0 \sum_{\tilde
  q,\tilde q' \atop 
  {\nu=x,y,z}} B^\nu_0(-\tilde q) \Pi^{\nu z}_{00}(\tilde q,\tilde q')
\delta m_{\tilde q'+{\bf Q}} .
\end{eqnarray}
$\Pi^{\nu\nu'}_{\mu\mu'}$ is the HF current-current correlation
function for $h_0=0$ and an 
AF order parallel to the $z$ axis (i.e. ${\bf\Omega}^{\rm cl}_{\bf
  r}=(-1)^{\bf r} \hat {\bf z}$). It is given in Appendix
\ref{app:hfcorr2}. 
Since $B^\nu_{0{\bf r}}$ is slowly varying, we can evaluate
$\Pi^{\nu\nu'}_{\mu\mu'}(\tilde q,\tilde q')$ at 
$\tilde q=0$ in order to obtain the result to second order in $h_0$,
$\partial_\mu$ and $\delta m$. Since $\Pi^{\nu z}_{00}(\tilde q=0,\tilde
q')=0$ (Appendix \ref{app:hfcorr2}), $\langle S_{h_0} S_{\delta m}
\rangle$ vanishes. A similar calculation shows that $\langle S_p S_{\delta m}
\rangle=\langle S_l S_{\delta m} \rangle=0$. We therefore conclude
that amplitude fluctuations decouple in the limit of a weak magnetic
field.  

The contribution due to $S_{h_0}^2$ is given by 
\begin{eqnarray}
\langle S^2_{h_0} \rangle &=& h_0^2 \int_0^\beta d\tau d\tau' 
\sum_{{{\bf r},{\bf r}'} \atop {\nu,\nu'=x,y,z}} 
B^\nu_{0{\bf r}}(\tau) \Pi^{\nu\nu'}_{00}({\bf r},\tau;{\bf r}',\tau')
B^{\nu'}_{0{\bf r}'}(\tau') \nonumber \\  
&=& h_0^2 \sum_{{\tilde q,\tilde q'} \atop {\nu,\nu'=x,y,z}}  
B^\nu_0(-\tilde q)
\Pi^{\nu\nu'}_{00}(\tilde q,\tilde q')  B^{\nu'}_0(\tilde q') .
\end{eqnarray} 
Since $B^\nu_{0{\bf r}}$ is slowly varying, we can evaluate
$\Pi^{\nu\nu'}_{00}(\tilde q,\tilde q')$ at $\tilde q=\tilde q'=0$: 
\begin{equation}
\langle S^2_{h_0} \rangle = h^2_0 \Pi^{xx}_{00} \int_0^\beta
d\tau \sum_{\bf r} \sin^2\theta_{\bf r} .
\end{equation}
Here and in the following, we use the notation
$\Pi^{\nu\nu'}_{00}\equiv \Pi^{\nu\nu'}_{00}(\tilde q=0,\tilde q'=0)$.  
We have used $\Pi^{\nu\nu'}_{00}\propto
\delta_{\nu,\nu'}(\delta_{\nu,x}+\delta_{\nu,y})$.
A similar calculation gives
\begin{eqnarray}
\langle S_{h_0} S_p \rangle &=& h_0 \Pi^{xx}_{00} \int_0^\beta d\tau
\sum_{\bf r} \frac{i}{2} \sin^2\theta_{\bf r} \dot \varphi_{\bf r} , 
\nonumber \\
\langle S_{h_0} S_l \rangle &=& h_0 m_0 \Pi^{xx}_{00} \int_0^\beta d\tau
\sum_{\bf r}\sin\theta_{\bf r} (-\cos\psi_{\bf r} l^x_{\bf r} +
  \sin\psi_{\bf r} l^y_{\bf r} ) .
\end{eqnarray}
To express $\langle S_{h_0}^2\rangle$, $\langle S_{h_0}S_p\rangle$ and
$\langle S_{h_0}S_l\rangle$ in terms of ${\bf n}$ and ${\bf l}$, we
use
\begin{eqnarray}
({\bf h}_0\times {\bf n_r})^2 &=& h^2_0 \sin^2\theta_{\bf r} , \nonumber \\
{\bf h}_0 \cdot ({\bf n_r} \times \dot {\bf n}_{\bf r}) &=& h_0^2 \dot
\varphi_{\bf r} \sin^2\theta_{\bf r} , \nonumber \\
L^z_{\bf r} &=& \sin\theta_{\bf r} (-\cos\psi_{\bf r} l^x_{\bf r} +
\sin\psi_{\bf r} l^y_{\bf r} ) . \label{Lz}
\end{eqnarray}
The last result follows from  ${\bf L_r}={\cal R}_{\bf r}{\bf l_r}$ and 
$l^z_{\bf r}=0$, where ${\cal R}_{\bf r}$ is given by
\begin{equation}
{\cal R}_{\bf r} = 
\left (
\begin{array}{ccc}
 \cos\theta_{\bf r}\cos\varphi_{\bf r}\cos\psi_{\bf r} 
-\sin\varphi_{\bf r} \sin\psi_{\bf r}
& 
-\cos\theta_{\bf r}\cos\varphi_{\bf r}\sin\psi_{\bf r}
-\sin\varphi_{\bf r} \cos\psi_{\bf r} & \sin\theta_{\bf r} \cos\varphi_{\bf r}
\\ 
\cos\theta_{\bf r} \sin\varphi_{\bf r} \cos\psi_{\bf r} 
+ \cos\varphi_{\bf r}\sin\psi_{\bf r}
& - \cos\theta_{\bf r}\sin\varphi_{\bf r}\sin\psi_{\bf r}
+\cos\varphi_{\bf r}\cos\psi_{\bf r}  &
 \sin\theta_{\bf r} \sin\varphi_{\bf r}
\\
-\sin\theta_{\bf r}\cos\psi_{\bf r} & \sin\theta_{\bf r} \sin\psi_{\bf
  r} & \cos\theta_{\bf r}  
\end{array}
\right ) .
\end{equation}
We therefore obtain
\begin{eqnarray}
\langle S^2_{h_0}\rangle  &=& \Pi^{xx}_{00} \int_0^\beta d\tau
\sum_{\bf r}  ({\bf  h}_0\times {\bf n_r})^2 , \nonumber \\
\langle S_{h_0} S_p \rangle &=& \frac{i}{2} \Pi^{xx}_{00}
\int_0^\beta d\tau \sum_{\bf r} {\bf h}_0 \cdot ({\bf n_r} \times
    \dot {\bf n}_{\bf r}) , \nonumber \\
\langle S_{h_0} S_l \rangle &=& m _0 \Pi^{xx}_{00}
\int_0^\beta d\tau \sum_{\bf r} {\bf h}_0 \cdot {\bf L_r} .
\end{eqnarray}

\end{widetext}


\begin{thebibliography}{99}
\bibitem{note7} It seems easier to determine the coefficients $\Pi_{\rho\rho}$,
$\Pi_{|\Delta||\Delta|}$ and $\Pi_{\rho|\Delta|}$ by working
  directly in the attractive Hubbard model and introducing 
  Hubbard-Stratonovich fields both in the particle-particle and
  particle-hole channels as done in Ref.~\onlinecite{DePalo99}. This
  approach, however, suffers from the 
  difficulties mentioned after Eqs.~(\ref{transformation}) in
  Sec.~\ref{sec:O2}.
\bibitem{note3} Note that the $m$ field in Sec.~\ref{sec:O3} plays the
  role of the $m^{\rm HS}$ field introduced in Sec.~\ref{sec:O2}. See
  the discussion after Eq.~(\ref{decomp1}).
\bibitem{note1} Note that in Ref.~\onlinecite{KB2}, the
  contribution $S_{\delta m}$ due to amplitude fluctuations was not
  considered. It is well known that for $h_0=0$, amplitude and
  direction fluctuations decouple in the long wavelength limit, as
  follows from 
  $\langle S_p S_{\delta m}\rangle =0$ and $\langle S_l S_{\delta
  m}\rangle =0$ (Appendix \ref{app:O3}).
\bibitem{Kostyrko92a} The mapping onto the Heisenberg model and the
  collective mode energy [Eq.~(\ref{dispersion_sc})] have also been
  obtained in Ref.~\onlinecite{Kostyrko92}.
\end{thebibliography}
\end{document}